\documentclass{aa} 

\usepackage{graphicx}
\usepackage{txfonts}
\usepackage{natbib}
\usepackage{longtable}
\bibpunct{(}{)}{;}{a}{}{,} % to follow the A&A style

\usepackage[normalem]{ulem}
\usepackage{bm}
\newcommand{\micron}{\ensuremath{\mu}m}

\newcommand{\teff}{\ensuremath{T_\textrm{eff}}\xspace}
\newcommand{\re}{\ensuremath{R_\textrm{e}}\xspace}
\newcommand{\logg}{\ensuremath{\log(\rm g)}\xspace}
\newcommand{\mh}{[M/H]\xspace}
\newcommand{\kms}{km\,s\ensuremath{^{-1}}\xspace}%
\newcommand{\msun}{\ensuremath{M_{\sun}}\xspace}
\newcommand{\vlos}{\ensuremath{V_\textrm{LOS}}\xspace}
\newcommand{\dlos}{\ensuremath{d_\textrm{LOS}}\xspace}
\newcommand{\slos}{\ensuremath{\sigma_{\textrm{LOS}}}\xspace}
\newcommand{\sgra}{Sgr~A$^\star$\xspace}
\newcommand{\sgr}{Sgr~A$^\star$\xspace}

\newcommand*\mean[1]{\overline{#1}}
\newcommand{\col}{\ensuremath{H-K_{S}}\xspace}
\newcommand{\ak}{\ensuremath{A_{K_S}}\xspace}

\begin{document}

 \title{A spectroscopic map of the Galactic centre}

 \subtitle{Observations and resolved stars }

 \author{A. Feldmeier-Krause 
 \inst{1}\fnmsep
 \inst{2} 
 \and
 N. Neumayer\inst{2}
  \and
 A. Seth\inst{3}
  \and
 G. van de Ven\inst{1}
  \and
 M. Hilker\inst{4}
  \and
 M. Kissler-Patig\inst{5}
  \and
 H. Kuntschner\inst{4}
  \and
 N.~L{\"u}tzgendorf\inst{6}
 \and
 A. Mastrobuono-Battisti\inst{7}
 \and
 F. Nogueras-Lara\inst{4}
 \and
 H. B. Perets
 \inst{8}
 \and
  R. Sch{\"o}del
 \inst{9}
 \and
  A. Zocchi
 \inst{1}
 }

 \institute{Department of Astrophysics, University of Vienna, T\"urkenschanzstrasse 17, 1180 Wien, Austria\\
 \email{anja.krause@univie.ac.at}
 \and
 Max Planck Institute for Astronomy, K\"onigstuhl 17, D-69117 Heidelberg, Germany
\and 
Department of Physics and Astronomy, University of Utah, Salt Lake City, UT 84112, USA
\and
European Southern Observatory, Karl-Schwarzschild-Strasse 2, 85748 Garching bei München, Germany
\and 
ESA—ESAC—European Space Agency, Camino Bajo del Castillo s/n, 28692 Villafranca del Castillo, Madrid, Spain
\and 
European Space Research and Technology Centre, Keplerlaan 1, 2200 AG Noordwijk, The Netherlands
\and
Dipartimento di Fisica e Astronomia “Galileo Galilei”, Univ. di Padova, Vicolo dell’Osservatorio 3, Padova 35122, Italy
\and
Technion – Israel Institute of Technology, Haifa, 3200002, Israel
\and
Instituto de Astrofísica de Andalucía (CSIC), Glorieta de la Astronomía s/n, 18008 Granada, Spain
 }

 \date{Received December 12, 2024; accepted March 10, 2025}

% \abstract{}{}{}{}{} 
% 5 {} token are mandatory
 
 \abstract
 % context heading (optional)
 {The Galactic Centre region contains a dense accumulation of stars, which can be separated into two components: A mildly flattened and extremely dense nuclear star cluster (NSC), and a surrounding, more extended and more flattened, nuclear stellar disc (NSD). 
  Previous studies have collected a few thousand spectra of the inner NSC, and also the outer NSD, and measured line-of-sight velocities and metallicities. Until now, such measurements exist only for a few 100 stars in the region where the stellar surface density transitions from being dominated by the NSC into being dominated by the NSD.} 
   % aims heading (mandatory)
 {   We want to study the stellar population from the centre of the NSC out to well beyond its effective radius, where the NSD dominates. In this way, we can investigate whether and how the mean properties and kinematics of the stars change systematically.
}
 % methods heading (mandatory)
 {  We conducted spectroscopic observations with Flamingos-2 in the $K$-band via a continuous slit-scan. The data extend from the central NSC into the inner NSD, out to $\pm$32\,pc from \sgra along Galactic longitude $l$. 
Based on their CO equivalent width we classify the stars as hot or cool stars. The former are massive, young stars, while almost all of the latter are older than one to a few gigayears.  
 Applying full-spectral fitting, we measure the overall metallicity \mh and line-of-sight velocity \vlos for \textgreater 2\,500 cool stars, increasing 
existing samples outside of the very centre by a factor of 3 in terms of number of stars, and  by more than an order of magnitude in terms of covered area. 
We present the first continuous spatial maps and profiles of the mean value of various stellar and kinematic parameters. }
 % results heading (mandatory)
 {We identify hot, young stars across the field of view. Some stars appear to be isolated from other hot stars, while others accumulate within 2.7\,pc of the Quintuplet cluster, or the central parsec cluster. 
The position-velocity curve of the cool stars shows no dependence on \mh, but it depends on the colour of the stars. The colour may be a tracer of the line-of-sight distance, and thus distinguish stars located in the NSC from those in the NSD. 
 A subset of the cool stars has high velocities \textgreater150\,\kms, they may be associated with the bar or tidal tails of star clusters. }
 % conclusions heading (optional), leave it empty if necessary 
% {conclusions}
 {}

 \maketitle
%
%-------------------------------------------------------------------

\section{Introduction}

The Galactic Centre is an extremely dense environment, with stellar densities several orders of magnitude higher than in the Galactic disc, a content of a few percent of the Milky Way's molecular gas, and a star formation density that is more than two magnitudes  higher than elsewhere in our galaxy. These properties make the Galactic Centre a region of special astrophysical interest.
Because of extreme crowding and interstellar extinction, the Galactic Centre is still far less explored by the great spectroscopic surveys than other parts of our Galaxy. 
Due to the high extinction, stellar populations can only be studied in the infrared. However,  their intrinsic stellar colour differences are significantly smaller than the extreme, differential reddening the stars suffer on the line-of-sight to the Galactic Centre, which impedes e.g. colour-magnitude diagram analyses. To study the stellar population in the Galactic Centre we therefore require infrared spectroscopy. Unfortunately, such data are sparse for the Galactic Centre, either targeted on individual bright stars \citep[e.g.][]{2021A&A...649A..83F,2022ApJS..259...35A}  or concentrated on the inner parsec(s) \citep[e.g.][]{2014A&A...570A...2F,2015ApJ...809..143D,2016ApJ...821...44F,2022ApJ...932L...6V}.
We are therefore missing several pieces of the puzzle to understand the formation and evolution of the inner part of the Milky Way.

The Galactic Centre stellar content is sometimes referred to as nuclear bulge which can be separated into two components: The more extended nuclear stellar disc (NSD) and the nuclear star cluster (NSC) within \citep{2002A&A...384..112L}.

The NSD appears as a thick disc, that extends out to distances $r$$\sim$200\,pc or more, with a radial scale length of $\sim$90\,pc and a scale height of $\sim$28-45\,pc 
\citep{2002A&A...384..112L,2013ApJ...769L..28N,2014A&A...566A..47S,
2020A&A...634A..71G,2022MNRAS.512.1857S}. 
The flattening $q$ (minor/major axis ratio) is $\sim$0.35, and the total mass derived from dynamical modelling is of the order of a few $10^8$ to $10^9$\,\msun\ \citep{2020MNRAS.499....7S,2022MNRAS.512.1857S}.

The NSC on the other hand is less flattened ($q$=0.66-0.8) and smaller with an effective radius $R_e\sim$5\,pc \citep{2002A&A...384..112L,2014A&A...566A..47S,2016ApJ...821...44F,
2020A&A...634A..71G}. Its total mass is of the order of a few $10^7$ \msun\ \citep{2014A&A...570A...2F,2015MNRAS.447..948C,2016ApJ...821...44F,2017MNRAS.466.4040F}.
The NSC contains the nearest supermassive black hole to us, \sgra. 

The stars in the innermost $\sim$1\,pc region have been monitored over decades at the highest available spatial resolution \citep[e.g.][]{2009ApJ...703.1323D,2014ApJ...783..131Y,2017ApJ...847..120H,2022ApJ...932L...6V}, and we have a deep knowledge of the stellar types and their 3-dimensional kinematics.
The more extended NSC and NSD are less well understood. Due to their large extent, observations are usually seeing-limited and hence restricted to brighter stars compared to the innermost $\sim$1\,pc \citep{2017MNRAS.464..194F,2018A&A...610A..83N,
2020MNRAS.494..396F,2021A&A...649A..83F,2022MNRAS.513.5920F}.

Most stars in the Galactic Centre are observed as red giant stars and are several Gyr old, though age estimates can differ by a few Gyrs  \citep{2003ApJ...597..323B,2011ApJ...741..108P,2020A&A...641A.102S,2020NatAs...4..377N,2023ApJ...944...79C}. 
Hot young stars have also been discovered (e.g., O and B type main sequence, supergiant stars, Wolf Rayet stars, and emission line stars). These young, massive stars appear to be separated into two groups. On the one hand, there are three massive ($\gtrsim10^4$\,\msun) clusters of young stars:  The central parsec cluster, located in the very centre of the NSC, the Arches and the Quintuplet clusters, located on the east side of the NSD. Their stars are only a few Myr old \citep{1999ApJ...514..202F,2004ApJ...611L.105N,2006ApJ...643.1011P,2012A&A...540A..14L,2013ApJ...764..155L,2018A&A...617A..65C,2018A&A...618A...2C}. 
On the other hand, several dozens of apparently isolated young, massive stars have been detected throughout the NSD
\citep[e.g.][]{1999ApJ...510..747C,2010ApJ...710..706M,2010ApJ...725..188M,2011MNRAS.417..114D,2021A&A...649A..43C,2022MNRAS.513.5920F}, and the census of hot stars in the Galactic Centre is far from complete. 
There have been attempts to identify hot stars via their narrow-band photometry \citep{2009A&A...499..483B,2011MNRAS.417..114D,2013A&A...549A..57N,2018MNRAS.476.3974P,2023ApJ...951..148N,2024A&A...689A.190G}. This method allows to access larger areas and fainter stars at lower observational costs compared to spectroscopy. However, the hot star identification is less reliable, as lower mass, intermediate-age stars can be mis-identified as hot young star candidate \citep{2016A&A...588A..49N}, and hence  spectroscopy is required to confirm the stellar type.

The several Gyr old red giant stars are the most numerous group of stars that are accessible for spectroscopy.
Their metallicity distribution is broad, ranging from sub- to super-solar values, with super-solar mean metallicities \citep{2015ApJ...809..143D,2017AJ....154..239R,2017MNRAS.464..194F,2020MNRAS.494..396F,2020ApJ...894...26T,2021A&A...649A..83F,2022MNRAS.513.5920F}.
There is some evidence that the mean metallicity decreases from the inner NSC to the outer NSD and the bulge beyond \citep{2021A&A...650A.191S,2022MNRAS.513.5920F,2023A&A...680A..75N}, and that the age decreases as a function of distance from \sgra\ in the NSD \citep{2023A&A...671L..10N,2024A&A...681L..21N}. 
However, only limited data are available in the transition region, where the NSC stellar density drops to a value below the NSD, and the NSD stars become dominant. \cite{2022MNRAS.513.5920F} present data of two fields located 20\,pc away from \sgra. Still, there is no continuous spectroscopic coverage, and metallicity gradients and the velocity curve are based on fields located several parsecs apart. A continuous coverage of this region is of great interest to constrain the gravitational potential. There is also some debate about whether the NSC and the NSD are different entities or part of the same structure \citep{2023A&A...680A..75N}.

In this study, we present the so far largest continuous spectroscopic data covering the NSC and the NSD out to 32\,pc to the east and to the west of \sgra along Galactic longitude, extending about 2--3\,pc along Galactic latitude. We extract the spectra of the brightest stars, identify several hot stars, and measure the line-of-sight velocity \vlos\ and metallicity \mh of \textgreater2\,500 cool stars. We show the first continuous data on stellar metallicity and kinematics from the centre of the NSC, across the transition region to the NSD, and out to distances where the NSD dominates fully.

This paper is organised as follows: we present the data set including data reduction in Sect. \ref{sec:data}, and we describe the analysis steps in Sect. \ref{sec:analysis}. We present our hot star candidates in Sect. \ref{sec:hot} and our red giant star kinematic and stellar population measurements in Sect. \ref{sec:ltresult}. We discuss our findings in Sect. \ref{sec:discussion}, and conclude in Sect. \ref{sec:conclusions}.

%--------------------------------------------------------------------
\section{Spectroscopic data}
\label{sec:data}
\subsection{Observations}
\begin{figure*}
 \centering
 \includegraphics[width=18cm]{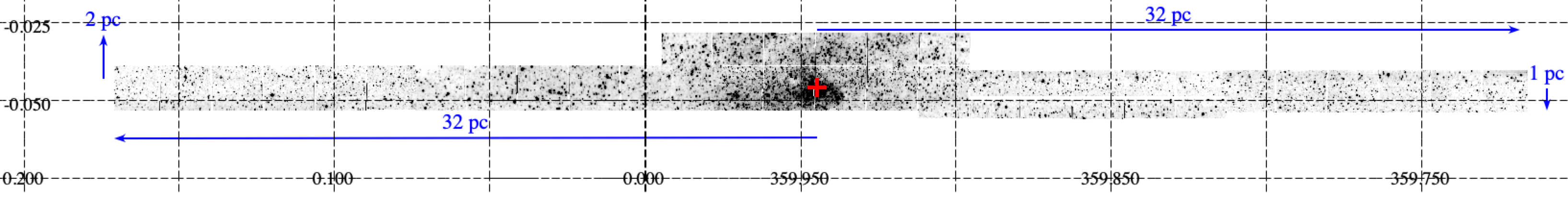}
 \caption{Spatial coverage of our observations. The data extend $\sim$32\,pc to the Galactic east and west of \sgra (marked as a red plus symbol), about 1\,pc to the Galactic north and south, except for the centre region, which extends about 2\,pc to the Galactic north. The image is a white light reconstructed image mosaic of our spectroscopic scans. }
 \label{fig:fov}
 \end{figure*}

We observed the Galactic Centre on five nights (June 24, 25, 26, 27, 29, 2015) with Flamingos-2 \citep[F2,][]{2004SPIE.5492.1196E}, a near-infrared imaging spectrograph at the Gemini South telescope. The observed five regions (Inner West, Outer West, Inner East, Outer East, and Centre) are centred on \sgra, and extend 32 pc to the Galactic east and 32 pc to the Galactic west; see also Fig.~\ref{fig:fov}. We list further details on the observed regions in Table \ref{tab:regions}.

To cover a continuous region we resorted to long slit scans. F2 is a multi-object spectrograph (MOS) that uses custom masks. Since the standard F2 long slit masks are only 4\farcm4 long, we designed a special mask that resembles a 6\arcmin\space long slit. This was achieved by cutting six $\lesssim$1\arcmin\space small slits or slitlets, aligned in a single row, with five small stabilising connectors that were not cut.  Our slitlets are 1 pixel wide (0\farcs18). The data were observed with the $K$-long filter ($\sim$1.906--2.472 \micron\space with 80\% transmission) and R3K grism, with a maximum spectral resolution of $R$=3\,400. 
We used the dark readout mode with 8 reads per exposure.

The observing strategy was as follows: For acquisition we usually took two images (6\arcmin\ field-of-view), to be used for astrometric calibration. After inserting the slit mask with the slit aligned parallel to the Galactic plane, we took a series of five short (2\,s) dark exposures to flush the detector and weaken the afterglow or persistence signal left from bright stars in the acquisition images. Then we started a series of usually 10 or 20 exposures of 300\,s each, as listed in Table ~\ref{tab:regions}. The exact number of exposures per series varies from 6--22, as we sometimes had to stop a series early due to reaching zenith or approaching high airmass. Throughout the five nights, we observed 50 exposures per region, except for the central region, which extends further North, where we observed 87 exposures combined.
During the observations, we drift-scanned the telescope slowly from Galactic north to Galactic south, with a rate of 1\arcsec\space per 300\,s, meaning each exposure covers a different $\sim$6\arcmin$\times$1\arcsec\space region of the sky. After an exposure series, i.e. four times per night, we made an offset to a dark sky field and took a series of four sky exposures. We observed early A-type dwarf stars as telluric standard stars (HD171296, HD175892) every 2-3 hours with the standard 1-pixel-wide long-slit mask and by offsetting along the slit with up to five exposures per star to account for the varying spectral resolution along the slit direction of the detector.

\begin{table}
 \caption{List of F2 spectroscopic observations, sorted by time. }
 \label{tab:regions}
\begin{tabular}{@{}lllll@{}}
\noalign{\smallskip}
\hline
\noalign{\smallskip}
Region&RA &Dec& Night & No. of \\
&hh:mm:ss&dd:mm:ss&&exp.\\
 \noalign{\smallskip}
\hline
\noalign{\smallskip} 
Inner West&17:45:27.3&-29:04:32.1&2015-06-24&10\\
Inner West&17:45:28.2&-29:04:39.8&2015-06-24&19\\
Inner West&17:45:29.9&-29:04:52.1&2015-06-24&21\\
Outer West&17:45:13.0&-29:09:35.0&2015-06-24&6\\
\hline 
Inner East&17:45:50.2&-28:56:10.6&2015-06-25&10\\
Inner East&17:45:51.2&-28:56:19.4&2015-06-25&20\\
Inner East&17:45:52.2&-28:56:27.5&2015-06-25&13\\
Inner East&17:45:52.9&-28:56:34.2&2015-06-25&7\\%
\hline
Outer West&17:45:13.4&-29:09:37.4&2015-06-26&10\\
Outer East&17:46:04.6&-28:51:10.3&2015-06-26&20\\
Outer East&17:46:06.6&-28:51:27.8&2015-06-26&20\\
Outer East&17:46:06.8&-28:51:28.9&2015-06-26&10\\
\hline
Outer West&17:45:14.1&-29:09:42.4&2015-06-27&12\\
Outer West&17:45:15.2&-29:09:51.0&2015-06-27&22\\
Centre&17:45:39.2&-29:00:18.6&2015-06-27&20\\
Centre&17:45:40.0&-29:00:25.3&2015-06-27&6\\
\hline 
Centre&17:45:40.6&-29:00:27.5&2015-06-29&10\\
Centre&17:45:41.3&-29:00:34.0&2015-06-29&15\\
Centre North&17:45:37.7&-29:00:07.2&2015-06-29&20\\
Centre North&17:45:36.5&-28:59:56.9&2015-06-29&16\\%
\hline 
\end{tabular}
\tablefoot{For each exposure series, we denote the location of the region (relative to \sgra in Galactic coordinates), the absolute right ascension (RA) and declination (Dec) coordinates, the night of the observation, and the number of exposures taken.}
\end{table}

\subsection{Data reduction}
\label{sec:dr}
The data were reduced with a combination of the \textsc{gemini} \textsc{IRAF} package, the ESO tool \textsc{molecfit}, and custom-made \textsc{IDL} and \textsc{python} scripts. 

\subsubsection{Basic reduction steps}
We reduced the dark, flat, and arc exposures using \textsc{IRAF} and created Masterdarks, Masterflats, and wavelength solutions. The flats and arcs were cut into six pieces, one for each slitlet, as indicated by the MOS mask (using \textsc{f2cut}). Also, the short and long wavelength ends, where the transmission is below $\sim$80\%, were cut off. 
To create the Masterflat, we used \textsc{nsflat}, for the arc solution \textsc{nswavelength} and a list of Argon lines in vacuum. 

We applied dark subtraction on the object and sky data frames with the standard \textsc{gemarith} tool. Then we subtracted the persistence signal, which originates from the acquisition images, from the object spectral exposures, 
for details see Appendix \ref{sec:pers}.
Next, the data frames were cut along the slit direction using the MOS masks (\textsc{f2cut}) into six pieces, 
and divided by the flat fields (\textsc{nsreduce}), which were cut in the same way. The four exposures of each sky series were combined into one Mastersky using the ``crreject'’ algorithm in \textsc{gemcombine}, thus rejecting the brightest pixels of the set, which can be cosmic rays (CRs) or stars in the sky field. Next, we removed any remaining CRs or hot pixels with L.A. cosmic \citep{2001PASP..113.1420V}. We kept two versions of each file, one with CR rejection, and one without. The reason is that L.A. cosmic sometimes over-corrected bright stars. We used the uncorrected files for extracting bright stars, and the CR-corrected files for unresolved faint stars, which will be analysed in a separate paper (Feldmeier-Krause et al. in prep.). We rectified all files and applied the wavelength calibrations derived from the arc files (\textsc{nstransform}). We also rectified the s-distortion, which we derived with \textsc{nssdist} from the many stars on the data themselves. 

\subsubsection{Sky line wavelength calibration correction}
\label{sec:lcalcorr}
We took arc exposures each night, but in the course of a night, small wavelength shifts can occur. For this reason, we refined our wavelength calibration using the sky lines on our exposures and cross-correlating the spectra with a reference exposure.

As reference exposure, we chose the last exposure taken on the night of 2015-06-26, as it was taken closest in time to the arc exposure of that night. For each exposure and slitlet, we normalised the 2-dimensional spectral frame by the median flux of each pixel row along the slit and then summed the flux to obtain a 1-dimensional spectrum per exposure and slitlet that is dominated by the sky rather than bright stars. We then cross-correlated this sky-dominated spectrum with the reference sky-dominated spectrum in 14 wavelength regions ranging from 1.945--2.425\,\micron, each 0.03\,\micron\space wide. The shifts were usually \textless2.5 pixels ($\sim$8.82\,\AA) and varied only on sub-pixel scales as a function of wavelength (usually $\sim$0.3\,\AA\ from 2.0--2.4\,\micron).
For each exposure, we fitted a second-order polynomial to the shifts as a function of wavelength. The mean of the standard deviation of the fit residuals is 0.26\,\AA. During the spectrum extraction (Sect. \ref{sec:sextr}), we used this polynomial to compute the corrected wavelength calibration for each exposure and slitlet and resampled the spectra on the corrected wavelength scale.

\subsubsection{Sky subtraction}
For sky subtraction we used the \textsc{IDL} code \textsc{skysub} by \cite{2007MNRAS.375.1099D}, with some adjustments. \textsc{skysub} uses the 2-dimensional science frames and the Mastersky frame taken close in time. The two files are cross-correlated to align the wavelength scales, and the best scaling factors for the OH sky lines in different wavelength segments are found to correct for changes in the sky emission. Both the skylines and thermal background can then be subtracted. We ran this procedure for each of the six individual slitlets per exposure. In principle, the scaling factors should be the same for the six slitlets, as they were observed in the same exposure, at the same time. However, each slitlet has a different distribution of stars, and a high number of stars can compromise the background estimation. We therefore combined the six scaling factors of each exposure to a single one, using the median. This way, we ensure that our sky correction is robust and we have no slitlets with strong outliers. The scaling factors should vary smoothly as a function of time for subsequently taken exposures. We modelled the scaling factors with a low degree polynomial for each of the series of 6--22 exposures. Then we applied these scaling factors to the sky exposures to create the optimal sky and subtract it from the data. With this approach, we ensure the sky residuals are comparable for subsequent exposures, as we remove potential outliers and minimise any bias that can lead to over- or under-subtracted sky. 

\subsubsection{Telluric correction}
Each of our telluric observations is a series of several exposures, dithered along the slit. We reduced the data in the standard way for long-slit data with the \textsc{gemini IRAF} package. In brief, the data were dark subtracted, flat fielded, and the sky subtracted using the closest one or two exposures in time in a telluric series. The up to five exposures per series were rectified, wavelength-calibrated, and extracted individually. 
We applied the \textsc{molecfit\_model} recipe of the ESO tool \textsc{molecfit} \citep{2015A&A...576A..77S,2015A&A...576A..78K}
with the framework \textsc{esoreflex} on each individual extracted telluric spectrum of a series. This tool models the atmosphere at the time of the observations by fitting specified wavelength regions of the observed telluric spectrum. 

We used the \textsc{molecfit} instrument setting "ANY" and had to change the format of the telluric spectra to make them readable for \textsc{molecfit}. In particular, we had to add several fits header keywords. We fitted three different molecules, H$_2$O, CO$_2$, and CH$_4$, and used a similar wavelength range as recommended for the instrument KMOS (ESO), which has a similar spectral resolution and wavelength coverage to our data, but sometimes had to slightly adjust the wavelength range to improve the results\footnote{The fitted wavelength regions usually cover 1.976-2.010, 2.041-2.060, 2.192-2.21, 2.269-2.291, 2.308-2.335, 2.360-2.379, 2.412-2.440, 2.444-2.457\,\micron, with slight variations if it improves the fit.}. 

The results provided by \textsc{molecfit\_model} include the instrumental full width at half maximum (FWHM) and atmospheric parameters. We found better results with a variable kernel, increasing with wavelength. 
For each series of a telluric, we computed the error-weighted mean value for the atmospheric parameters. As the telluric observations in a series were taken immediately after each other, close in time, we expect only small variations of the atmospheric parameters from exposure to exposure. By taking the error-weighted mean, we ensure our atmospheric parameters are robust. The instrumental FWHM however does indeed vary from exposure to exposure (with a range of $\sim$3 to $\sim$4.5 pixel), because each exposure was taken on different regions of the slit, and thus fell on different regions of the detector. The minimum of the FWHM is near the centre of the detector. For each telluric series, we linearly interpolated the FWHM to the middle positions of the six slitlets of the science data. Then, we computed the atmospheric transmission spectrum using the \textsc{molecfit\_calctrans} recipe of the \textsc{esorex} command line tool for each science exposure and extension. For each exposure, we used the error-weighted mean atmospheric parameters from the telluric series closest in time and the airmass at the time of the science observation. Each exposure was divided into six extensions; for each of them, we used the instrumental FWHM at that detector position to create the telluric model. Then we divided the 2-dimensional spectral frames by their respective telluric model.

\begin{figure}
\resizebox{\hsize}{!}{\includegraphics{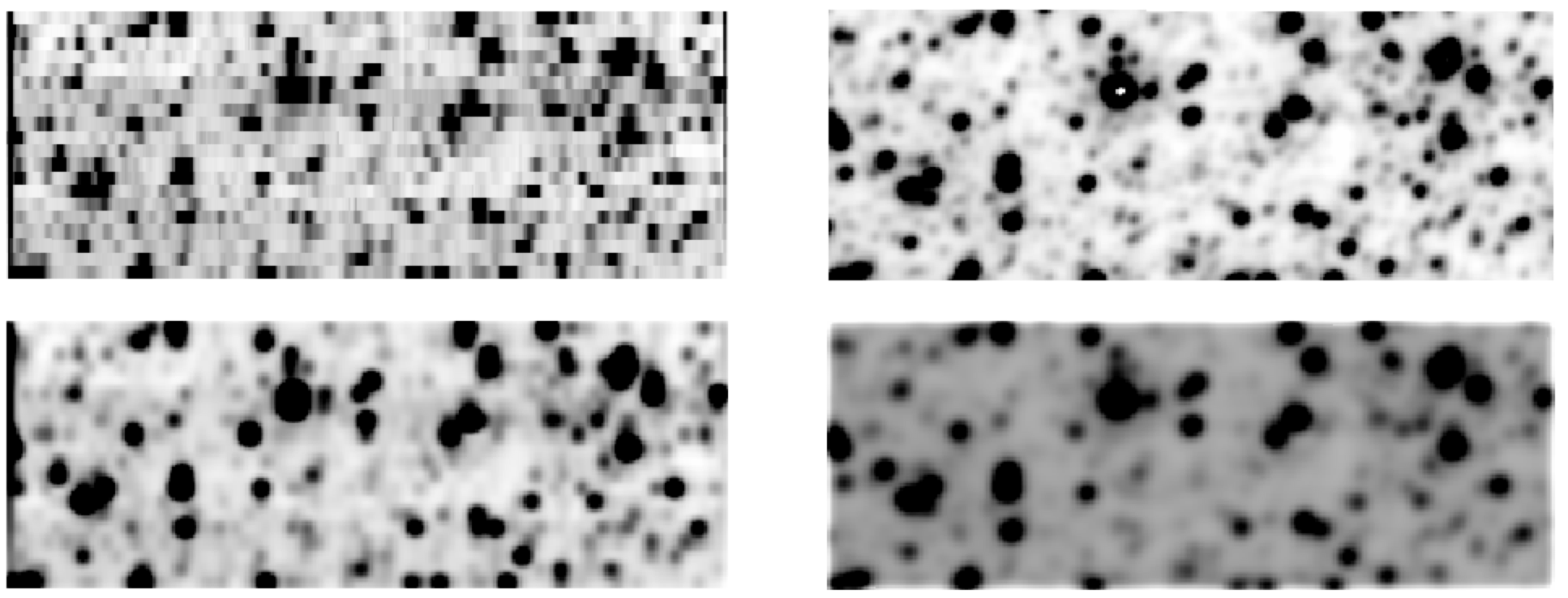}}
 \caption{Top left: reconstructed image of the data from 20 subsequent spectra; Top right: VVV $K_S$-band image cutout of the same region, resampled to the pixel scale of 0.18\,arcsec\,pixel$^{-1}$; Bottom row: same as top row but convolved with a Gaussian PSF with a FWHM of 1\arcsec. The images cover $\sim$59\arcsec$\times$22\arcsec\space (0.86$\times$2.3\,pc).}
 \label{fig:recim}
\end{figure}
\subsection{Astrometric calibration}

To identify the location of a star on the 2-dimensional spectral frames, we require an astrometric calibration. 
We use the Vista Variables in the Vía Láctea (VVV) survey \citep{2012A&A...537A.107S} $K_S$-band image (b333) as an astrometric reference. The image has a pixel scale of $\sim$0.34\,arcsec\,pixel$^{-1}$ and covers our entire field of view (FOV). 
We constructed stitched images from our spectroscopic observations and cross-correlated them with the reference image. This was done as follows: 

We have 20 series of observations, with 6-22 exposures per series. These were taken without interruptions due to sky or telluric observations, hence, they cover continuous regions. Since each series contains the data from six slitlets, we constructed 20$\times$6=120 stitched images by summing the flux in the wavelength range of 2.05--2.29\,\micron. The stitched images conserve the pixel scale along the slit (0.18\,arcsec\,pixel$^{-1}$), which is in the Galactic east-west direction. With the F2 pixel scale, each exposure covers six pixels in the north-south direction (6$\times$0.18\,arcsec$\approx$1\farcs1,), taking into account the drift. 
Each stitched image extends over $\sim$59\arcsec\space ($\sim$ 2.3\,pc) along the Galactic east-west direction and over $\sim$6\farcs5--24\arcsec\space ($\sim$0.25-0.9\,pc) along the Galactic north-south direction, depending on the number of exposures. An example of such a stitched image is displayed on the top left of Fig. \ref{fig:recim}.

We know the approximate position of each stitched image, and we registered their centre position on the VVV image. We cut out the regions covered by the stitched images from the VVV image, leaving an additional 30 pixels ($\sim$10\arcsec) on each side. We resampled those cutouts to the same pixel scale as the F2 data (0.18\,arcsec\,pixel$^{-1}$). Both the VVV cutouts and F2 images were convolved with a Gaussian point spread function (PSF), with the FWHM of 1\arcsec, and then cross-correlated to find the remaining small shifts, usually only a few F2 pixels. We then updated the headers of the stitched F2 images with the new astrometry. We show a reconstructed F2 image, a VVV cutout of the same region, and the convolved image versions in Fig.~\ref{fig:recim}. A complete mosaic of all the reconstructed images is shown in Fig.~\ref{fig:fov}.

\subsection{Star extraction}
\label{sec:sextr}
To extract the spectra of bright stars, we used a star catalogue with information on the coordinates and $JHK_S$ band photometry. 
We used the GALACTICNUCLEUS (GNS) catalogue \citep{2019A&A...631A..20N} with minor adjustments. 

For our extraction method, completeness is more important than photometric precision and accuracy. As bright stars can be saturated in GNS data, the photometry of $\sim$25 stars in the F2 FOV was replaced with SIRIUS IRSF \citep{2003SPIE.4841..459N,2006ApJ...638..839N} photometry, which was also used for the photometric calibration of the GNS catalogue. For another $\sim$400 stars in the central 40\arcsec$\times$40\arcsec, we replaced $HK_S$ photometry with deep adaptive optics imaging from the instrument NACO at the ESO VLT \citep{2020A&A...641A.102S}, as these data have a superior spatial resolution. Another 67 bright stars ($J$\textless17\,mag or $H$\textless15.3\,mag) have no $K_S$ photometry in either catalogue (possibly due to saturation), and we used the available $J$ or $H$ photometry to estimate it. We assume that the intrinsic colour $(H-K_S)_0$=0\,mag, and that the stars are located in the Galactic Centre. Hence we can use an extinction map (see Sect. \ref{sec:extmap} for details), assume an extinction coefficient $\alpha_{HK}$=2.23 \citep[and $\alpha_{JH}$=2.44,][]{2020A&A...641A.141N}, and get a $K_S$ estimate from $K_S\approx H-A_H+A_{K_S}$.  These stars will later be considered stars with unknown status (see Section \ref{sec:member}). Still, it is important to include these bright stars in the catalogue used for the spectrum extraction to ensure that they are accounted for and do not contaminate the spectra of nearby fainter stars.

The procedure to extract stars is as follows: For each of the 120 stitched images, we selected stars in the photometric catalogue that are located in the region of the stitched image. Knowing where the brightest stars are, we can derive the seeing in each stitched image. We fit a Gaussian function at the location of the $\sim$50 brightest stars per stitched image, along the slit directions. The mean value of the FWHM is used as seeing. The median seeing of our data is 0.8\arcsec. 

Next, we selected stars up to $K_S$<15\,mag, located again in the region of the stitched images, 
and up to 2\arcsec\space beyond. We sort them by magnitude, starting with the brightest star.
For each of these stars, we create an artificial image of the star in the sky, in an array of the same size and sampling as the F2 stitched image, with an FWHM corresponding to the previously measured seeing. The sum of the artificial images resembles the stitched image in terms of size, but it has the F2 sampling of 0.18\,arcsec\,pixel$^{-1}$ in both dimensions. 

For each star, again starting with the brightest one, we selected the spectral exposures that likely contain the flux of said star. We first considered the exposure that covers the position where the star is located, but due to the seeing, a star can contribute flux to several exposures. For this reason, we also considered the exposures taken before and after, i.e. three exposures per star. For stars fainter than $K_S$=13\,mag, we used only the primary exposure that covered the location of the star and the closest adjacent exposure (taken either before or after). 

Starting with the primary exposure, we performed a Gaussian fit to obtain the exact location of the star on the slit, and its Gaussian $\sigma$. The extraction window was, by default, 6$\sigma$ wide. Using the knowledge of the position and brightness of other stars in the field, we checked where other stars contributed more flux than the target star, and we reduced the extraction window accordingly if this was the case. The same procedure was repeated with the adjacent exposures that also contained the light of the star, but the Gaussian fit result of the primary exposure was used as an initial guess in the Gaussian fit. To extract a 1D spectrum, we computed the total flux in the extraction window. We extracted 30\,000 spectra of stars with $K_S$\textless14\,mag. We resampled each spectrum to its respective corrected wavelength calibration (see Sect. \ref{sec:lcalcorr}).

During the extraction process, we created masks of bright stars and foreground stars, and we used those to create data cubes of the unresolved faint stars. These data will be shown in a separate publication (Feldmeier-Krause et al. in prep.).

\subsection{Spectral resolution}
\label{sec:res}
The spectral resolution of the F2 spectrograph varies across the detector, both as a function of position on the slit, and more significantly as a function of wavelength\footnote{see \url{https://www.gemini.edu/sciops/instruments/flamingos2/Ksmapmay14.jpg}}. We measured the resolution by fitting a Gaussian function with width $\sigma_{\rm LSF}$ to the sky emission lines on the dedicated sky exposures after the rectification step. We used 14 sky emission lines in the range 2.001--2.252\,\micron. At shorter and longer wavelengths, there are no isolated emission lines suitable for a Gaussian fit. We performed these fits also as a function of the spatial slit direction. As expected, we found that the variation of the spectral resolution as a function of slit position at a given sky emission line is relatively small ($\sigma_{\rm R(x)}$$\sim$200, or 0.22\,\AA) compared to the variation along wavelength ($\sigma_{\rm R(\lambda)}$$\sim$700, or 0.96\,\AA) at a given slit position. Yet, as there is some variation along the slit, we decided to derive six different spectral resolution functions for the six different slitlets, resulting in a resolution variation for a slitlet $\sigma_{\rm R(x)}$$\sim$90, or 0.1\,\AA. In each slitlet region, we made a 2-degree polynomial fit to the spectral resolution as a function of wavelength. 
The spectral resolution is highest at wavelengths of 2.1--2.2\,\micron\space with $\sigma_{\rm LSF}$$\sim$2.6\,\AA\space or $R\sim$3\,400.

\section{Analysis}
\label{sec:analysis}

\subsection{Spectral indices}
\label{sec:ppxf}

\begin{figure*}
\centering
\includegraphics[width=17cm]{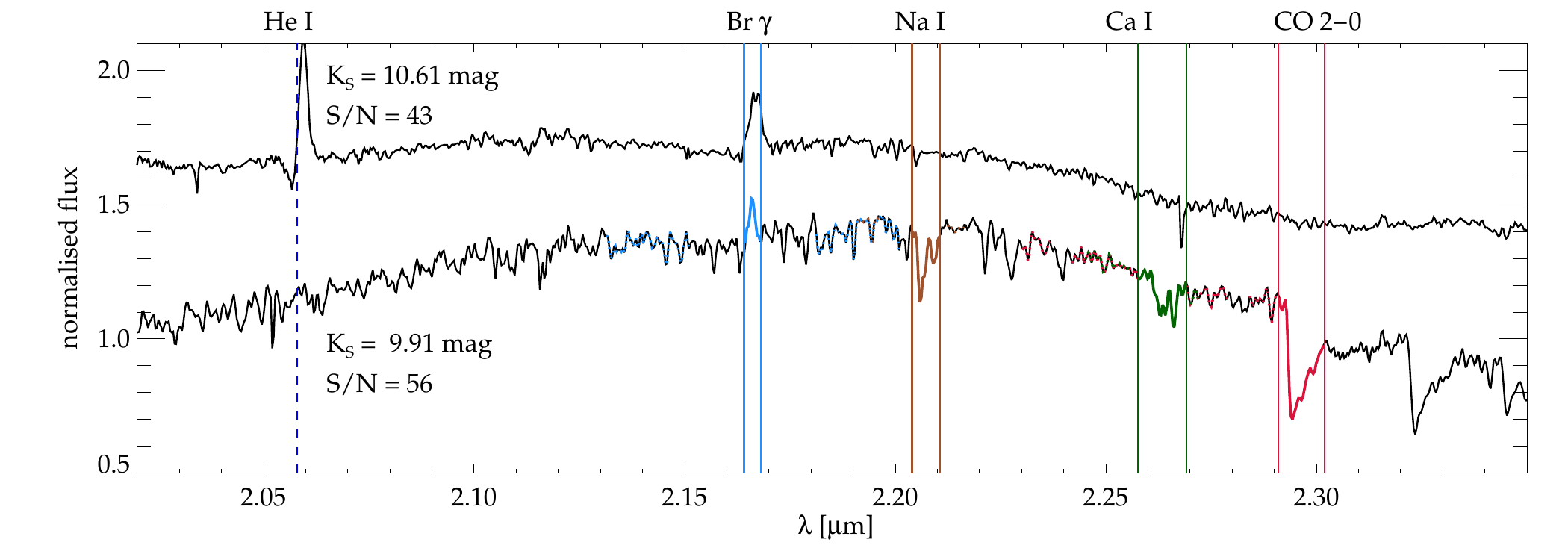}
 \caption{Example spectra of a hot star (top) and a cool star (bottom). Both spectra are normalised by their median flux, and a small offset is added to improve visibility. The vertical lines indicate several spectral features used for the analysis, labelled on the top. Both spectra have Br~$\gamma$ emission due to surrounding gas (the hot star has even \ion{He}{I} emission), but only the cool star has strong \ion{Na}{I}, \ion{Ca}{I} and CO 2-0 absorption features, and plenty of other metal lines. The regions used to measure the spectral indices are marked by different colours in the cool star spectrum, solid lines for the feature, and dotted lines for the pseudo-continuum regions. We annotate the $K_S$ and S/N (computed using the \textsc{pPXF} fit residual) for each star.}
 \label{fig:arbspec}
\end{figure*}

\begin{figure*}
 \centering
 \includegraphics[width=6cm]{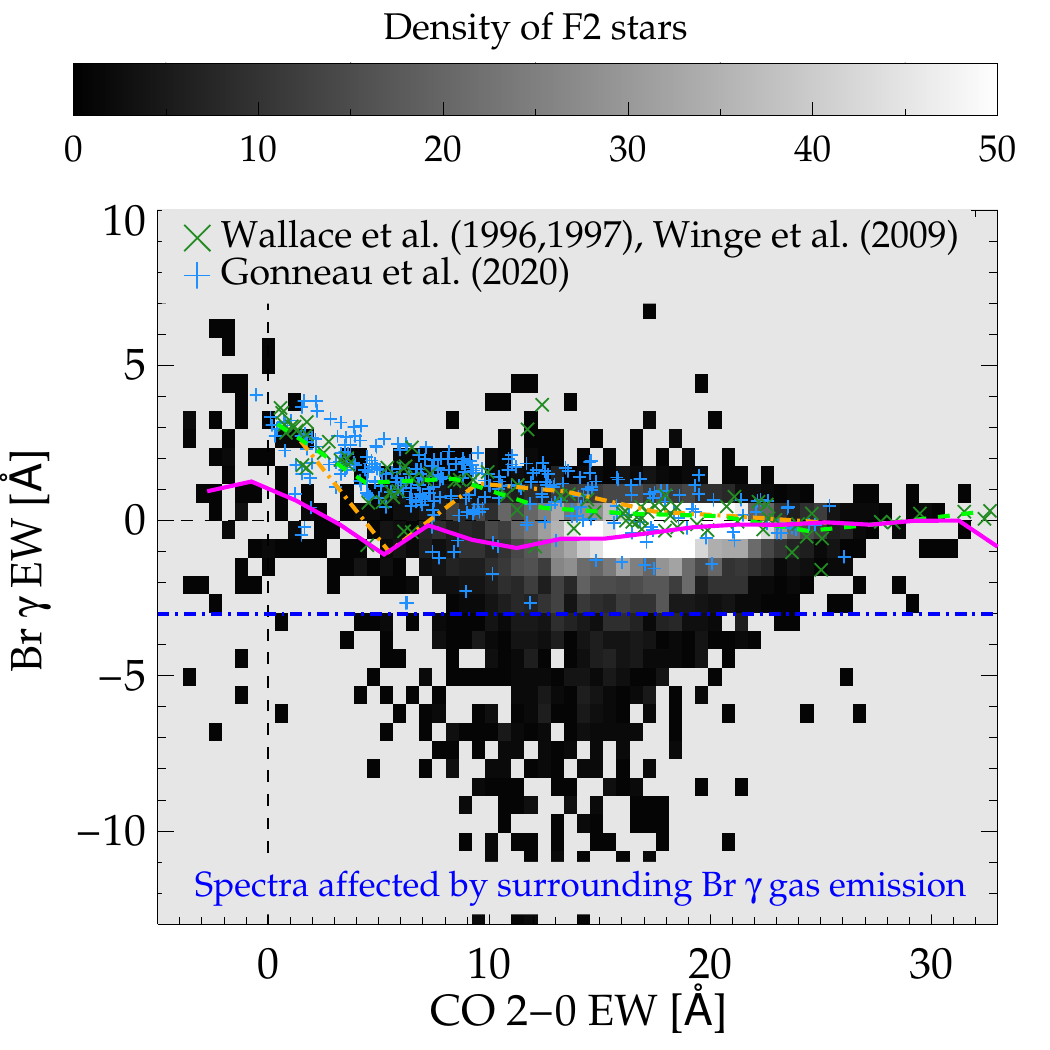}
  \includegraphics[width=6cm]{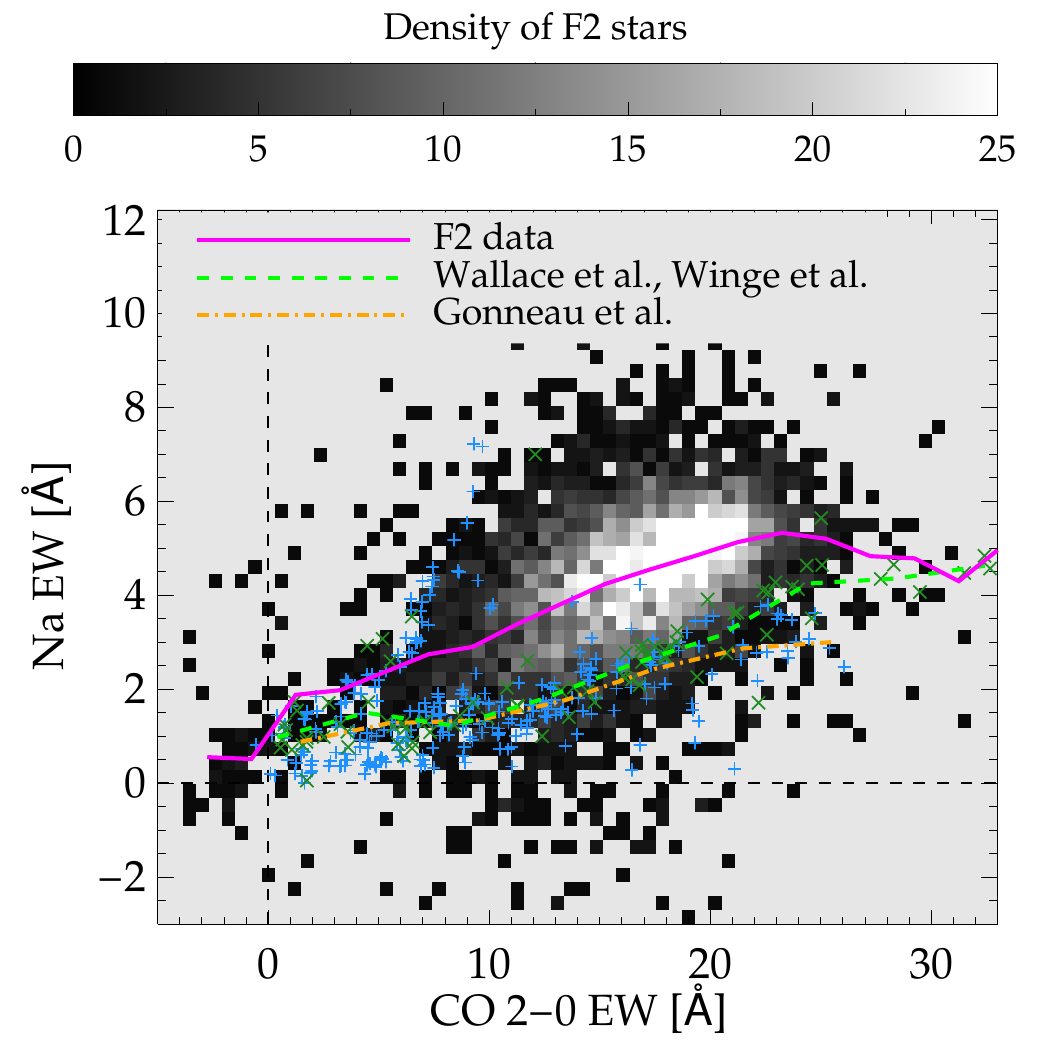}
 \includegraphics[width=6cm]{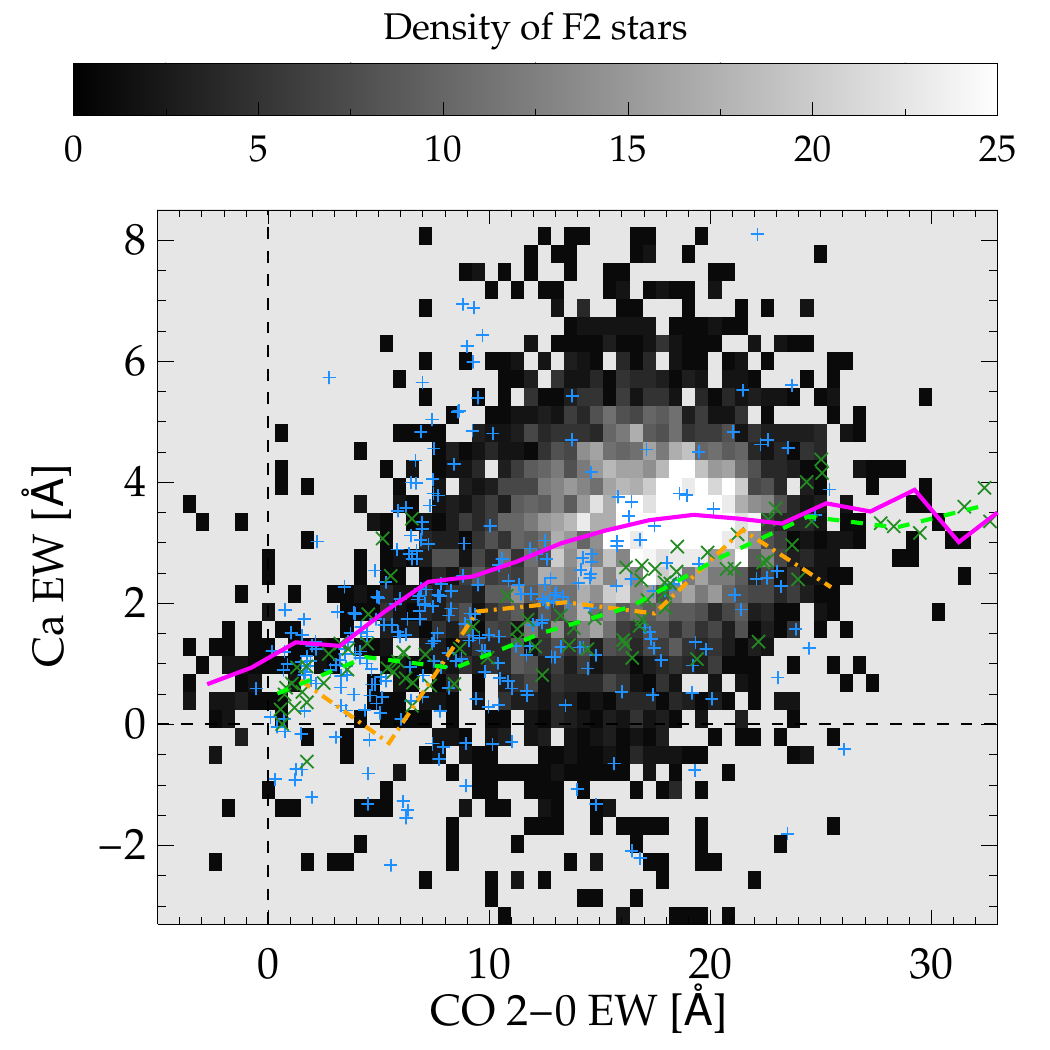}

 \caption{Spectral equivalent width (EW) of Br~$\gamma$, \ion{Na}{I}, and \ion{Ca}{I} as function of the EW of CO 2-0. The density maps show measurements on our extracted spectra after a quality cut on the required S/N. The green  x-symbols are $\sim$100 stars in the \cite{1996ApJS..107..312W,1997ApJS..111..445W} and \cite{2009ApJS..185..186W} spectral libraries, and the blue plus-symbols are \textgreater400 Milky Way stars from \cite{2020A&A...634A.133G}. Magenta solid lines, green dashed (for \cite{1996ApJS..107..312W,1997ApJS..111..445W,2009ApJS..185..186W},  and orange
dot-dashed lines (\citealt{2020A&A...634A.133G}) show the robust mean as function of $EW_{\rm CO}$. Stars with a Br $\gamma$ measurement below the dot-dashed horizontal blue line in the left panel are affected by Br $\gamma$ gas emission at the centre of the NSC. }
 \label{fig:index3}
 \end{figure*}

As a first step in the spectral analysis, we derive the line-of-sight velocity \vlos, correct the spectrum to the rest frame, and measure spectral indices, which help to distinguish hot OB (early-type) stars from cool KM (late-type) giant stars. 

We measure \vlos\space using \textsc{pPXF}  \citep[version 5.2.1,][]{2017MNRAS.466..798C}, a full-spectral fitting code, in the wavelength range 2.15--2.3155\,\micron. This region includes  \ion{H}{I} Brackett (Br) $\gamma$, a \ion{Na}{I} doublet, a \ion{Ca}{I} triplet, and the CO 2-0 band head. We use the high-resolution spectral library of late-type stars provided by \cite{1996ApJS..107..312W}, and a set of KMOS B-type dwarf stars \citep{2020MNRAS.494..396F}, and convolve the templates to the spectral resolution of the data (as measured in Sect. \ref{sec:res}). \textsc{pPXF} assigns weights to the template spectra, and the linear combination gives an optimal template for each fitted spectrum. Uncertainties are estimated by adding random noise to the spectra and repeating the fits in 80 realisations. The median \vlos uncertainty for all spectra is 11\,\kms, and when we apply a quality cut on the required S/N\textgreater 20, it is 6\,\kms.

After applying the \vlos\space shift to the spectra, we measured the following spectral indices (alternatively called equivalent width $EW$): Br $\gamma$, which can be in absorption or emission in early-type stars 
\citep[as defined by][]{2021A&A...649A..83F}; the CO 2-0 band head ($\sim$2.2935\,\micron); the \ion{Na}{I} doublet (2.2062 and 2.209\,\micron); and the \ion{Ca}{I} triplet (2.2614, 2.2631, 2.2657\,\micron), which are seen in absorption in cool late-type stars \citep[as defined by][]{2001AJ....122.1896F}. We show example spectra in Fig.~\ref{fig:arbspec}; spectral features are indicated as vertical lines, and the spectral regions of the index measurements are highlighted by different colours.

We repeated the same measurements on the subset of Milky Way spectra of the X-SHOOTER spectral library \citep[XSL DR2,][]{2014A&A...565A.117C,2020A&A...634A.133G}, convolved to the median spectral resolution function of the F2 data, to have a calibration sample. However, in these data, the change from one to the next Echelle order is at the location of the CO and \ion{Ca}{I} spectral index continua, sometimes causing biased index measurements. Therefore, we also used the spectral libraries of \cite{1996ApJS..107..312W,1997ApJS..111..445W} and \cite{2009ApJS..185..186W}. These data revealed that $EW_{\rm CO}$ of giant stars at the F2 spectral resolution is \textless 25\,\AA, but supergiants can have \textgreater30\,\AA\ (as found at slightly higher spectral resolution by \citealt{2017MNRAS.464..194F}).

We show the spectral index distributions of the spectral libraries and our data in Fig. \ref{fig:index3}. There is a larger spread in our data because of multiple reasons: Our data has lower S/N than the spectral libraries, even after applying an S/N cut (\textgreater 20). There can be sky residuals in the region of the CO feature, causing very low or high values; Spectra were not cosmic-ray corrected, to prevent the over-correction of bright star spectra (see Sect. \ref{sec:dr}); in particular the Br~$\gamma$ index is affected by interstellar gas emission, causing very low negative values. We note that for a given $EW_{\rm CO}$, the F2 data tend to have higher $EW_{\rm Na }$ and $EW_{\rm Ca}$, something that was already noted by \cite{1996AJ....112.1988B} and \cite{2017MNRAS.464..194F} for Galactic Centre stars, and hints at enhanced chemical abundances compared to Milky Way disc stars.

\subsection{Stellar parameters}
\label{sec:spfit}
As the next step, we fit the stellar spectra extracted in Section \ref{sec:sextr} with the full spectral fitting code \textsc{StarKit} \citep{2015zndo.....28016K,2015ApJ...809..143D} using the PHOENIX spectral library of synthetic spectra \citep{2013A&A...553A...6H} as templates. \textsc{StarKit} interpolates the template spectra and applies Bayesian sampling \citep[Multinest v3.10, pymultinest v2.11;][]{2008MNRAS.384..449F,2009MNRAS.398.1601F,2014A&A...564A.125B,2019OJAp....2E..10F} to obtain the best-fitting parameters. We fit the total metallicity \mh, effective temperature \teff, surface gravity \logg, and, in addition, the line-of-sight velocity \vlos. We ignore that the stars in our sample can have a range of chemical abundances \citep{2020ApJ...894...26T} and fit only \mh, i.e. the overall metallicity, as a parameter. This means the synthetic model spectra are computed with [$\alpha$/H]=\mh, and [$\alpha$/Fe]=0\,dex. 

We use the same constraints, limits, and bounds as in our previous works \citep{2017MNRAS.464..194F,2020MNRAS.494..396F,2022MNRAS.513.5920F}. In detail, the template spectra sample a grid with the ranges \mh\space = [$-$1.5\,dex, +1.0\,dex], $\teff$\space = [2\,300 K; 12\,000 K], and $\logg$\space = \,[0.0\,dex, 6.0\,dex], 
and with step sizes of $\Delta$\mh\space = 0.5\,dex, $\Delta\teff$\space = \,100\,K, and $\Delta\logg$\space = 0.5\,dex. Before the fit, we convolve the template spectra to the spectral resolution of the data, as measured in Sect. \ref{sec:res}. As fitting bounds, we use information obtained in Sect. \ref{sec:ppxf}, in particular, the value and uncertainty of \vlos\space in a Gaussian prior. 
The CO index measurement can be used to limit the uniform prior bounds of \logg. Giants in the spectral libraries have always $EW_{\rm CO}$\textless25\,\AA, and only supergiants have $EW_{\rm CO}$\textgreater25\,\AA, For $\sim$20 stars with $EW_{\rm CO}$\textgreater25\,\AA that are also brighter than $K_{S,0}$=10\,mag (extinction corrected), we use 0.0\,dex\,\textless\,$\logg$\,\textless 2.0\,dex; for all other stars we use a more generous 0.0\,dex\,\textless\,$\logg$\,\textless 4.0\,dex, as they may be giants or supergiants. The priors for \teff\space and \mh\space are uniform within the ranges of the PHOENIX spectra. We fit the spectral region 2.09--2.29\,\micron, but exclude the regions around the \ion{Na}{I} doublet (2.2027--2.2125\,\micron) and \ion{Ca}{I} triplet (2.2575--2.2685\,\micron), as these are enhanced in Galactic Centre stars compared to normal Milky Way disc stars (as also shown in Fig. \ref{fig:index3}), and would bias our \mh\space to higher values. 

After fitting the spectra, we compute the residuals by subtracting the best-fit model from the data and subsequently estimate the signal-to-residual ratio. We discard fits with S/N$_r$\textless20, and also fits with large statistical uncertainties ($\sigma$(\teff)\textgreater 250\,K, $\sigma$(\logg)\textgreater 1\,dex, $\sigma$(\mh)\textgreater 0.25\,dex, $\sigma$(\vlos)\textgreater 10\,\kms), which indicate either a poor fit or hot star candidates (see Sect. \ref{sec:hot}). When we have several spectra and good-quality fits of the same star, we combine the stellar parameter measurements with a simple mean. We use either the sum of statistical uncertainties in quadrature or the standard deviation of the multiple measurements as our new statistical stellar parameter uncertainty, depending on which is larger. 

We compare our measurements with the literature in Appendix \ref{sec:litcom} and find no strong biases. While the statistical uncertainties for \teff\space and \logg\space are underestimated, the statistical uncertainty for \mh\space is a good approximation. 
We need to consider systematic uncertainties to estimate the total uncertainty. We estimate the systematic uncertainties caused by, e.g. the choice of the synthetic model grid or variations of the elemental abundances in the stars by following the procedure outlined in \cite{2022MNRAS.513.5920F} and summarised in Appendix \ref{sec:xslsk}. 

\subsection{Velocity corrections}
\label{sec:vcor}
The measured line-of-sight velocities \vlos\space are affected, e.g. by the motion of the Earth around the Sun, and also by the motion of the Sun around the Galactic Centre. 

We compute a barycentric correction, which takes into account the rotation of the Earth itself, the rotation around the Earth-Moon barycentre, and the rotation around the Sun by considering the individual coordinates of each star, the time of the observation, and the coordinates and altitude of the telescope with the \textsc{IDL} program \textsc{helcorr.pro}, which uses the algorithms of \textsc{IRAF} \textsc{noao.astutils.rvcorrect}. The barycentric correction ranges from -5.5 to -2.3\,\kms. 

Perspective rotation is an effect caused by the large extent of the data on the sky and the substantial motion of the Sun around the Galactic Centre. This causes a so-called perspective rotation and increases the difference between the motion of stars in the very east and very west by almost 2\,\kms. We compute the effect with the equations given by \cite{2006A&A...445..513V} and assuming a distance of 8.2\,kpc, a velocity of 220\,\kms\space of the Sun in the Galactic plane, and -7\,\kms\space perpendicular to it. The latter motion is negligible, causing only a perspective rotation of $\sim$0.003\,\kms\space in the FOV of the data. The correction for the motion in the Galactic plane is in the range of -0.95 to +0.95\,\kms.

\subsection{Extinction map}
\label{sec:extmap}
We create an extinction map using the photometric data of the GNS. We follow the procedure outlined in \cite{2018A&A...610A..83N} and also applied in \cite{2022MNRAS.513.5920F}. In brief, we selected all GNS stars in the region of our spectroscopic F2 data and several arcseconds beyond. Of these, we selected the likely red clump stars, which have a colour 1.3\,mag\textless\col\textless 2.6\,mag, and we applied colour-dependent magnitude cuts as shown in Fig. 2 of \cite{2022MNRAS.513.5920F}. Using equation (5) of \cite{2018A&A...610A..83N}, and assuming the same filter effective wavelengths, intrinsic colour for the red clump stars ($(H-K_S)_0$=0.089\,mag), and extinction coefficient ($\alpha_{JHK_S}$=2.3), we derived the extinction \ak for each red clump star. From these, we derived an extinction map (with pixel scale 0.1797\arcsec$\cdot$pixel$^{-1}$) by computing in each pixel the distance weighted mean \ak of the 15 closest red clump stars within 12 arcsec. Our extinction map has a mean value of \ak=2.1\,mag with a standard deviation of 0.22\,mag. The values of \ak range from $\sim$1.5--2.7\,mag.

\subsection{Galactic Centre membership classification}
\label{sec:member}
\begin{figure}
 \resizebox{\hsize}{!}{\includegraphics{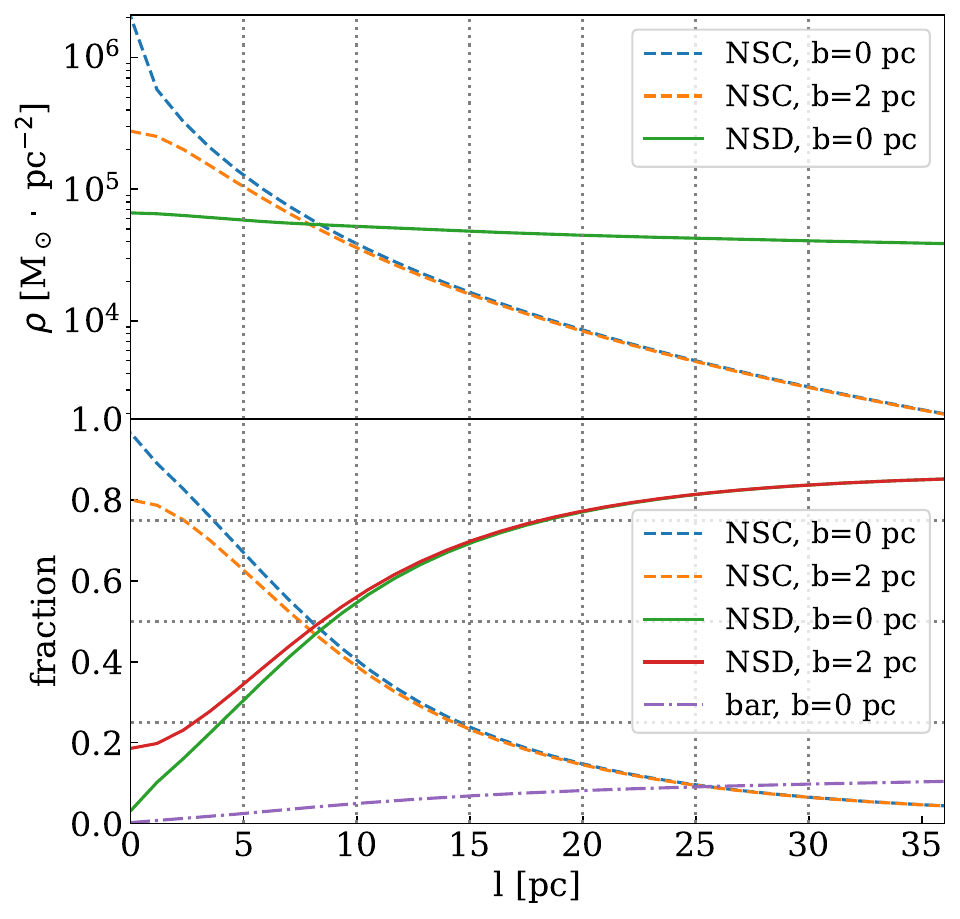}}
 \caption{Stellar surface density profile in the Galactic Centre. Top panel: NSC profile from \cite{2015MNRAS.447..948C}, NSD profile from \cite{2022MNRAS.512.1857S} as a function of $l$. Bottom panel: Fraction of the NSC and NSD density profiles, if the bar density profile is constant, and set to 20\% of the NSD density profile at $l$=35\,pc and $b$=0\,pc.
} \label{fig:fracprof}
 \end{figure}
 \begin{figure*}
 \includegraphics[width=8.5cm]{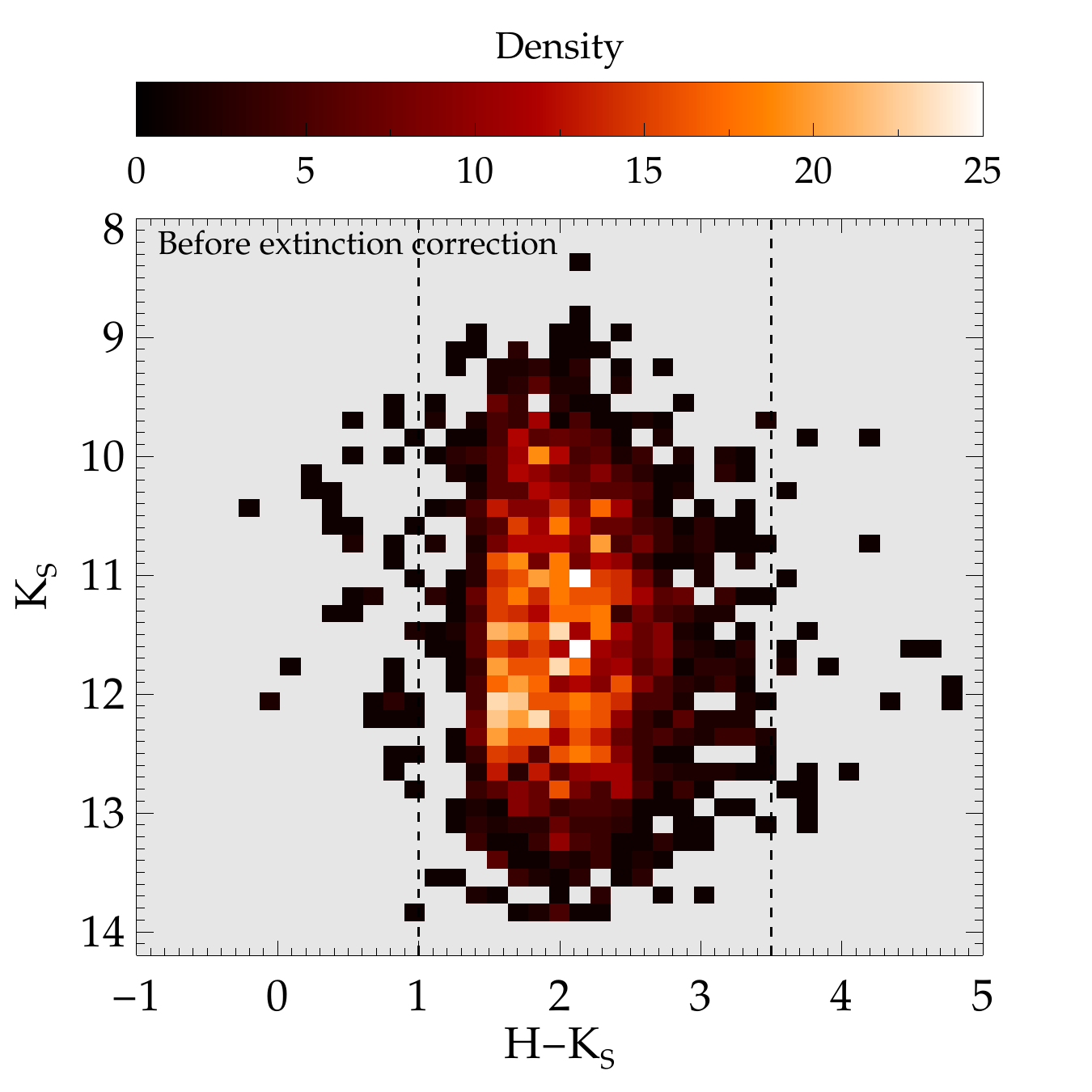}
\includegraphics[width=8.5cm]{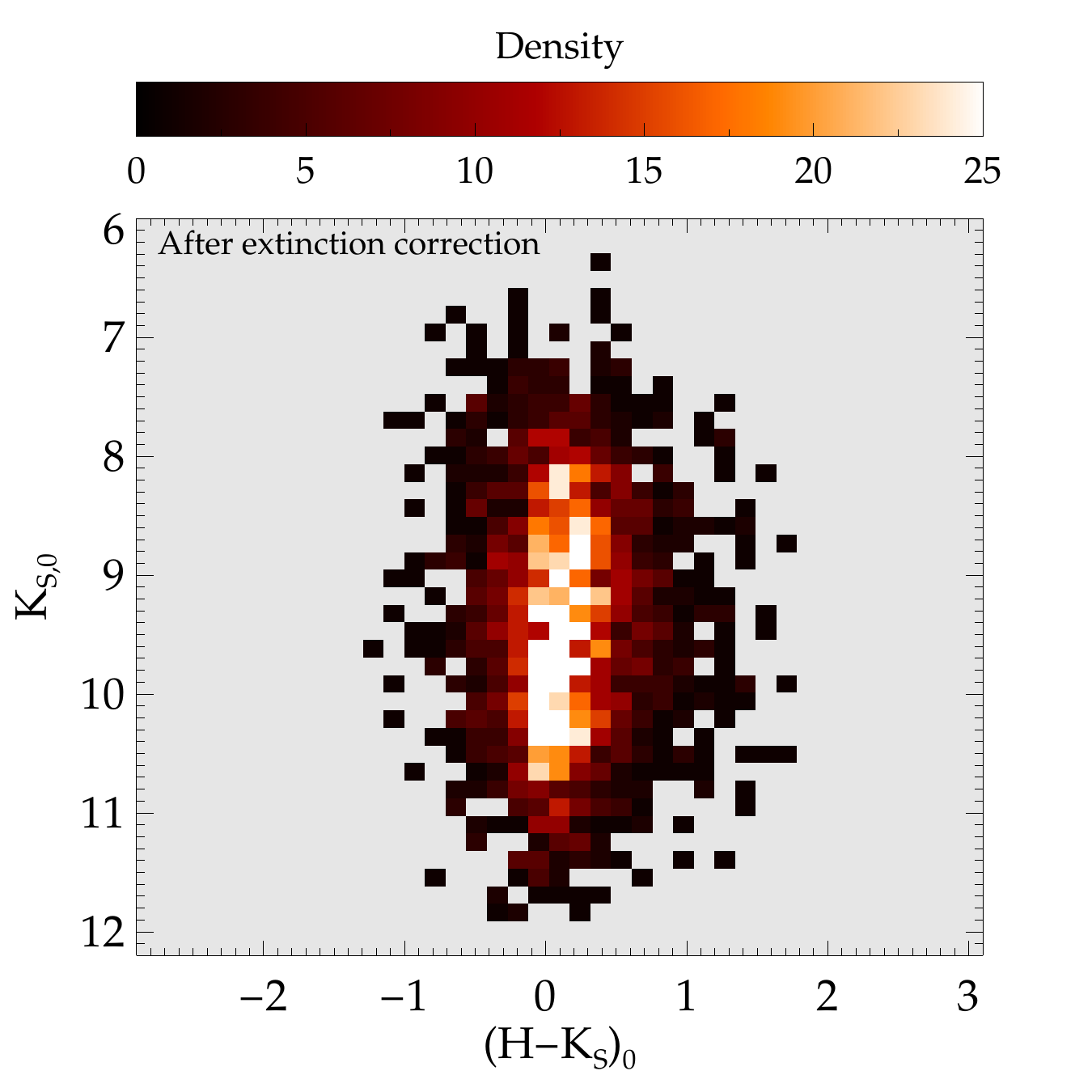}
 \caption{Colour-magnitude diagram of the sample with stellar parameter fit. Left panel: Observed \col vs. $K_S$ diagram for all stars with stellar parameter fit. The vertical dashed lines enclose the stars classified as being located in the Galactic Centre. Right panel: Extinction corrected $(H-K_S)_0$ vs. $K_{S,0}$ diagram of stars with stellar parameter fit and classified as being located in the Galactic Centre.}
 \label{fig:cmd}
\end{figure*}
We use simple colour cuts to classify stars that are likely located within the Galactic Centre (GC) structures NSC and NSD, or foreground stars, or background stars. In detail, we identify a star as a foreground star if \col$\leq$ 1\,mag \citep[as in ][]{2021A&A...649A..43C}, and as a background star if \col$\geq$ 3.5\,mag. Stars for which either $H$ or $K_S$ are missing are classified as unknown. We refer to the combination of these three groups of stars (foreground, background, unknown) as non-GC stars. Our classification criteria are rather inclusive compared to other studies, which classified stars as foreground stars if \col$\lesssim$ 1.3\,mag \citep[e.g.][]{2016ApJ...821...44F,2022MNRAS.513.5920F}. These studies have smaller fields of view, with less variation of the foreground extinction. Using the stricter criterion on the foreground only removes 70 stars with stellar parameter measurement, which is less than 3\% of the sample, and has no influence on our results.

We show the colour-magnitude density diagram of the cool late-type stars in Fig.~\ref{fig:cmd}, the left panel before extinction correction versus the right panel after extinction correction in the Galactic Centre.

 We estimate the contribution of the bar, the NSD, and the NSC as a function of longitude in our field of view using the \textsc{Agama} package \citep{2018ascl.soft05008V,2019MNRAS.482.1525V}. \cite{2022MNRAS.512.1857S} give estimates of the bar contribution in various regions of the NSD beyond our observed fields. In the fields closest to our data, though still several parsec away, the surface density contribution of the bar compared to the total surface density of the bar and NSD is 19--26\%. We estimate that the bar contribution is $\lesssim$20\% in our field. 
We further use the stellar density profiles of \cite{2015MNRAS.447..948C} for the NSC and \cite{2022MNRAS.512.1857S} for the NSD (see Fig.~\ref{fig:fracprof}, top panel), and compare the projected surface density profiles (i.e. integral of density along the line-of-sight) at the location of our data. Assuming a spatially constant bar contribution, we find that the value of Galactic longitude where the NSC surface density contributes \textgreater50\% of the total (NSC+NSD+bar) density is at $l\lesssim$7.7\,pc for $b$=0\,pc, and $l\lesssim$7.2\,pc for $b$=2\,pc (see Fig.~\ref{fig:fracprof}, bottom panel). Further out, the NSD dominates the projected stellar surface density. However, we made several assumption: We assume that the bar contribution is a simple extrapolation from \cite{2022MNRAS.512.1857S}, we do not consider the orientation of the bar, and we neglect potential observational biases caused by varying extinction. Nonetheless, these estimates help us to understand where the NSC or the NSD likely dominate our sample.

 \begin{figure*}
 \centering
 \includegraphics[width=18cm]{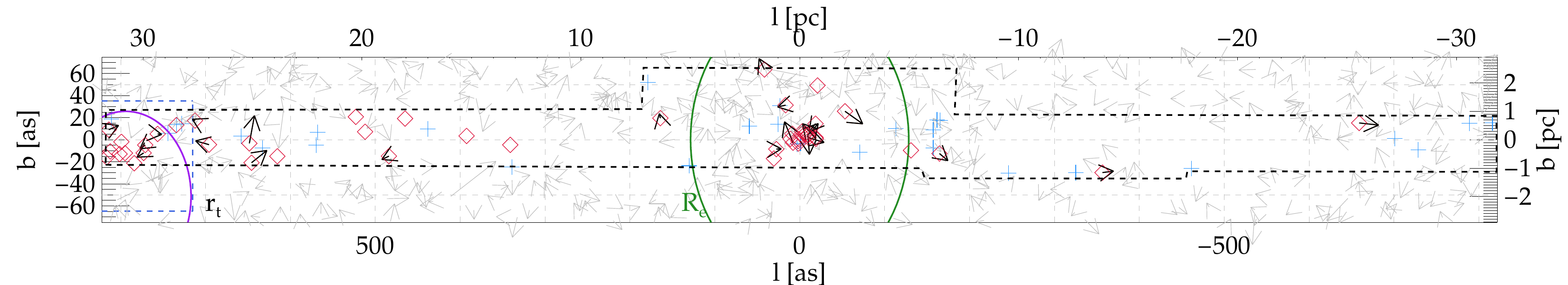}
 \caption{Proper motions of stars in the field of view. 
 Red-coloured diamonds denote hot star candidates we classified as being located in the Galactic Centre (GC), as listed in Table \ref{tab:hotcand}, and blue cross-symbols denote those hot stars we classified as non-GC stars or stars with unknown status. The black arrows indicate the proper motions of hot star candidates from \cite{2022A&A...662A..11S}. The arrow lengths are multiplied by a factor of 3\,000 for better visualisation. The grey arrows are a subset (2.5\%) of the \cite{2022A&A...662A..11S} proper motions to illustrate the distribution of proper motions in this region. Black dashed lines denote the approximate outline of our FOV, blue dashed lines the region shown in Fig.~\ref{fig:hotpmq}. The purple circle denotes Quintuplet's tidal radius $r_t\sim$3\,pc \citep{2019ApJ...877...37R}, the green solid circle denotes the NSC \re=5\,pc. The x-axis and y-axis have different scales, therefore, the circles and proper motions appear elongated along the y-axis. }
 \label{fig:hotpm}
 \end{figure*}
\begin{figure}
 \resizebox{\hsize}{!}{\includegraphics{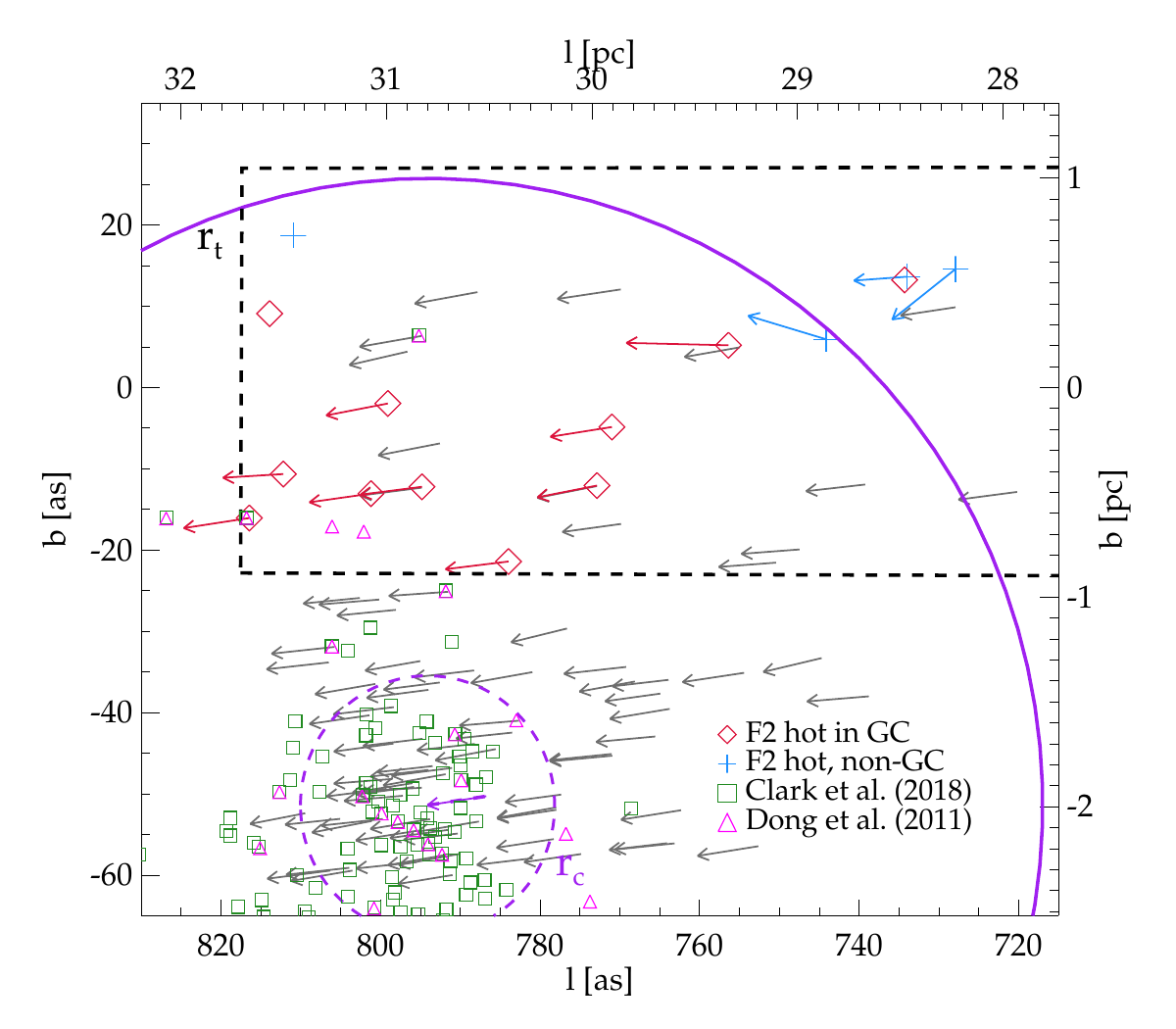}}
 \caption{Proper motions of stars in the region close to the Quintuplet cluster. 
 Red-coloured diamonds denote hot star candidates we classified as being located in the GC, and blue cross-symbols denote those stars we classified as non-GC stars or stars with unknown status. The red arrows indicate the proper motions of our hot star candidates from \cite{2022ApJ...939...68H}. The arrow lengths are multiplied by a factor of 3\,000 for better visualisation. The grey arrows are a subset of the \cite{2022ApJ...939...68H} proper motions, showing 33\% of the stars with \textgreater 80\% cluster probability. Black dashed lines denote the approximate outline of our FOV. The dashed purple circle denotes Quintuplet's core radius $r_c$, the solid purple circle its tidal radius $r_t$ \citep[adopted from][]{2019ApJ...877...37R}, and the purple arrow pointing at the centre denotes the direction of the orbit of Quintuplet \citep[Fig. 6 in][]{2022ApJ...939...68H}. Green coloured squares denote spectroscopic hot Quintuplet stars from \cite{2018A&A...617A..65C}, magenta triangles are Paschen $\alpha$ candidates from \cite{2011MNRAS.417..114D}.}
 \label{fig:hotpmq}
\end{figure}

\section{Hot star candidates}
\label{sec:hot}

The strength of CO absorption decreases with increasing \teff\space \citep{1986ApJS...62..501K,2017MNRAS.464..194F}. A spectrum without CO absorption indicates a hot and, given the brightness of our sample, young and massive star. We classify a star as a hot star candidate if $EW_{\rm CO}\lesssim$5\,\AA. Sometimes noise or sky line residuals contaminate our CO measurement. Hence, we visually inspected the spectra and, in some cases, we classified stars as hot candidates even though the value of $EW_{\rm CO}$ exceeds the above threshold, as bad pixels affect the measurements. We list these 78 stars in Table~\ref{tab:hotcand}. 
This list includes stars with visible but weak CO absorption ($EW_{\rm CO}$\textgreater0\,\AA). Due to our poor spatial resolution, it may be possible that the detected weak CO absorption is caused by contamination of nearby late-type stars rather than from the star itself. Hence, we list these stars as potential hot stars. If the CO absorption is intrinsic to the star, it is likely a rather warm giant.

We verified our hot star candidates by comparison with the literature, notably with the early-type candidates of \cite{2015A&A...584A...2F,2022MNRAS.513.5920F}, but also the late-type stars in \cite{2017MNRAS.464..194F,2020MNRAS.494..396F}.
\cite{2015A&A...584A...2F} identified \textgreater100 young stars in the central 4\,pc$^2$ around \sgra, and we identified 15 candidates in this region. We matched stars with a maximum distance of 0\farcs3 and found 15 matches in the \cite{2015A&A...584A...2F} data. We have two additional candidate stars with $EW_{\rm CO}\approx$4.6\,\AA\space in the same FOV, but they were classified as late-type stars in \cite{2017MNRAS.464..194F}. We mark these stars with a footnote in Table~\ref{tab:hotcand}. These stars are fainter ($K_S\sim$13\,mag) than the rest of our matches. As \cite{2015A&A...584A...2F} and \cite{2017MNRAS.464..194F} have a better spatial resolution (seeing limited) than we do (\textgreater1\arcsec\space along latitude), their classification is less likely to be contaminated by background sources. We also matched two hot stars (within even \textless0\farcs1) with \cite{2022MNRAS.513.5920F}, which cover two 4\,pc$^2$-sized fields located about 20\,pc east and west of \sgra. Our matches correspond to the 2 brightest out of the 9 hot stars in \cite{2022MNRAS.513.5920F}. 
This comparison shows us that we can identify bright hot stars reliably. 

We show the spatial distribution of the hot star candidates in Fig.~\ref{fig:hotpm}. As noted in other spectroscopic studies \citep{2014A&A...570A...2F,2015A&A...584A...2F,2015ApJ...808..106S}, we find that the young stars are concentrated in the central $\sim$1\,pc region around \sgra. Beyond this region, hot stars are rather sparse. 

At a $\sim$20-30\,pc distance from \sgra, we see a higher density of hot stars in the Galactic east compared to the west. These stars may be related to the Quintuplet cluster, one of the young star clusters in the Galactic Centre ($\sim$4\,Myr, \citealt{1999ApJ...514..202F,2012A&A...540A..14L}). Quintuplet is located only 30\arcsec\space ($\sim$1.2\,pc) to the south of our FOV, and we have several hot star candidates just north of it (at $\sim$30\,pc east of \sgra). 
If the stars are associated with Quintuplet, their proper motions should point in the same direction as the proper motion of the cluster. If some of these stars used to be associated but were ejected (see Sect.~\ref{sec:dischot}), their proper motion should point away from the current or former position of the Quintuplet cluster. Quintuplet is on an orbit around the Galactic Centre and moves mostly to the Galactic east and slightly towards the south.  

We matched our hot star candidates with the proper motion catalogue of \cite{2022A&A...662A..11S}. We list the 23 matches (within $\leq$0\farcs2) in Table~\ref{tab:pmhot}. The proper motions are depicted as arrows in Fig.~\ref{fig:hotpm}. All of these stars are classified as GC star via their \col colour (\textgreater 1.34\,mag).
Some stars (e.g. F2\_26652917-28857695\_Ks12.95, F2\_26652484-28854546\_Ks12.20, F2\_26643866-28966084\_Ks12.46) appear to have proper motions directing them away from Quintuplet's orbit, so they may have been ejected. 

We also have 9 matches within $\leq$0\farcs2 with the proper motion catalogue of \cite{2021MNRAS.500.3213L}, but only four are classified as GC stars (see Table~\ref{tab:pmhotlib}). Two stars also have proper motions in \cite{2022A&A...662A..11S}, but the results are not consistent, which may be caused by the different reference frames used by these studies. \cite{2021MNRAS.500.3213L} data are in the absolute Gaia reference frame, while \cite{2022A&A...662A..11S} are only relative proper motions.

In addition, we matched our data with the Quintuplet proper motion catalogue of \cite{2022ApJ...939...68H}, which covers only the very east of our data. However, the proper motions are more precise than those of \cite{2022A&A...662A..11S}, and in the Gaia reference frame. We obtain 12 matches (coordinates match within $\leq$ 0\farcs2), which we list in Table \ref{tab:pmhothosek} and show in Fig. \ref{fig:hotpmq}. From the nine stars that we classify as GC stars (marked by a red diamond), eight move roughly parallel to Quintuplet and with a similar amplitude. Only one of these stars was reported as a spectroscopic hot star in the stellar census of Quintuplet stars by \cite{2018A&A...617A..65C} and is also a Paschen $\alpha$ source in \cite{2011MNRAS.417..114D}. Their proper motions, projected location within the tidal radius, and their spectral types make these eight stars likely Quintuplet members.

\section{Late-type star stellar parameters and kinematics}
\label{sec:ltresult}
 \begin{figure}
 \centering
 \includegraphics[width=\columnwidth]{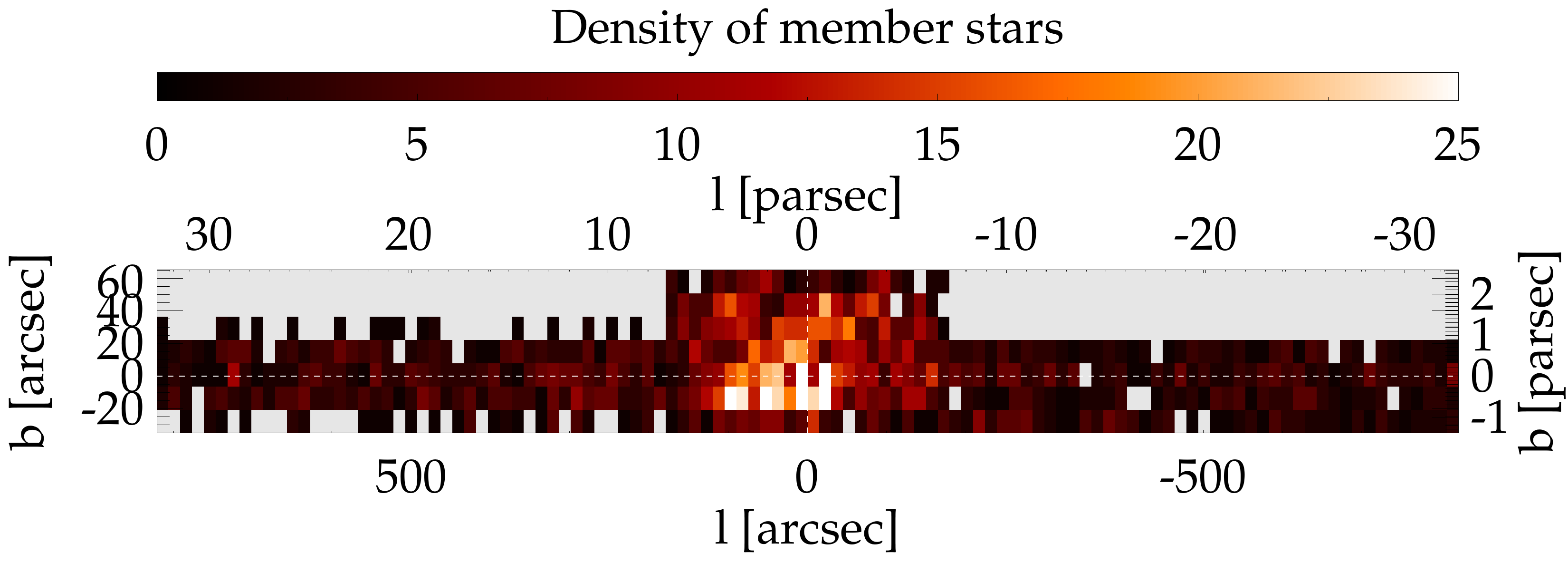}
 \caption{Density of late-type GC stars with stellar parameter measurements, as a function of Galactic longitude and Galactic latitude, centred on \sgra. We have the largest density in the centre of our field, where the stellar density is also the highest. The higher density in the inner 5\,pc east of \sgra compared to inner 5\,pc west of \sgra is likely caused by the lower extinction of the east region (see also Fig.~\ref{fig:vorK}, bottom panel), and is already visible in the photometric catalogue. Note that the vertical axis is stretched relative to the horizontal axis of the plot to improve visibility. }
 \label{fig:latedens}
 \end{figure}

 \begin{figure*}
 \centering
 \includegraphics[width=17cm]{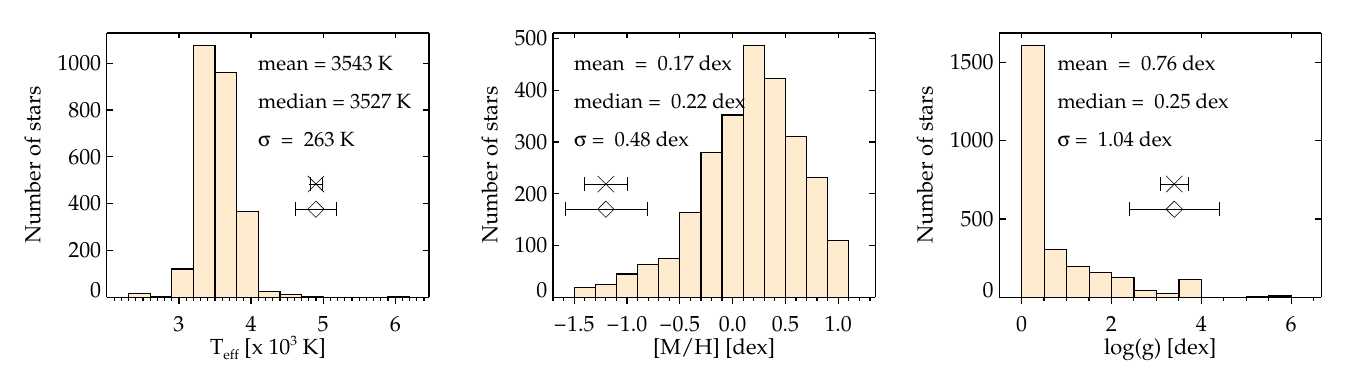}
 \caption{Stellar parameter distributions of GC stars. From left to right panel: effective temperature \teff, overall metallicity \mh, and surface gravity \logg. We denote the mean, median, and standard deviation of the distributions on each panel, and we show the mean statistical and total uncertainty with a cross and diamond symbol. }
 \label{fig:spdist}
 \end{figure*}

After deselecting stars with low S/N and poor fits (Sect. \ref{sec:spfit}), we obtain stellar parameters for 2\,715 stars, of which we classified 2\,580 as GC stars. We show their density distribution across our observed FOV in Fig.~\ref{fig:latedens}. Similar to what we see in Fig.~\ref{fig:fov}, the stellar density is highest in the centre, around \sgra.

\subsection{Stellar parameter distributions}
\label{sec:spdist}
We show the stellar parameter distributions of the 2\,580 GC stars for \teff, \mh, and \logg\space in Fig. \ref{fig:spdist}. 
The values of \teff\space (left panel) are mostly 3\,000--4\,000\,K and consistent with M-type red giant stars. The surface gravity \logg\space (right panel) can hardly be constrained by our data but is also consistent with red giant stars.

The \mh\space distribution (middle panel) has a wide range, covering values from -1.5\,dex to +1.0\,dex, which includes the entire range of the model grid. Most stars have super-solar overall metallicity (\mh\textgreater 0\,dex). About one out of four stars even has \mh\textgreater 0.5\,dex. Our method derives the overall metallicity \mh, meaning that all elements are considered in the measurement (not only Fe), and [$\alpha$/Fe]=0\,dex and thus [$\alpha$/H]=\mh in the models. 
We note that very high iron abundances of [Fe/H]\textgreater0.5\,dex have not yet been found in high-spectral resolution data \citep{
2018ApJ...855L...5D,2020ApJ...894...26T,2025ApJ...979..174R}, and we could not calibrate if our method works at such high values. It is possible that high elemental abundances (as indicated by high $EW_\text{Ca}$ and $EW_\text{Na}$ for a given $EW_\text{CO}$, see Fig.~\ref{fig:index3}) push the overall metallicity \mh\space measurement of Galactic Centre stars to high values. Although we excluded the spectral regions around the \ion{Ca}{I} and \ion{Na}{I} lines from our \mh fit, other elements and lines in the spectra can indicate super-solar abundances \citep{2018ApJ...855L...5D,2020ApJ...894...26T}. Thus, a value of \mh\textgreater0.5\,dex is not only caused by iron. Nonetheless, such stars can be considered stars with super-solar metallicity.

\subsection{Stellar evolutionary stages}
The extinction corrected $K_{S,0}$ photometry (see colour-magnitude diagram, right panel in Fig. \ref{fig:cmd}) 
 helps us to estimate the luminosity class of the stars. According to \cite{2003ApJ...597..323B}, red supergiant stars located in the Galactic Centre are normally bright ($K_{S,0}$=5.7\,mag), but can be as faint as $K_{S,0}$=7.5\,mag, which applies to only 50 of our GC stars. 
However, stars with $K_{S,0}$\textgreater5.7\,mag are most likely red giant branch (RGB) stars, and we conclude that the majority of the stars in our sample are normal red giant stars. All of them are brighter than the red clump, which is centred at $K_{S}$$\sim$15.8\,mag \citep[for a mean extinction of $A_K$$\sim$2.6\,mag, ][]{2020A&A...641A.102S}. 

A few stars in our sample must be asymptotic giant branch (AGB) stars. They tend to be younger than RGB stars ($\sim$1--2\,Gyr, depending on their mass). 
The time a star spends on the AGB is about 40 times shorter than their time as an RGB star \citep{2011spug.book.....G}. Hence, we expect $\sim$65 AGB stars in our sample. Via their variability, we identified 26 (39 including non-GC, coordinates match within \textless0\farcs3) AGB Miras from the catalogue of \cite{2009MNRAS.399.1709M}, accounting for 41\% of the expected AGB stars. In total, we found that 72 (100) of the stars in our GC (full) sample are listed as variable stars in \cite{2009MNRAS.399.1709M}. Not for all of them a period could be obtained, hence, it is unclear what causes the variability and if they are AGB Miras.

We also matched our sample with the variable star catalogue of \cite{2013MNRAS.429..385M} and found two matches (coordinates match within \textless0\farcs2), which happen to be among the 4 warmest stars in our sample with \teff\textgreater 4\,700\,K \citep[F2\_26637872-29052971\_Ks10.57 and F2\_26638443-29048691\_Ks10.26 correspond to stars 18 and 20 in][] {2013MNRAS.429..385M}. These stars are classical Cepheids, i.e. pulsating supergiants that evolved from intermediate-to-high-mass stars (4--10\,\msun).

\subsection{Position-velocity diagram and high-velocity stars}
\label{sec:highv}

\begin{figure}
 \resizebox{\hsize}{!}{\includegraphics{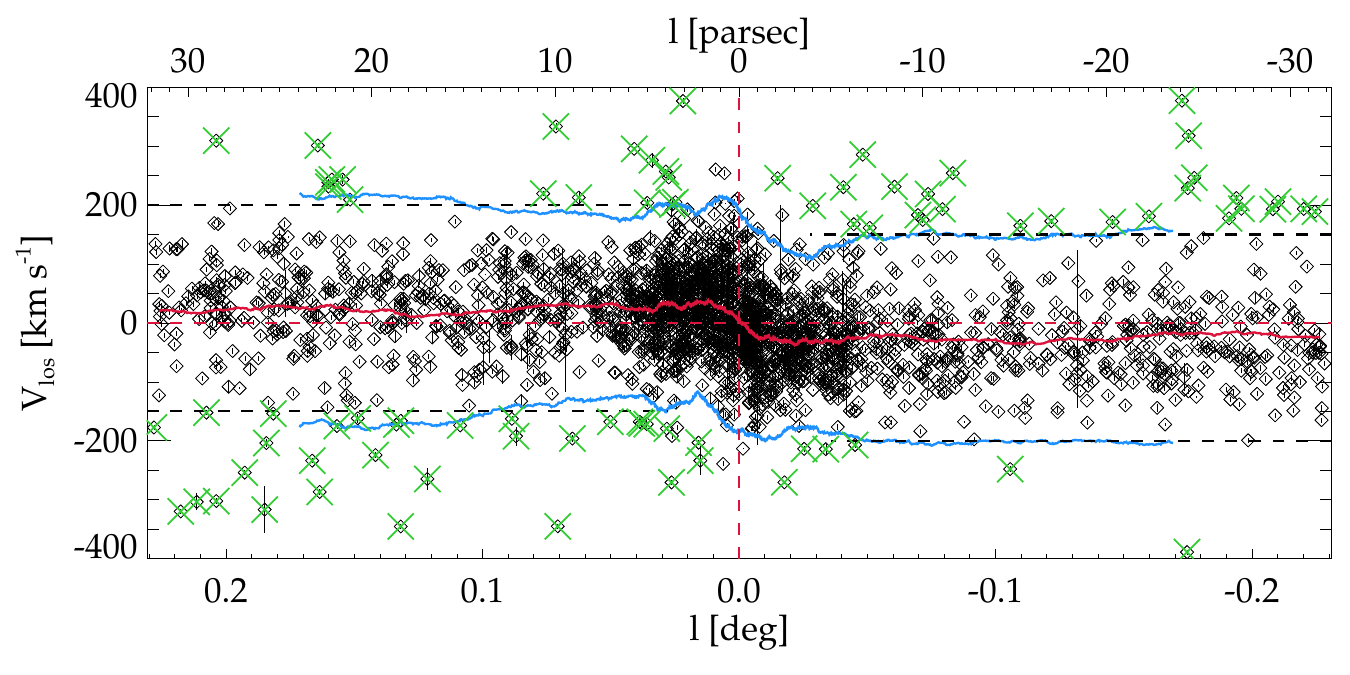}}
 \caption{Position-velocity plot of Galactic Centre late-type stars along Galactic longitude, centred on \sgra.
 Each diamond symbol denotes a star, and green x-symbols denote stars that we consider high-velocity stars. The red line denotes the moving average \vlos\ of 150 stars, the blue lines denote the moving robust $\sigma_r\times$2.5, which is close to our cuts to classify high-velocity stars, shown as dashed horizontal lines. }
\label{fig:lvel_disc}
 \end{figure}

We find on average rotation of the stars in the same sense as the Milky Way disc, with stars receding in the Galactic east and approaching in the Galactic west. 
We show 2\,580 stars (after the colour cut to remove non-GC stars) on a position-velocity diagram in Fig.~\ref{fig:lvel_disc}. The red line denotes the moving average \vlos as a function of Galactic longitude $l$, computed using the nearest 150 stars. This curve is roughly symmetrical about $l$=0, and flat beyond the inner $\sim$100\arcsec (3.9\,pc), with an absolute value close to 25\,\kms.

The position-velocity diagram reveals several stars that move apparently in the opposite direction, and stars with large \vlos, exceeding 150\,\kms. 
We classify a star as a high-velocity star if it satisfies one of the following conditions: 
\begin{enumerate}
 
\item located at a distance of $r$\textgreater50\arcsec\space from \sgra, and |$V_\text{LOS}$| \textgreater$V_\text{cut,1}$=200\,\kms,
\item at $r$\textgreater100\arcsec\space from \sgra, and \vlos\textgreater$V_\text{cut,2}$=150\,\kms, if the star is located \textgreater100\arcsec\space to the west of \sgra (i.e. counter-rotating in the West), or
\item at $r$\textgreater100\arcsec\space from \sgra, and \vlos\textless$-V_\text{cut,2}$= -150\,\kms, if the star is located \textgreater100\arcsec\space east of \sgra (i.e. counter-rotating in the East). 
\end{enumerate}

The values of $V_\text{cut,1}$ and $V_\text{cut,2}$ are chosen to be symmetrical about the flat \vlos curve, and after visual inspection of the position-velocity curve. For comparison, the overall moving robust $\sigma$ times a factor of 3 is $\sim$200\,\kms, the exact value depends slightly on the number of stars used for the moving average.

Even though these high-velocity stars are only a small number compared to the size of our data set, we excluded them as possible contaminants from further analysis. We discuss their potential origin in Sect. \ref{sec:highvdisc}.

 \subsection{Stellar parameter maps}
 \label{sec:SPmaps}
 \begin{figure*}
 \centering
 \includegraphics[width=17cm]{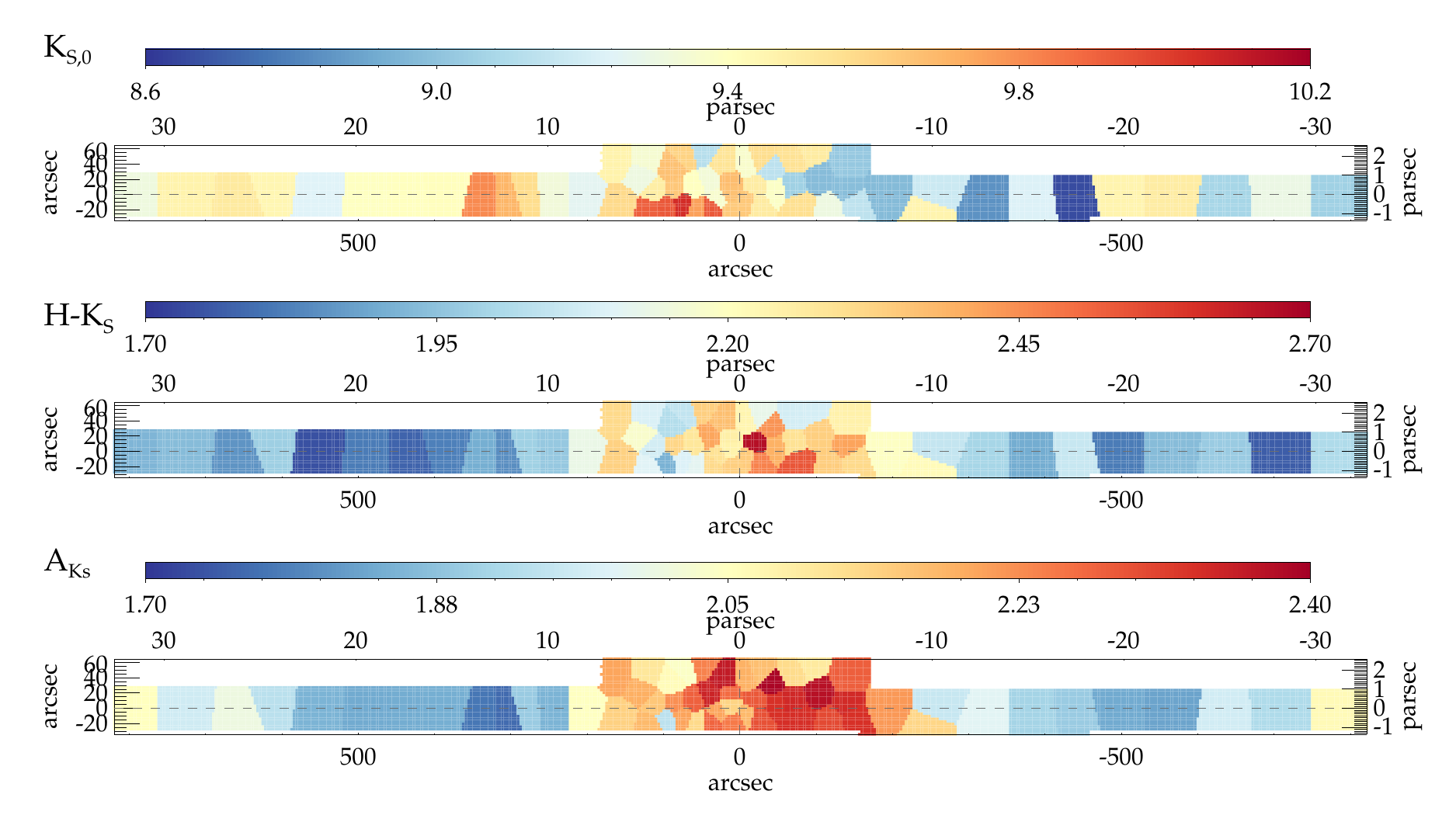}
 \caption{Mean photometric properties in different Voronoi bins, top: $K_{S,0}$, middle \col, bottom $A_{K_{S,0}}$. The mean $K_{S,0}$ indicates that the stars in the central region are slightly fainter than stars in the outer regions, especially in the west. The stars in the centre are slightly redder (higher mean \col), and in this region, the extinction \ak is higher. High-velocity stars and foreground stars were excluded. Each bin contains $\sim$40 stars. }
 \label{fig:vorK}
 \end{figure*}

 \begin{figure*}
 \centering
 \includegraphics[width=17cm]{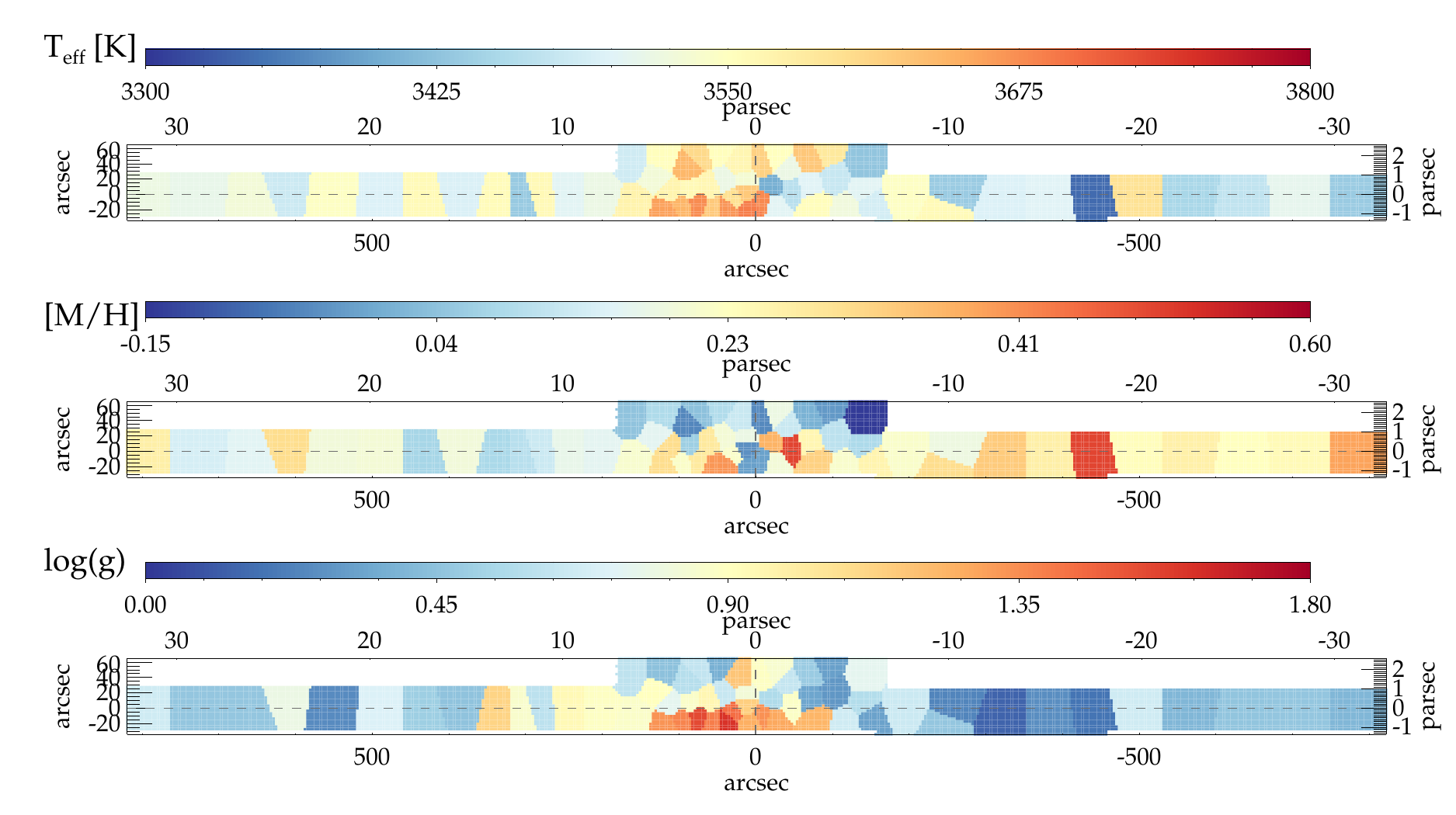}
 \caption{Mean effective temperature \teff, overall metallicity \mh, and surface gravity \logg. The regions with fainter stars in Fig.~\ref{fig:vorK} (top panel) have higher \teff\space and \logg, as expected. High-velocity stars and foreground stars were excluded. Each bin contains $\sim$40 stars.}
 \label{fig:vorSP}
 \end{figure*}
 \begin{figure*}
 \centering
 \includegraphics[width=17cm]{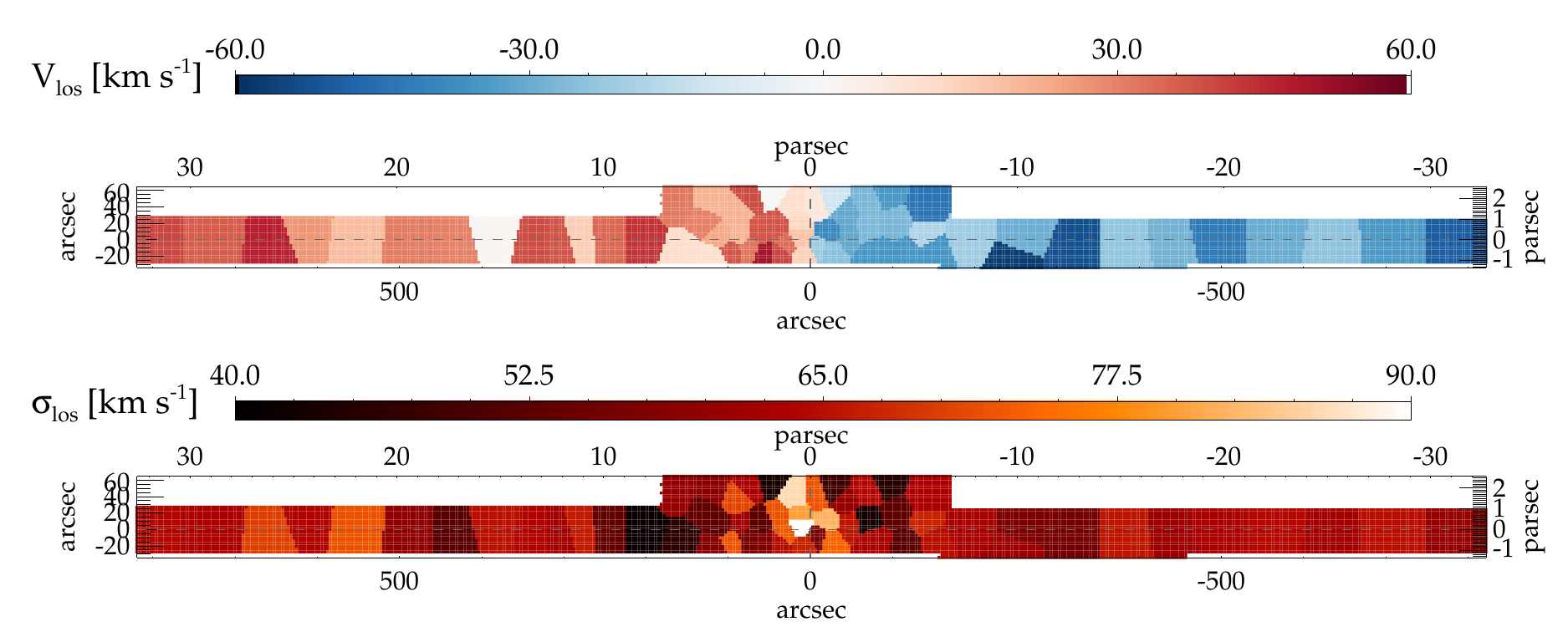}
 \caption{Mean line-of-sight velocity \vlos\ (top) and velocity dispersion \slos\ (bottom) in bins of $\sim$40 stars. High-velocity stars and foreground stars were excluded. }
 \label{fig:vsig}
 \end{figure*}

In this section, we study how the mean properties of the stars vary spatially. 
We apply Voronoi binning on the data (using the code provided by \citealt{2003MNRAS.342..345C}), such that nearby stars are grouped in bins of $\sim$40 stars each. We also tried different bin sizes (e.g. 30, 60, 100) but found consistent trends. A lower number of stars per bin naturally gives noisier maps, and a higher number of stars per bin washes out spatial differences. Then we computed the mean values of various parameters, e.g. extinction corrected $K_{S,0}$-band photometry, observed colour \col, extinction \ak (see Fig.~\ref{fig:vorK}), \teff, \mh, and \logg\space (see Fig.~\ref{fig:vorSP}). Maps of the median instead of mean value per bin show similar trends, i.e., regions with high or low values coincide. We also show the moving average profiles of 200 stars along Galactic longitude $l$ in Figs.~\ref{fig:lphotav} and \ref{fig:lspav}, which show consistent trends.

Our maps show a region near the centre, located southeast of \sgra\ (-1\,pc\textless $b$\textless0\,pc, 0\,pc\textless$l$\textless+8\,pc), where the stars are on average fainter, warmer, and have higher \logg\ than e.g. the opposing northwest side of \sgra\ (0\,pc\textless$b$\textless+1\,pc, $-8$\,pc\textless$l$\textless0\,pc), and also the northeast, and southwest sides. We compared the GNS catalogue and found that indeed, the southeast side of \sgra\ has a higher density of stars with 10\,mag\textless$K_S$\textless12\,mag compared to the west. 
The higher average values of several parameters (\teff, \logg, and $K_S$) in the southeast of \sgra\ agree with expectations from red giant star evolution. Along the red giant branch, \teff decreases slightly while the total luminosity increases and \logg decreases. 
Hence, if we detect on average fainter stars, it is natural that they have on average higher \logg and are warmer. However, our uncertainties of \teff and \logg are substantial, and quantitative comparison with isochrones is not meaningful. 
The inner -10\,pc\textless$l$\textless10\,pc region has higher extinction (\ak\textgreater 2.0\,mag), which also causes the redder \col. A redder colour can indicate a larger line-of-sight distance \dlos, as redder stars lie behind a larger amount of interstellar dust.

As shown in Fig.~\ref{fig:vorSP}, the mean metallicity, $\mean{\rm \mh}$, varies spatially. There appears to be an increase from the central $r\lesssim$1\,pc region around \sgra with $\mean{\rm\mh}$ = 0.06\,dex ($\sim$200 stars), towards the surrounding region (-5\,pc\textless $l$\textless 5\,pc and -1\,pc\textless $b$ \textless 1\,pc, $\sim$650 stars), with a $\mean{\rm\mh}$=0.27\,dex. Further, $\mean{\rm\mh}$ decreases towards the north at $b\gtrsim$1\,pc to 0.02\,dex ($\sim$500 stars). This trend can also be seen in Fig.~\ref{fig:lspav} top panel, where the blue dashed line, indicating the region $b$\textless1\,pc, has higher $\mean{\rm \mh}$ than the orange line, denoting $b$\textgreater1\,pc. 
All these regions are dominated by the NSC stars rather than the NSD stars according to the projected surface density (Sec.~\ref{sec:member}). 
The lower \mh found in the north is in agreement with the larger fraction of sub-solar \mh stars in the Galactic north compared to the south reported in \cite{2020MNRAS.494..396F}. While our data in the north extend even further than the data of this study, we have no coverage in the south at $b$\textless-1\,pc. 

At $|l|\sim$2--8\,pc, and $b\sim$0\,pc, $\mean{\rm \mh}$ decreases on both sides of the NSC. This decline does, however, not continue into the inner NSD, and from 10\,pc outwards, where the NSD is dominating the stellar density (Sect.~\ref{sec:member}), $\mean{\rm \mh}$ is rather constant on both sides. 
There appears to be a slight east-west asymmetry, with $\mean{\rm\mh}$= 0.15\,dex at $l$\textgreater 10\,pc ($\sim$450 stars, east) and $\mean{\rm\mh}$=0.26\,dex at $l$\textless-10\,pc ($\sim$400 stars, west). In these regions, the contribution of the NSC to the stellar surface density is below 50\%. At $l$\textgreater 10\,pc and $l$\textless-10\,pc, we find a similar level of foreground extinction ($A_K \sim$1.8--2.0\,mag), which is lower than in the NSC-dominated area (2.1-2.4\,mag).

The distribution of \mh\ is broad, with $\sigma_{\rm \mh}$=0.43--0.47\,dex. 
However, the variation of $\mean{\rm\mh}$ is likely too large to be caused by randomly drawing a finite number of stars from the entire sample. We tested this by drawing 1,000 random samples from the \mh\ distribution (after removing high velocity and foreground stars) of our data, with varying sample sizes (40 -- 200 stars). For each sample, we compute $\mean{\rm\mh}$, and the resulting value is always close to the overall $\mean{\rm\mh}$, 0.17\,dex, even if we draw only 40 stars. As expected, the standard deviation of the 1,000 simulated sub-samples decreases with increasing sample size. We computed how likely it is to obtain the values of $\mean{\rm\mh}$ stated above for the different regions and for the given number of stars: Obtaining $\mean{\rm\mh}$=0.06\,dex (as found in the central 1\,pc) when drawing 200 stars randomly from the data set is 3.5$\sigma$ below the expected value, while $\mean{\rm\mh}$=0.27\,dex (as in the surrounding region, for 200 stars) is 2.8$\sigma$ above. If we draw random samples only from the stars in the inner \textless8\,pc, we obtain -2.9$\sigma$ and +3.5$\sigma$, respectively. If we allow the individual \mh\ measurements to vary within their respective total uncertainties $\sigma_{\rm \mh}$, the significance is slightly lower, 2.2$\sigma$ and 2.5$\sigma$.

This suggests that the null hypothesis that there is no spatial variation of the \mh\ distribution (even only in the inner \textless8\,pc) can be discarded, and there is a real spatial variation of $\mean{\rm\mh}$ in the data. The east-west asymmetry is less significant, with only 1.6$\sigma$ and 0.9$\sigma$ (for 200 stars and if considering individual \mh uncertainties).
We discuss possible explanations for the variation of 
$\mean{\rm\mh}$ in Sect. \ref{sec:discmetvar}.

We further analysed the different regions with one-dimensional Gaussian mixture models (GMMs). We tested single and double Gaussian models and used the Bayesian information criterion (BIC) and the Akaike information criterion (AIC) to decide which is a better representation of the data. In all regions except the centre, a double Gaussian gives a better fit to the data. We list the resulting Gaussian centres, widths, and relative weights of the first double-Gaussian in Table \ref{tab:gmm} for the different regions (the corresponding histograms are shown in the appendix, Fig.~\ref{fig:gmm}). We obtained uncertainties by running Monte Carlo simulations with 1\,000 different data representations. We used the statistical uncertainty $\sigma_{\rm \mh}$ to draw modified values of \mh for each star in the regions and repeated the GMM analysis. The standard deviation of the 1\,000 runs is listed as the uncertainty of the GMM, and the mean as the value. Using the mean of the MC runs rather than the actual value increases [M/H]$_2$ and the Gaussian widths $\sigma_{\rm \mh_{1,2}}$.

The east and west regions are similar, though the east region has a larger weight at the low \mh Gaussian, and the second component is centred at lower \mh compared to the west region. In comparison to all other regions, the north region has the lowest \mh values for both Gaussian components, reflecting the lower values of $\mean{\rm \mh}$ in this region. The central \textless1\,pc region is the only one where a single Gaussian gives better results than a double Gaussian. 
The region surrounding the innermost $r$=1\,pc has indeed the highest \mh of all regions for both components, confirming what we see for $\mean{\rm \mh}$ in Fig.~\ref{fig:vorSP}. This is caused by a relatively large number of stars with very high \mh\textgreater 0.5\,dex, which may be overestimated (see Sect. \ref{sec:spdist}). The widths of the first components are similar everywhere ($\sigma_{\rm \mh_1}$=0.48--0.60\,dex), and broader than that of the second, higher \mh components ($\sigma_{\rm \mh_2}$=0.35--0.49\,dex). 
Overall, the GMM analysis confirms our findings of a spatially varying \mh distribution, with lower \mh in the north, the central $r$\textless 1\,pc, and higher \mh in the surrounding \textgreater1\,pc region.

%---------------------------
\begin{table*}
\caption{Gaussian mixture model of the \mh distribution in different regions.}
 \label{tab:gmm}
\begin{tabular}{@{}llcccccr@{}}
\noalign{\smallskip}
\hline
\noalign{\smallskip}
Region &Extent&[M/H]$_1$ & $\sigma_{\rm \mh_1}$& [M/H]$_2$ & $\sigma_{\rm \mh_2}$ &Weight$_1$ &N \\
 \noalign{\smallskip}
\hline
\noalign{\smallskip}
\hline
All&...                                                                                                & -0.18 $\pm$  0.03 &  0.55 $\pm$  0.03 &  0.39  $\pm$ 0.02&  0.38  $\pm$ 0.01  & 0.37 $\pm$  0.02  &  2498\\
East & $l$\textgreater 10\,pc                                                                            & -0.23 $\pm$  0.13 &  0.56 $\pm$  0.06 &  0.35   $\pm$ 0.03 &  0.49  $\pm$ 0.02  & 0.33 $\pm$  0.04  &   447\\
West & $l$\textless -10\,pc                                                                          & -0.24 $\pm$  0.10 &  0.55 $\pm$  0.05 &  0.46   $\pm$ 0.02 &  0.42  $\pm$ 0.01  & 0.25 $\pm$  0.04  &   391\\
North& $b$\textgreater1\,pc                                                                            & -0.32 $\pm$  0.10 &  0.60 $\pm$  0.08 &  0.19   $\pm$ 0.04 &  0.48    $\pm$ 0.02  & 0.33 $\pm$  0.05  &   511\\
Centre&$r$\textless 1\,pc                                                                              & 0.06 $\pm$ 0.02 &  0.49 $\pm$  0.02 &   ...              &    ...               & ...  &   197\\
Surrounding&$r$\textgreater 1\,pc, $\lvert l\lvert$\textless 5\,pc,  $\lvert b \lvert$\textless 1\,pc  & -0.06 $\pm$  0.04 &  0.48 $\pm$  0.03 &  0.59   $\pm$ 0.04 &  0.35    $\pm$ 0.02  & 0.47 $\pm$  0.04  &   648\\

\hline 
\end{tabular}

\end{table*}

\subsection{Stellar kinematic maps and variations}
\label{sec:skinvar}
The stars in the NSC and NSD rotate in the same sense as the rest of the Galaxy, with \sgra in the centre. In Fig. \ref{fig:vsig}, we show maps of the mean \vlos\ (top) and velocity dispersion \slos\ (bottom) in bins of $\sim$40 stars each. Foreground stars and high-velocity stars (Sects. \ref{sec:member}, \ref{sec:highv}) were not considered for these maps. The \slos\ map has a maximum in the innermost bin ($\sim$90\,\kms), this marks the typical \slos\ increase around a supermassive black hole. Further out, \slos\ is rather constant at $\sim$62\,\kms. 

 \begin{figure}
 \resizebox{\hsize}{!}{\includegraphics{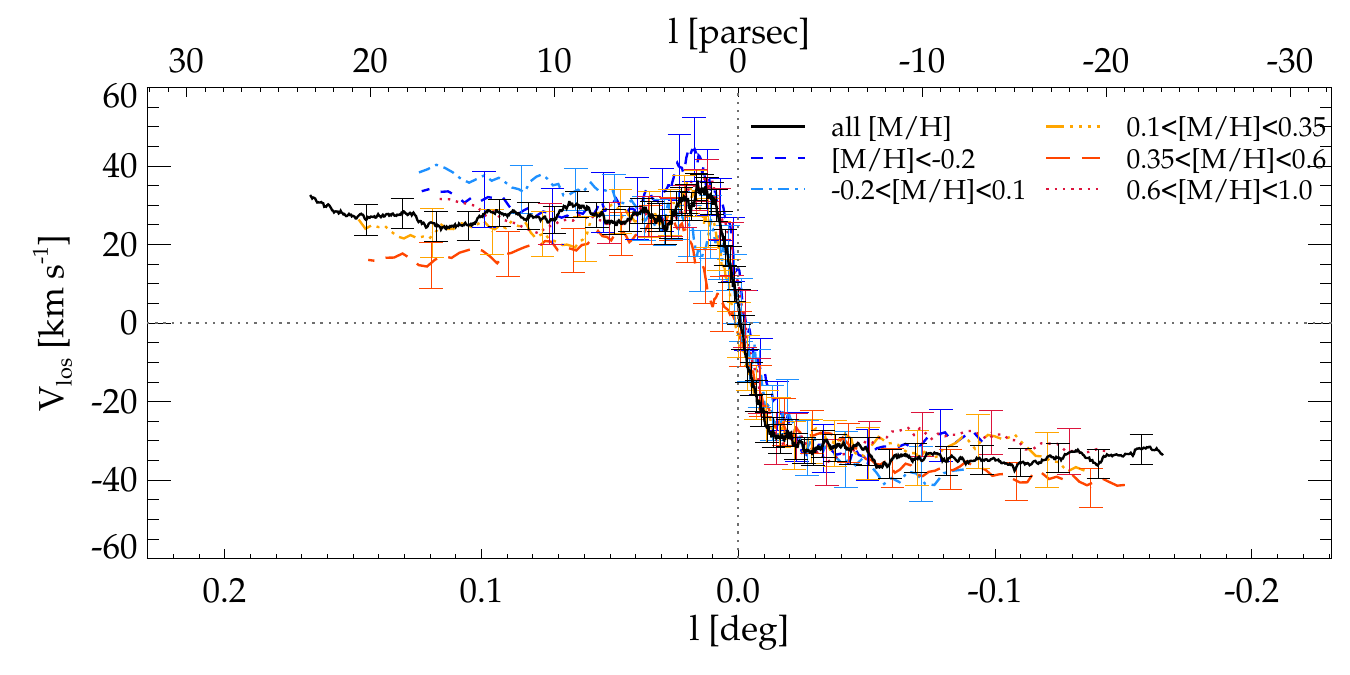}}
 \resizebox{\hsize}{!} {\includegraphics{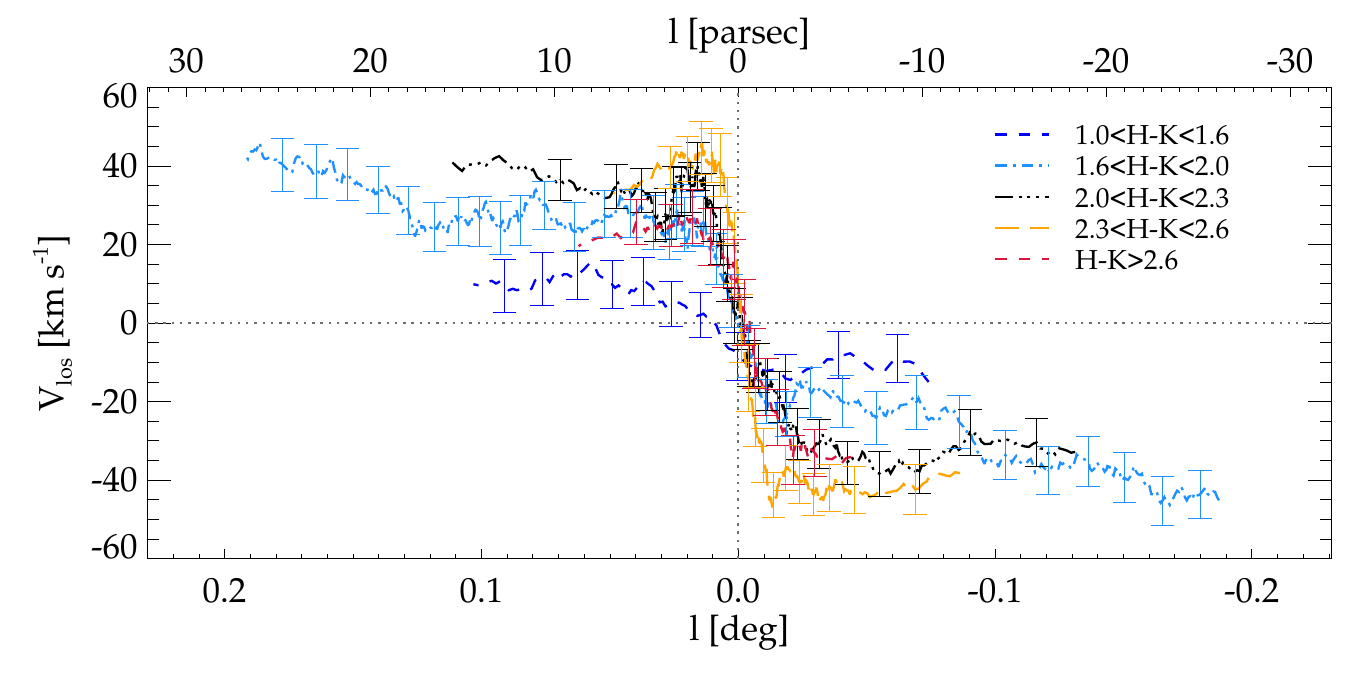}}
 \caption{Binned position-velocity plots of Galactic Centre late-type stars after removing high-velocity stars, along Galactic longitude, centred on \sgra.
 Top panel: We apply different bins in \mh, and show the position-velocity curves for different \mh\ bins (moving average of 100 stars). Different colours and lines denote the different \mh\ selection; see figure legend. Bottom panel: Same as top panel, but applying bins in \col. } 
\label{fig:lvel_mbin}
 \end{figure}
 
Since the stars have a broad \mh\ distribution, it is interesting to check if the stellar kinematics shows any dependence on \mh. We made position-velocity plots along Galactic longitude $l$, and binned the data according to their \mh, see Fig. \ref{fig:lvel_mbin}, top panel. In each \mh\ bin, the stars show a rotation signature. There is more variation in the Galactic east among different \mh bins, and the moving average value of \vlos\ has a standard deviation of up to $\sim$10\,\kms, while it is below 5\,\kms in the west. The velocity curves agree within their uncertainties in the west but show some deviation in the east, exceeding the uncertainties. As a consequence, the velocity curves are not perfectly symmetric about \sgra. We quantify this deviation from asymmetry by calculating the median difference of the absolute value of \vlos on the two sides and the robust standard deviation. 
The numbers tell us that the  0.35\,dex\textless\mh\textless0.6\,dex bin has the largest deviation from symmetry, with a median difference of 17.7\,\kms, while all other \mh bins have \textless5\,\kms. This \mh bin has the lowest velocity on the east side from all the \mh bins, which is the reason for the asymmetric \vlos curve.   We cannot detect a variation of the \vlos curve with \mh\ as seen by \cite{2021A&A...650A.191S} in the NSD, at larger distances ($l\lesssim$200\,pc) from \sgra than our data. \cite{2021A&A...650A.191S} found that stars with increasing \mh\ show a stronger rotation signal, but we see no significant differences in \vlos with \mh in the west, and differences in the east do not show such a trend in our data. These trends do not change if we use a stricter colour cut to exclude foreground stars (e.g. \col\textless1.3\,mag instead of 1.0\,mag), and exclude more stars as high-velocity stars (e.g. $\lvert$\vlos$\lvert$\textgreater 120\,\kms instead of 150\,\kms). 
The stars in the inner $r$\textless2\,pc have a velocity dispersion ranging from 71--85\,\kms.  Stars in the highest \mh bin have the lowest velocity dispersion value, and stars with \mh\textless0.1\,dex the highest. This may indicate slight differences in the projected distance of the stellar samples to \sgra.

We also make a position-velocity plot (running average of 100 stars) where we bin stars according to the observed \col colour (bottom panel of Fig. \ref{fig:lvel_mbin}), and there we see significantly larger variations. The stars with the bluest colour (212 stars, 1.0\,mag\textless\col\textless 1.6\,mag) show the weakest rotation signal, with $\lvert \vlos\rvert\lesssim$ 16\,\kms. Stars with 1.6\,mag\textless\col\textless 2.0\,mag (834 stars) do show stronger rotation (up to $\lvert \vlos\rvert\sim$45\,\kms), but the maximum $\lvert \vlos\rvert$ are at the outermost values of $l$. We observe a strong $\lvert \vlos\rvert$ peak at 46\,\kms in the inner $l$\textless4\,pc for the 2.3\,mag\textless\col\textless2.6\,mag bin (474 stars). The neighbouring colour bins, 2.0\,mag\textless\col\textless 2.3\,mag (635 stars) and 2.6\,mag\textless\col\textless 3.5\,mag (343 stars)  show indications for such a peak, but less pronounced.

Besides the variation of the maximum $\lvert \vlos\rvert$ and its position, we note that the stars with bluer \col are more extended along $l$, while the stars with redder \col can be found predominantly in the centre, in agreement with the mean \col map shown in Fig. \ref{fig:vorK}, middle panel. Stars with red colour (\col \textgreater 2.3\,mag) have the highest velocity dispersion in the inner $r$\textless2\,pc, \textgreater82\,\kms, stars with bluer colour (1.6\,mag\textless\col\textless 2.0\,mag) have 71\,\kms, and the stars in the bluest bin only 39\,\kms.  We discuss these results in Sect. \ref{sec:discposvel}.

\section{Discussion}
\label{sec:discussion}

\subsection{Hot star discoveries}
\label{sec:dischot}
We identified 78 hot star candidates, of which 48 are classified as GC stars based on their \col colour. Fifteen of these stars are located in the projected inner 1\,pc region around \sgra, and those stars match the brightest of the hot young stars extensively studied in the literature \citep[e.g.][]{1991ApJ...382L..19K,2006ApJ...643.1011P,2009ApJ...690.1463L,2015A&A...584A...2F,2022ApJ...932L...6V}. 

We discovered 31 hot stars that were, to the best of our knowledge, not yet reported in the literature as hot stars.  A larger number of these hot stars is located in the Galactic east, at $l$=30\,pc, with 11 (and an additional 4 non-GC) sources just a few pc north of the Quintuplet cluster of young stars. Only one of these stars is listed as a spectroscopically confirmed hot star by \cite{2018A&A...617A..65C}. We tested if these stars may be associated with the Quintuplet cluster by matching them with the proper motion catalogue of \cite{2022ApJ...939...68H}. We found that indeed eight stars have similar proper motions to Quintuplet. The most distant star is $\sim$2.7\,pc (projected on the sky) away from Quintuplet's centre, and the closest one only $\sim$1\,pc. 
\cite{2019ApJ...877...37R} identified 715 Quintuplet cluster members, as far out as 3.2\,pc. They adopt a tidal radius $r_t$=3\,pc, and a core radius $r_c$=0.62\,pc. Hence, the eight stars co-moving with Quintuplet are within its tidal radius and are likely associated with it. 
We found no match in the proper motion catalogue for one of the stars, as it is likely outside of its FOV. One star (F2\_26654034-28827679\_Ks11.74) is moving more northwards and is faster than the rest of the stars that co-move with Quintuplet. It may either not be part of the cluster, or some dynamical event (e.g. a close encounter, kick) may have changed its proper motion.

There are several rather isolated young stars in our FOV, with no association to the known young star clusters. Seventeen GC stars are located outside Quintuplets tidal radius or the NSC's \re. Only two of them were already reported as spectroscopic hot stars in \cite{2022MNRAS.513.5920F}. Nonetheless, their discovery is unsurprising, as several studies also found young, massive stars throughout the GC. Most of the known isolated hot stars were discovered thanks to Paschen (Pa) $\alpha$ excess, caused by emission lines \citep{2010ApJ...725..188M,2011MNRAS.417..114D,2015MNRAS.446..842D,2021A&A...649A..43C}. These stars are either Wolf Rayet stars, O hypergiants, or luminous blue variable stars. \cite{2021A&A...649A..43C} discuss that Pa $\alpha$ surveys can miss mid-O supergiants, and detect no O5-9 stars of luminosity class III to V. 
Our hot star candidates are not limited to emission line stars, and the stars in our sample can have Br$\gamma$ absorption. Thus, our sample likely includes different spectral types of stars, including O and B giants to main sequence stars. Indeed, 17 of our isolated hot stars in the Galactic centre (26 including stars close to the centre and Quintuplet) are in the range of $K_{S,0}=K_S-A_{K_S}$=10--12\,mag, which makes them probably  O3V - B0V stars  \citep{2012ApJ...746..154P, 2013ApJS..208....9P}. Such stars have no emission lines, which explains why the Pa $\alpha$ survey did not detect them.

Some of the isolated hot stars have proper motions from \cite{2022A&A...662A..11S} that are, however, neither parallel to Quintuplet's proper motion nor aligned with its orbit, as would be expected from a tidal tail \citep{2014A&A...566A...6H}. Also, \cite{2019ApJ...877...37R} did not detect tidal tails for Quintuplet, but \cite{2018MNRAS.478..183P} showed that tidal tails are barely detectable via massive stars, as most stars in the tidal tail are \textless 2.5\,\msun.   

A possible mechanism to eject a hot star from a star cluster is a three-body interaction of a single star with a binary or a multiple system. According to \cite{2011MNRAS.410..304G}, in more than 50\% of encounters of a massive star (12--20\,M$_{\sun}$) with a massive binary, the star is accelerated, typically to $\sim$90\,\kms. 
Slightly higher velocities can be reached if a binary system evolves into a supernova (SN) and ejects a star. Quintuplet and the central parsec cluster are old enough for this process, while Arches is too young. 
If three-body interactions and SNe ejections generate runaway kicks of $\sim$100\,\kms, a star can be displaced by $\sim$100\,pc within 1\,Myr \citep{2021A&A...649A..43C}. This is less than the age of the young clusters and makes them, therefore, potential sources for some of the isolated massive young stars found throughout the GC. 

Then, there is also the possibility that the isolated young stars formed in small groups that are already dissolved, or even in true isolation \citep{2015MNRAS.446..842D}. Stars from dissolved clusters or groups are expected to retain information on their kinematics and move with similar velocities. Such co-moving groups of stars were reported by \cite{2019A&A...632A.116S,2024A&A...683A...3M}. 
A better estimate of their kinematics and spectral type would be useful to constrain the origin of these isolated hot stars. With higher-quality spectra at a better spatial resolution than our data, it should be possible to constrain the spectral type and luminosity class of the stars and, in addition, measure their \vlos.

\subsection{High-velocity star origin}
\label{sec:highvdisc}

Our sample contains several stars that move with a high \vlos, some of them even in the opposite sense than the bulk of the stars. 
The high-velocity stars have lower \mh-values than the slower stars. The mean value of \mh\space of the 82 high-velocity stars is -0.21\,dex, the median even -0.27\,dex, significantly less than the mean and median of the other $\sim$2\,500 stars, +0.17\,dex and +0.22\,dex. The observed colours \col and $J-K_S$ are on average less for the fast stars, with mean values of 1.87 and 7.50\,mag versus 2.13 and 9.98\,mag, indicating that they may be on average closer to us along the line-of-sight, lying in front of some of the Galactic Centre dust. However, the distributions of \teff, \logg, or $K_0$ are not significantly different. 

 \begin{figure*}
 \centering
 \includegraphics[width=18cm]{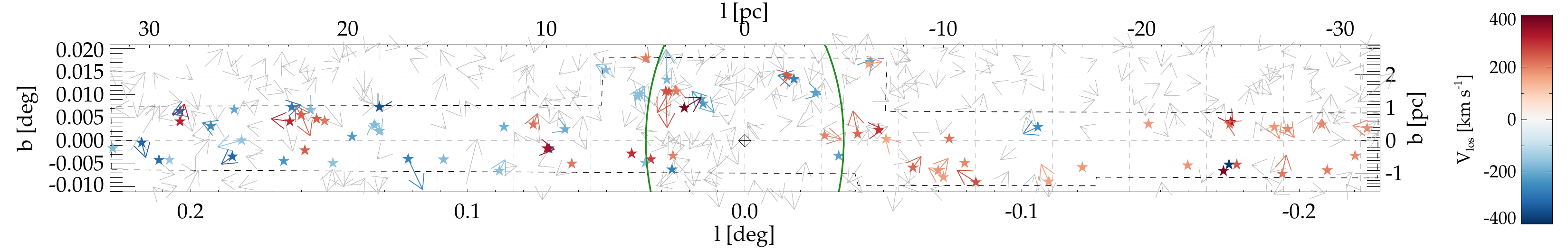}
 \caption{Spatial distribution and 3d motions of high-velocity stars in the FOV. 
 The blue-to-red coloured symbols denote high-velocity stars, colour-coded by their \vlos. The arrows denote the direction of their proper motion
 from \cite{2022A&A...662A..11S}, but the arrow length is multiplied by a factor of 3\,000 for better visualisation. The grey arrows are a subset (2.5\%) of the \cite{2022A&A...662A..11S} proper motions to illustrate the distribution of proper motions in this region. Black dashed lines denote the approximate outline of our FOV. The x-axis and y-axis have different scales, therefore, proper motions along the y-axis appear larger. The green circle denotes 1\,\re=5\,pc of the NSC.}
 \label{fig:extpm}
 \end{figure*}

One possibility for the origin of these stars is that they belong to the Galactic bar. For such stars, we expect the mean \vlos\ close to 0\,\kms\ at the location of our data, but also a large velocity dispersion of $\sim$120\,\kms\ \citep{2017MNRAS.465.1621P,2017MNRAS.470.1233P}. If a star is on an $x_1$ orbit in the bar, it moves along the bar's long axis, which is oriented at an angle of $\alpha$=28\degr--33\degr\ \citep{2015MNRAS.450.4050W}. If the star is moving outwards, the star's \vlos has the opposite sign than the \vlos of stars in the NSD and NSC, which are rotating around \sgra. In this case, the star would appear as counter-rotating. The star turns around and moves inwards after reaching the apocentre of its orbit, leading to an apparently co-rotating \vlos. Stars in the bar have lower \mh\ than stars in the Galactic Centre \citep{2019A&A...627A.152S}, which is also in agreement with our findings. Given that we expect a bar contribution of $\lesssim$10\% (averaged over the entire region, see Fig. \ref{fig:fracprof}, bottom panel), these high-velocity stars would account for $\gtrsim$31\% of the bar stars in the region. Thus, we expect even more unidentified bar stars in our GC star sample.

It is also possible that the stars are counter-rotating because they received a kick in a dynamical three-body interaction or in a supernova explosion. But then we would not expect a significantly different mean \mh and colour. While some high-velocity stars may have such an origin, this is likely not the case for all of them. 

Another possibility may be that high-velocity stars are the remnants of dissolved star clusters or the tidal tail of still existing star clusters, lost during a close pericentre passage in the Galactic Centre.  There may be up to 3-4 globular cluster interactions per 1 Gyr within \textless50\,pc \citep{2023A&A...674A..70I}. Over 8\,Gyr, a globular cluster can lose 70--90\% of its initial mass, and some of it may end up in the centre \citep{2024A&A...689A.178I}. This scenario would also be consistent with a lower average \mh\ and colour. However, globular clusters have typically even lower metallicities [Fe/H]\textless-0.3\,dex \citep{1996AJ....112.1487H}. 
If several of the stars belong to the same tidal tail, their proper motions should be pointing in similar directions. However, depending on the time of the infall, this information may already be washed out to some extent. 
The two-body relaxation timescale of stars in the NSC is $\sim$14\,Gyr \citep{2018A&A...609A..28B}, and likely longer at larger distances from \sgr\ \citep{2010ApJ...718..739M}.
However, massive perturbers (e.g. massive clumps of gas, star clusters or giant molecular clouds) can reduce the two-body relaxation time by several orders of magnitude \citep{2007ApJ...656..709P} in the regions beyond the central 1.5\,pc.

We matched the fast stars with the proper motion catalogue of \cite{2022A&A...662A..11S}. Requiring a distance of $\leq$0.2\arcsec, we found 52 matches. The spatial distribution and three-dimensional velocities of these high-velocity stars are shown in Fig.~\ref{fig:extpm}. In regions with a high density of high-velocity stars (e.g. at $l$=4\,pc, $b$=1.5\,pc or $l$=22\,pc, $b$=1\,pc), the stars do not move in the same direction and are possibly only close in projection. We compute the angular momentum in the line-of-sight direction, $L_z=l v_b - b v_l$, to separate stars moving in a projected clockwise and counter-clockwise direction around \sgra. The resulting values of $L_z$ are rather balanced, with 22 stars moving in clockwise and 30 in counter-clockwise direction. The distribution of $L_z$ has a median close to 0 (-33\,mas yr$^{-1}$ $\cdot$  arcsec ).  We conclude that we see no indication of a tidal stream, but this may be because there are several overlapping streams, and we detect too few stars per stream.

In summary, several explanations exist for the origin of high-velocity stars (kicks, cluster infall, bar interlopers). The lower value of \mh of the fast stars compared to the other stars favours either multiple cluster infalls or bar interlopers. As we expect a contamination of bar stars, several (though possibly not all) high-velocity stars are likely bar stars.

\subsection{The Galactic Centre \mh\ profile}

Our data probe stars of the NSC (most stars at $r\lesssim$ 7\,pc) and the inner NSD (most stars at $l\gtrsim$ 10\,pc). We find that
the $\mean{\rm{\mh}}$ varies spatially in the Galactic Centre, as illustrated by our continuous $\mean{\rm{\mh}}$ map in Fig.~\ref{fig:vorSP}. In addition, we show a \mh profile along longitude in Fig. \ref{fig:lspav} by computing a moving average of 200 stars. The error bars show the Poisson errors.  Due to the higher number of stars compared to the 2-dimensional map, the profile appears smoother. Also, neighbouring bins are correlated, as they contain a subset of the same stars. Nonetheless, we notice similar trends as described in Sect. \ref{sec:SPmaps}. 

The top panel also shows the latitude dependence at 8\,pc$\gtrsim l \gtrsim-8$\,pc): The red line (entire FOV) lies below the blue line (excluding the region $b$>1\,pc), hence $\mean{\rm{\mh}}$ is lower in the northern region (orange dot-dashed line) of the NSC. This finding is consistent with the higher fraction of sub-solar \mh stars detected (albeit in a smaller FOV) by \cite{2020MNRAS.494..396F}. Neglecting the northern region weakens a $\mean{\rm{\mh}}$ minimum on the east side at l$\sim$3.5\,pc.

In a further attempt to remove bar stars, we apply a stricter colour cut (\col\textgreater 1.7\,mag). This cut is motivated by the colour-\mh plot in the bottom panel of Fig.~\ref{fig:lspav}, which shows $\mean{\rm \mh}$ as a function of \col. 
The colour of a star can be used as a proxy for the line-of-sight distance \dlos \citep{2023A&A...680A..75N}. As stars with shorter \dlos suffer from less extinction, their colours are less reddened than stars located at larger \dlos.  Hence, stars with a bluer colour (lower value of \col) are more likely to be located on the near side of the NSD (where the bar contamination is higher). In contrast, stars with a redder colour are more likely to be located within the NSC (where the bar contamination is negligible) or behind it at the far side of the NSD. This is only true statistically and not on a star-by-star basis. The extinction changes on arcsecond scales and is, therefore, different for each star and each line-of-sight, as there can be dark clouds at different \dlos with a different size and density. Hence, the relation of \col to \dlos is not necessarily linear.

\cite{2023A&A...680A..75N} used the colours \col and $J-K_{S}$ and
the \mh measurements of \cite{2017MNRAS.464..194F,2020MNRAS.494..396F} within $\sim$1\,\re of the NSC. They find increasing \mh with colour and interpret this as a \mh gradient along the line-of-sight. 
Our colour-\mh plot also shows that 
at lower \col (i.e. bluer colour, less extinction, shorter \dlos), $\mean{\rm \mh}$ tends to be lower in all regions,  confirming the trend reported in \cite{2023A&A...680A..75N}. 
The east and west regions have a strong $\mean{\rm \mh}$ dependence on \col at blue colours, falling to $\mean{\rm \mh}\sim-0.1$\,dex (east) and $\sim0.05$\,dex (west). The NSD data of \cite{2021A&A...649A..83F} does not even reach such low values at $l\sim$200\,pc. However, their bulge control field at $b$=100\,pc has $\mean{\rm \mh}\sim0.07$\,dex, and it is to 95\% dominated by the bar \citep{2022MNRAS.512.1857S}. Therefore, bar stars may dominate the east and west regions for \col$\lesssim$1.7\,mag. We note that the colour-\mh relation in the central region is globally low. The NSC dominates the stellar surface density in this region. There are only a few stars with blue colours that may be associated with the NSD or the bar, and $\mean{\rm \mh}$ is below the inner NSD value ($\mean{\rm \mh}$$\sim$0.2-0.3\,dex) we found in the east and west regions.

We exclude stars with \col$\textless$1.7\,mag as possible bar contaminants and, in addition, stars in the north  ($b$\textgreater1\,pc), and show the resulting $l$-\mh profile as a black dot-dashed line in the top panel of Fig.~\ref{fig:lspav}. This colour cut is barely noticeable at $\lvert l \rvert$\textless 3\,pc, but becomes significant further away from \sgra, especially on the east side, where it leads to an increase of $\mean{\rm \mh}$ by $\sim$0.1\,dex. 
These cuts weaken some fluctuations, but the  $l$-\mh profile is still not monotonic. 
It has a maximum at $\lvert l\rvert$\textgreater2\,pc with a decrease towards $\gtrsim$1~\re (5\,pc). After a minimum at $\sim$0.2\,dex, the $\mean{\rm{\mh}}$ profile is increasing further out at $\sim$7.5--14\,pc (west) and $\sim$13-16\,pc (east), and becomes rather flat further out (Fig.~\ref{fig:lspav}), where the NSD dominates. 
The higher $\mean{\rm{\mh}}$ we obtain in the west compared to the east may be due to different bar contributions on the two sides. As the near side of the bar is located on the Galactic east, and the far side on the west, the bar stars on the east are on average closer to us, therefore brighter and more likely to be observed. However, the models of \cite{2022MNRAS.512.1857S} predict only slightly higher bar contamination ($\sim$1\%) at the east side of the NSD data of \cite{2021A&A...649A..83F}. If the difference in bar contamination is similar to our F2 data, the east-west asymmetry is rather not caused by the bar.

On the middle panel, we see the more extended $\mean{\rm{\mh}}$ profile derived from the KMOS data of \cite{2021A&A...649A..83F} for the NSD, and in addition \cite{2017MNRAS.464..194F,2020MNRAS.494..396F,2022MNRAS.513.5920F}. We have excluded stars with extreme velocities (as for the F2 data, Sect. \ref{sec:highv}),  which explains the small differences to Fig. 11 of \cite{2022MNRAS.513.5920F}. Although the purple data points of \cite{2021A&A...649A..83F} extend to low values of $l$, the stars of their sample are located at larger $b$ than our data, which means they are part of the NSD with low NSC contribution. The data of \cite{2022MNRAS.513.5920F} overlap spatially with our data; the $\mean{\rm{\mh}}$ at the west side is lower than our data but agrees on the east side. 
Our $\mean{\rm{\mh}}$ at the outer edge of our data is higher than the NSD data of \cite{2021A&A...649A..83F}. We did not exclude blue stars from these data. The NSD  data cover a large FOV with varying extinction, making it hard to tell which stars are likely bar stars. The bar contamination fraction in the NSD data is, however, significant,  ranging from 26\% ($\lvert l \rvert \sim$50\,pc) up to 60\% ($\lvert l \rvert \sim$200\,pc).  This explains the overall lower $\mean{\rm{\mh}}$ compared to our inner NSD data.

\subsection{Is the central mean metallicity variation real?}
\label{sec:discmetvar}
\begin{figure}
 \resizebox{\hsize}{!}{\includegraphics{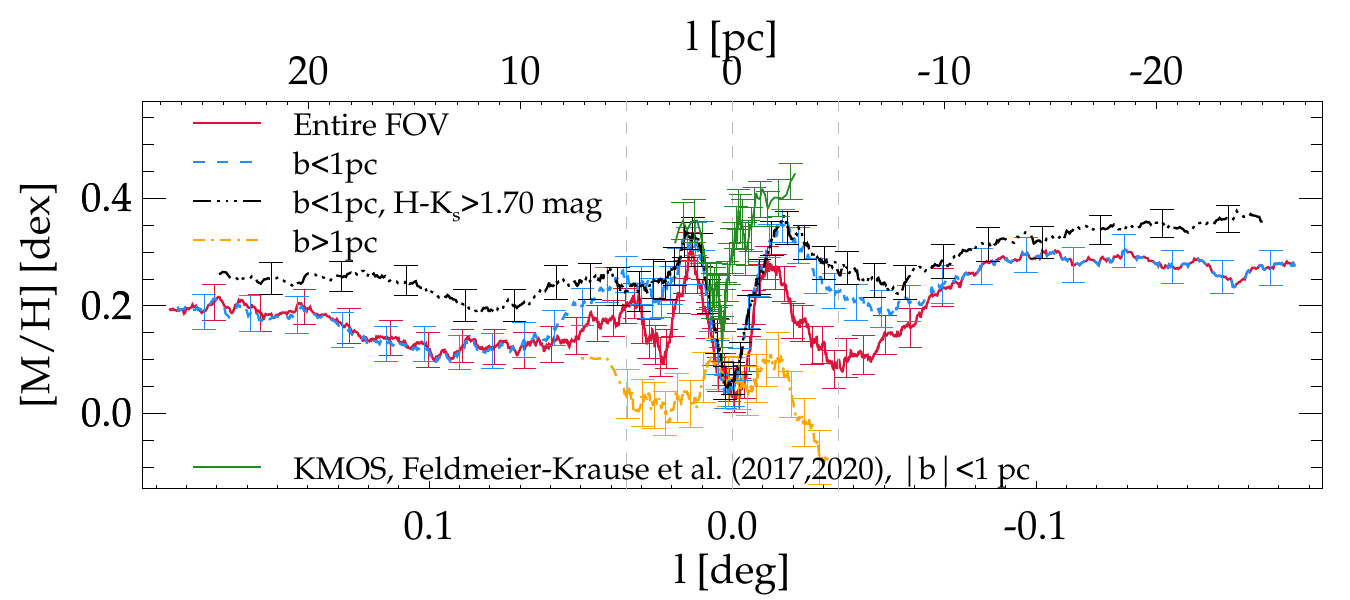}}
  \resizebox{\hsize}{!}{\includegraphics{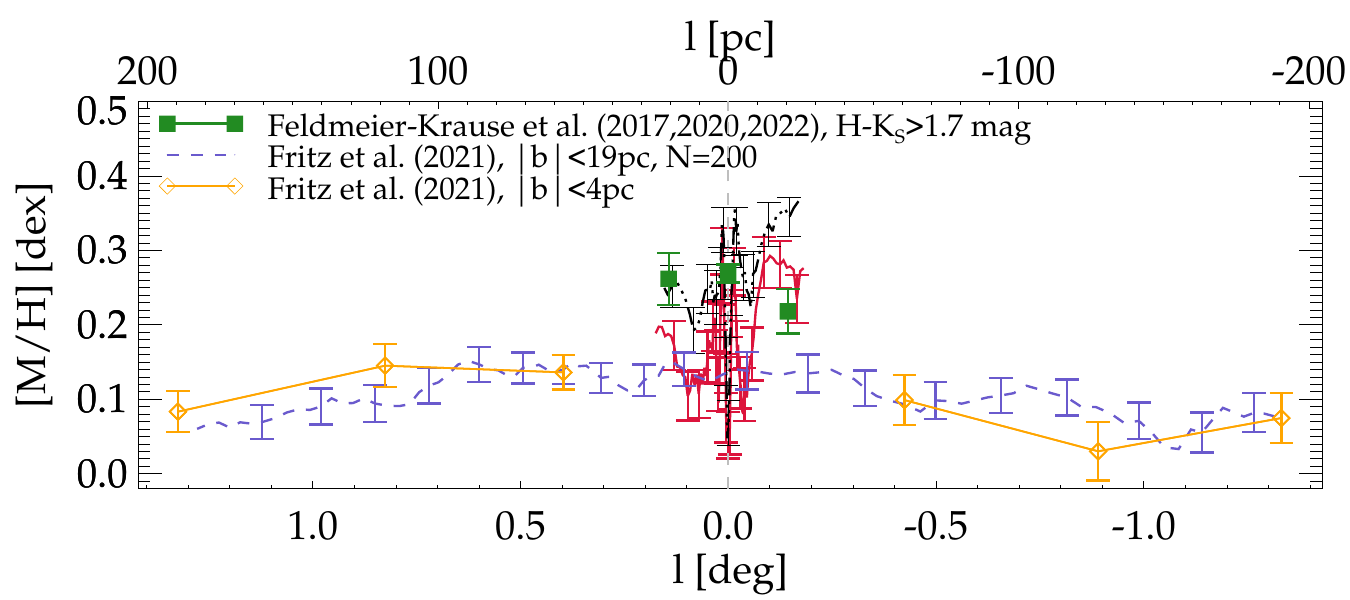}}
   \resizebox{\hsize}{!}{\includegraphics{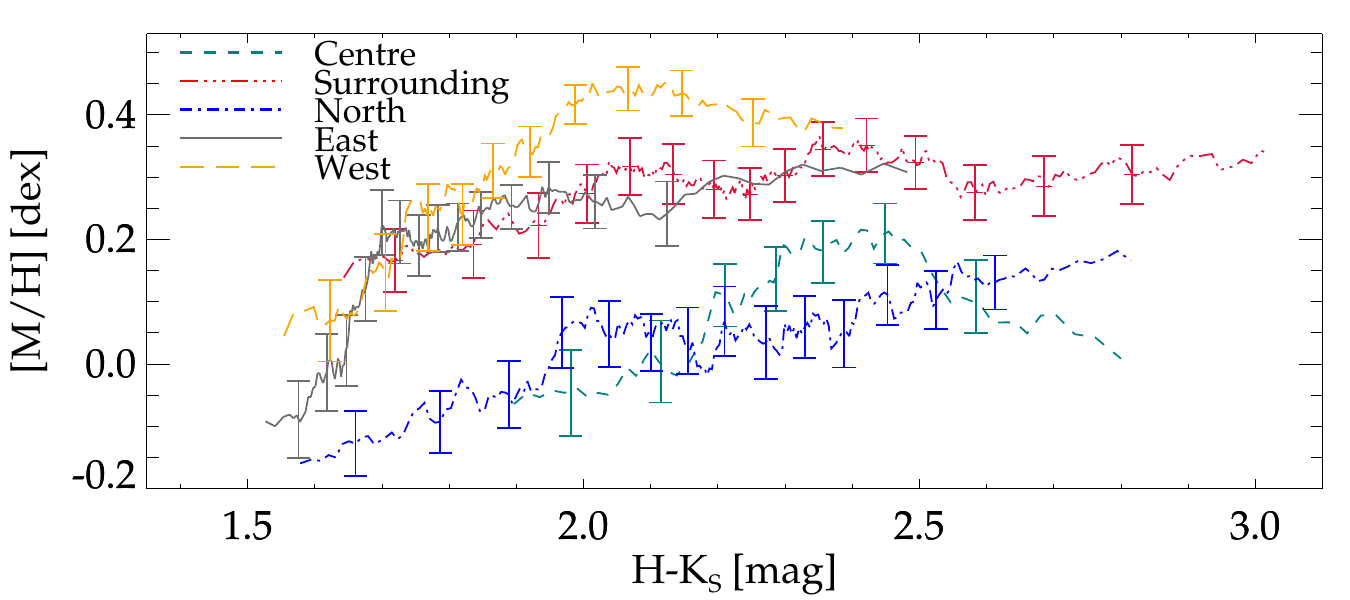}}
 \caption{Moving average \mh profile of Galactic Centre late-type stars. Top panel: Profile along Galactic longitude, centred on \sgra. Red lines denote the average over the entire FOV,  blue dashed lines for stars at $b$\textless 1\,pc, i.e. without stars located in the northern region of the FOV (shown as orange dot-dashed line), and black dot-dashed line excludes, in addition, stars with \col\textless1.7\,mag. The average is computed on 200 stars, uncertainties are Poisson errors. The green line denotes the moving average of \mh of 100 stars from \cite{2017MNRAS.464..194F,2020MNRAS.494..396F}. 
 Vertical dashed lines denote the NSC \re=5\,pc, and the centre at 0\,pc. Middle panel: Same as top panel, but including NSD data from \cite{2021A&A...649A..83F}, shown as purple dashed and solid orange line with diamond symbols,  and \cite{2022MNRAS.513.5920F}, shown together with the \cite{2017MNRAS.464..194F,2020MNRAS.494..396F} data as green square symbols. To aid visibility, we plot error bars only every $\sim$20\,pc. Bottom panel: Profile along observed \col, for different regions in our FOV, and averaged over 50--100 stars, to account for the lower and different number of stars per field.
} \label{fig:lspav}
 \end{figure}

Our $\mean{\rm\mh}$ map indicates an increase of the mean \mh, from roughly the inner $r\leq$1\,pc ($\sim$200 stars) to the surrounding region ($r$\textgreater1\,pc, $b$\textless1\,pc, -5\,pc\textless$l$\textless 5\,pc, $\sim$650 stars), by $\sim$0.21\,dex (see Fig.~\ref{fig:vorSP}, middle panel and Fig.~\ref{fig:lspav}, top panel). The number of stars to compute $\mean{\rm\mh}$ is large, and our simulations when drawing 1\,000 times subsamples of 200 stars obtain that such different values are unlikely (\textgreater 2.2$\sigma$ from expectations), when drawn from the same distribution. Also, the Gaussian mixture models in these regions are different. They favour a single Gaussian in the centre and a double Gaussian in the surrounding region with two rather high \mh values compared to other regions (e.g. north, east, or west, see Table \ref{tab:gmm}).

We find a similar trend of a lower $\mean{\rm\mh}$ in the inner 1\,pc also in the KMOS data of \cite{2017MNRAS.464..194F} and \cite{2020MNRAS.494..396F}. We combine these two data sets, deselect foreground stars, and obtain an increase of $\mean{\rm\mh}$ by $\sim$0.09\,dex (from $r\leq$1\,pc to stars in the surrounding region defined as $r$\textgreater1\,pc, $b$\textless1\,pc, -4.6\,pc\textless$l$\textless 4.6\,pc; 558 and 462 stars, respectively), see also green line in  Fig.~\ref{fig:lspav}, top panel. The KMOS data do not cover the same stars as our F2 data but have some overlap (see Table~\ref{tab:litcomparison}). The data of \cite{2017MNRAS.464..194F}, which cover only the inner $r\leq$1.5\,pc, have the most complete sample, reaching fainter magnitudes than \cite{2020MNRAS.494..396F}, which is yet deeper than our data. 
The absolute values of the F2 $\mean{\rm\mh}$ are lower than the KMOS data by 0.2\,dex and 0.08\,dex. We investigate whether there may be a systematic offset between the data sets in Appendix \ref{sec:litcom}, by comparing the common stars, and find reasonable agreement, the median $\Delta$\mh $\sim$0.07-0.09\,dex is of the same order as the $\mean{\rm\mh}$ offset of the data sets in the surrounding regions (0.08\,dex, Table \ref{tab:litcomparison}).
The offset within the central $r\leq$1\,pc is larger. A possible explanation is the higher completeness of the \cite{2017MNRAS.464..194F} data, resulting in a 2.8 times larger number of stars in the inner 1\,pc.

We speculate that the $\mean{\rm\mh}$ decrease in the inner parsec may be a projection effect. Assuming there is a monotonous negative \mh gradient with increasing distance from \sgra, from the NSC towards the NSD, then we may observe a drop of $\mean{\rm\mh}$ towards the inner projected 1\,pc, if (some of) these stars are only close in projection. In other words, their line-of-sight distance to \sgra could be (on average) higher, such that their de-projected, real distance to \sgra exceeds that of the stars at a projected radius 1\,pc\textless $r\leq$4.5\,pc. This means if we have a larger fraction of NSD or bar stars in the inner parsec region compared to the surrounding region, this could cause the low central $\mean{\rm\mh}$.
However, given the low value of $\mean{\rm\mh}$ in the central $r$\textless 1\,pc that is not even obtained at the NSD fields of \cite{2021A&A...649A..83F}, this appears unlikely.

To test the possibility of a projection effect, we compare the stellar kinematics, in particular, the line-of-sight velocity dispersion \slos. Closer to \sgra, \slos has to increase, as also the map in Fig. \ref{fig:vsig} indicates. However, the value of \slos in the KMOS data is higher than in the F2 data, by $\sim$20\,\kms (99$\pm$3\,\kms KMOS, 76$\pm$4\,\kms F2, central pc) to $\sim$8\,\kms (82$\pm$3\,\kms and 75$\pm$2\,\kms in surrounding region). Thus, the stars in the KMOS data are possibly indeed closer to \sgra in 3D distance than the F2 data. They are also closer in projected distance (median 0.64\,pc for KMOS and 0.70\,pc for F2). This may explain the differences in $\mean{\rm\mh}$ between the datasets and why the F2 data have a more extreme $\mean{\rm\mh}$ drop towards the centre. 

On the other hand, we cannot exclude that the low spatial resolution affects our measurements. Despite our careful extraction, the spectra of some stars may be contaminated by the flux of nearby stars. This is especially possible in the dense inner pc region and may bias the measured \vlos to the mean \vlos in the region and cause an underestimated \slos. 
It may also influence our \mh\ measurements and cause the central $\mean{\rm\mh}$ to decrease. 
In summary, we cannot exclude that the drop of the $\mean{\rm\mh}$ is (at least partially) a projection effect or caused by the low spatial resolution.

However, if the $\mean{\rm\mh}$ decrease is real, what could have caused it? Some mechanisms have been proposed that may affect the stellar atmosphere of red giant stars by removing their outer envelope in the central $r\lesssim$0.5\,pc region.
These include tidal disruption by the SMBH, interaction with a (former) jet, and collisions with dense clumps in a gas disc \citep[e.g.][]{2012ApJ...757..134M,2016ApJ...823..155K,2020MNRAS.492..250A,2020ApJ...903..140Z,2024arXiv240917773K}, though the first two were shown to affect mainly stars within $\lesssim$0.1\,pc. Increasing the mass loss and removing the envelope of a red giant alters various properties of the star, including \teff and its brightness. The strength of the effect depends on various parameters, including the number of encounters, the distance, the surface density of the gas clouds, and the mass and radius of the star \citep{2020MNRAS.492..250A,2021MNRAS.505.3314M}.  
 It is unclear if and how the spectrum of a red giant star changes under increased mass loss, but it was shown that red supergiant spectra alter depending on the mass loss rate \citep[e.g.][]{2021MNRAS.508.5757D}. It may be possible that increased mass loss in the inner $r\textless$0.5\,pc affects the spectra such that they appear to have a lower \mh.

\subsection{Colour-dependent position-velocity diagram reveals NSC and NSD}
\label{sec:discposvel}

We have shown the position-velocity diagram in Sect. \ref{sec:ltresult}, with \vlos as a function of Galactic longitude $l$. After removing contaminants, we compared the position-velocity diagram for stars with different values of \mh, and we found overall good agreement for stars with sub-solar, roughly solar, or super-solar \mh, but differences for varying observed colour \col. 

\cite{2021A&A...650A.191S} analysed the NSD data compiled by \cite{2021A&A...649A..83F}. They also study the position-velocity diagram as a function of \mh, and at the scales of their data ($\pm$1.5\degr$\sim\pm$200\,pc), they see significant differences. Stars at lower \mh show only weak rotation, and with increasing \mh, the velocity increases, and the stars rotate faster around \sgra. The stars in their lowest \mh bin (\textless -0.5\,dex) may even be counter-rotating and thus be remnants of accreted and disrupted star clusters or Galactic bar stars on $x_1$ orbits. We deselected stars with extreme \vlos from our sample to decrease the bar contribution. We do not see a \mh dependent position-velocity curve in our NSC and NSD-dominated data. A possible reason is that our sample of stars is closer to \sgra and stars with varying \mh are better mixed via dynamical processes. Further, the contamination fraction from the bar is significant in the data used by \cite{2021A&A...650A.191S} \citep[$\sim$20--62\%,][]{2022MNRAS.512.1857S}, which can cause these changes, whereas the bar contribution is much lower in our FOV ($\sim$0 to $\lesssim$20\%).

We see a dependence of the position-velocity diagram on the colour \col. Stars with bluer colours have a less pronounced increase of the velocity curve, increasing beyond the inner 5\,pc, and they extend to larger values of $l$. Stars with a redder colour (in particular at \col=2.3--2.6\,mag) have pronounced peaks of the velocity curve at $\lesssim$5\,pc. 
As noted above, the colour can indicate line-of-sight distance \dlos. Stars with bluer colours tend to be closer to us, and stars with redder colours tend to be farther away. However, this mainly holds in regions along a small FOV, as the extinction, the main reason for the colour change, varies on scales of arcseconds. Although there is not a perfect 1:1 correlation of colour and \dlos, this, albeit noisy, relation can help us understand the colour dependence of the position-velocity diagram.

We interpret the bottom panel of  Fig.~\ref{fig:lvel_mbin} as follows: Stars with redder colours lie deeper in the GC and, therefore, belong preferentially to the NSC, while bluer stars are more likely to form part of the more extended NSD. This explains the concentration of red stars around \sgr\ and their rapidly changing velocity profile (as a function of distance from \sgr). NSD stars lie at greater distances from \sgr, where the gravitational potential changes less rapidly and hence, they show a flatter velocity profile.
The bluest stars have the weakest rotation signal, they 
are more likely to be closer to us and thus a part of the near side of the NSD. In a future publication, we will present chemo-dynamical models of the data that will help to constrain the kinematic properties of the stars in the NSC and inner NSD.

%=========================================
\section{Conclusions}
\label{sec:conclusions}

We have presented extended spectroscopic slit scan data of the Galactic Centre region, ranging from the NSC to the inner part of the NSD, out to $\pm$32\,pc along Galactic longitude, and $\sim\pm$1\,pc along Galactic latitude. We extracted spectra of several thousand stars, predominantly red giant stars, and measured \vlos and \mh for \textgreater2\,500 of them.

We identified 78 hot star candidates, of which we classify 48 as likely being located in the Galactic Centre; 31 of these are classified as hot stars for the first time. Several of the stars likely belong to the Quintuplet cluster ($\sim$9) or the central parsec cluster ($\sim$12). But $\sim$20 hot stars are rather isolated, and their origin remains an open question. Further, we detected $\sim$80 stars with extreme velocities ($\lvert V_\text{LOS}\lvert$\textgreater 150\,\kms), and we attribute these stars mainly to the Milky Way bar or tidal tails from globular clusters.

We presented the first map of the mean \mh in the region beyond 1~\re of the NSC. We see a spatial \mh variation, with a decrease to the centre of the NSC, and we discuss the possibility that this is a projection effect caused by low spatial resolution or caused by high stellar mass loss in this region. We further detect another $\mean{\rm \mh}$ minimum where the NSD stellar density starts to dominate over the NSC at $\sim$10\,pc (the exact location varies by a few pc in the east and west), followed by a slight increase and a flat profile in the region dominated by the NSD. We note a
higher $\mean{\rm \mh}$ in the Galactic west compared to the east of the NSD. 

Finally, we show the first continuous position- \vlos diagram at these scales and how it varies with \mh and colour \col. Using colour, we can separate the position-\vlos contribution of stars in the NSC and of stars belonging to the NSD (i.e. at larger \dlos and $\lvert l \lvert$). In the future, we will use the data for detailed discrete dynamical models and to constrain the mass distribution and orbit distribution of the inner $\lvert l \lvert$ \textless 32\,pc of the Galactic Centre.

\begin{acknowledgements}
We thank the Gemini Observatory staff for their support during the planning and execution of the observations and advice on data reduction. 
We thank David Rupke for providing a general-purpose library for IFU data cubes. We also thank the anonymous referee for constructive comments and suggestions. 
AFK acknowledges funding from the Austrian Science Fund (FWF) [grant DOI 10.55776/ESP542].
Based on observations obtained at the international Gemini Observatory, a program of NSF NOIRLab, which is managed by the Association of Universities for Research in Astronomy (AURA) under a cooperative agreement with the U.S. National Science Foundation on behalf of the Gemini Observatory partnership: the U.S. National Science Foundation (United States), National Research Council (Canada), Agencia Nacional de Investigaci\'{o}n y Desarrollo (Chile), Ministerio de Ciencia, Tecnolog\'{i}a e Innovaci\'{o}n (Argentina), Minist\'{e}rio da Ci\^{e}ncia, Tecnologia, Inova\c{c}\~{o}es e Comunica\c{c}\~{o}es (Brazil), and Korea Astronomy and Space Science Institute (Republic of Korea). Data were processed using the Gemini IRAF package. IRAF is distributed by the National Optical Astronomy Observatory, which is operated by the Association of Universities for Research in Astronomy (AURA) under cooperative agreement with the National Science Foundation \citep{1993ASPC...52..173T}. This research has made use of the VizieR catalogue access tool, CDS,
Strasbourg, France \citep{10.26093/cds/vizier}. The original description 
of the VizieR service was published in \citet{vizier2000}.
This research made use of Montage. It is funded by the National Science Foundation under Grant Number ACI-1440620, and was previously funded by the National Aeronautics and Space Administration's Earth Science Technology Office, Computation Technologies Project, under Cooperative Agreement Number NCC5-626 between NASA and the California Institute of Technology.
This research made use of NumPy \citep{harris2020array}, SciPy \citep{Virtanen_2020}, matplotlib, a Python library for publication quality graphics \citep{Hunter:2007}, Astropy, a community-developed core Python package for Astronomy \citep{2013A&A...558A..33A,2018AJ....156..123A, 2022ApJ...935..167A}. The acknowledgements were compiled using the Astronomy Acknowledgement Generator. 
 
 \end{acknowledgements}
 %----------
% for the bibliography, at the end
\bibliographystyle{aa} % style aa.bst
\bibliography{bibtex.bib} % your references Yourfile.bib

 %----------
\begin{appendix}

 \section{Persistence removal}
\label{sec:pers}

Our 2-dimensional spectral frames of the Galactic Centre contain the afterglow from the brightest stars in the acquisition images taken earlier. This is called a persistence signal. To remove the persistence, we follow a similar procedure as described in \cite{2014A&A...570A...2F}, though some changes were necessary owing to the different observing strategy, instrument, and the weaker level of persistence in our data thanks to the series of Dark exposures taken after the acquisition images. In short, we used the two acquisition images per observing block and the 2-dimensional spectral frames to measure the strength of the persistence signal in the individual exposures. Then, we modelled the decrease of the persistence signal with time and used this model to subtract the persistence signal from the 2-dimensional spectral frames. 

We measure the persistence signal on the dark subtracted spectra, however, the persistence signal is much weaker than the flux from the sky and stars. For this reason, we required a region that is free from these sources. To subtract these sources, we first removed the distortion of the data. 
We measured the curvature of the stars in the 2-dimensional spectrum on one of the exposures using \textsc{IRAF} \textsc{identify} and \textsc{reidentify}, and we applied the transformation on both the spectra and acquisition images to create rectified files. 

To measure the persistence signal and how it changes in time, we created a cutout of each rectified spectrum. We used a region with only weak sky lines, and cut out the regions along the slit (between slitlets) that stabilise the mask. Then, we subtracted the median flux along the wavelength axis, thus subtracting the sky lines. Removing the flux from stars is more complicated, as the stellar spectra have absorption lines, which leave residuals when simply subtracting the median flux along the spatial slit axis. Instead, we cut out the regions with bright stars in each spectrum. These stars are identified as follows: we computed the median flux along the slit, and fit a robust third-degree polynomial to the median flux. Then we applied 3$\sigma$ clipping and measured the mean and standard deviation of the residual. We cut out regions where the residual exceeds 1 $\sigma$, which is due to bright stars. This leaves a part of the 2-dimensional spectrum that is rather uniform, and the persistence signal is easily detectable. We call this 2-dimensional spectrum $S_\text{cut}$. 

From the two rectified acquisition images, we created a mask $M$, which shows the pixels where the flux of either one of the images is higher than a certain threshold. We tried different threshold values and compared the results of the persistence model. We found little variation around values of 34\,000-36\,000 counts and settled on values of 34\,000, 35\,000, and 36\,000 counts, depending on the exposure times of the images (2, 3, or 5 s). We cut the mask in the same way as the 2-dimensional spectrum to obtain $M_\text{cut}$.

On this mask, we applied a Gaussian smoothing filter $G_\text{smooth}$ before subtracting it from $S_\text{cut}$. We used \textsc{mpfit2dfun.pro} to fit the best smoothing parameters. Those are the amplitude $A$, the FWHM, and an eccentricity parameter $\epsilon$ to account for an elliptical FWHM. We minimised the residual spectrum $R_\text{cut} = S_\text{cut} -G_\text{smooth} \ast M_\text{cut}$. This is done for each spectral frame in each observing block. While the FWHM and $\epsilon$ are rather stable, the amplitude $A$ decreases with time for each series of $\sim$10--20 spectra, following a power law $A(t)=C \times t^{-b}$, where $t$ is the UT time difference in seconds between the mean UT of the two acquisition images causing the persistence and the respective spectral frame. We then computed the value of $A$ at the appropriate time for each 2-dimensional spectral frame and used the mean values for FWHM and $\epsilon$ to apply $G_\text{smooth}$ on the uncut mask $M$, which we subtract from the un-rectified, dark subtracted spectral frames. This gives us dark and persistence-corrected 2-dimensional spectral frames.

\section{Stellar parameter uncertainties}
\subsection{Comparison with other data}
\label{sec:litcom}
Our data overlap with the spectroscopic data analysed in \cite{2017MNRAS.464..194F,2020MNRAS.494..396F,2022MNRAS.513.5920F}, and we compared our stellar parameter results. We matched the stars of our data and the literature catalogues if they are within 0\farcs4 of each other, their $K_S$ photometry differs by no more than 0.5\,mag, and their \vlos\space by no more than 20\,\kms. With these three criteria, we hope to exclude most mismatches. We obtained 104, 284, and 65 matches of our data with the aforementioned papers. The stellar parameters were obtained similarly to ours, using the same code and spectral grid. However, the spectral resolution of the KMOS (VLT) instrument used in FK17, FK20, and FK22 is slightly higher (R$\sim$4~000), and the spatial resolution is seeing limited, hence less affected by contamination of nearby stars compared to our 1\arcsec\space slit scan data. 

We compared the median stellar parameter differences of \teff, \logg, and \mh, and the ($\sigma$-clipped) standard deviation of the parameter difference divided by the squared sum of the statistical error (see Table~\ref{tab:litcomparison}). The median differences are close to zero and not strongly biased. The distributions of the differences normalised with the statistical uncertainties are, however, broader than 1$\sigma$, indicating that the statistical uncertainties are underestimated. There is indeed a trend that this is more severe for the stars located in the dense and crowded NSC (FK17 and FK20) in comparison to the data at 20\,pc distance (FK22), where the stellar number density is lower. We see the largest deviation for \teff\space (1.6$\sigma$--6$\sigma$) and \logg\space (1.5$\sigma$--4.0$\sigma$), meaning that the statistical uncertainties are underestimated, whereas the uncertainties of \mh\space are reasonable (1$\sigma$--1.9$\sigma$).

From this comparison, we conclude that our statistical uncertainties for \teff\space and \logg\space are underestimated in the central region, our statistical uncertainty for \mh\space is reasonable, and we have no strong systematic differences to the slightly higher spatial resolution data in the literature.

\begin{table}
\caption{List of matches with late-type star catalogues. }
 \label{tab:litcomparison}

\begin{tabular}{@{}lrrr@{}}
\noalign{\smallskip}
\hline
\noalign{\smallskip}
Parameter&FK17 &FK20& FK22 \\
 \noalign{\smallskip}
\hline
\noalign{\smallskip}
Number of matched stars & 104 & 284& 65 \\
\hline 
median $\Delta$\teff\space [K] & 73& 11 &-87\\
$\sigma$($\Delta$\teff/stat.unc.) & 5.9& 6.0 &1.6\\
\hline 
median $\Delta$\logg\space [dex] & 0.17& 0.08 &0.06\\
$\sigma$($\Delta$\logg/stat.unc.) & 4.0& 3.5 &1.5\\
\hline 
median $\Delta$\mh\space [dex] & -0.09& -0.07 &0.04\\
$\sigma$($\Delta$\mh/stat.unc.) & 1.9& 1.9 &1.0\\

\hline 
\end{tabular}
\tablebib{\citet[FK17]{2017MNRAS.464..194F}; \citet[FK20]{2020MNRAS.494..396F}; \citet[FK22]{2022MNRAS.513.5920F}}
\end{table}

 \subsection{Systematic uncertainties}
\label{sec:xslsk}
 
 There are likely systematic uncertainties in our stellar parameter measurements, due to e.g. an incomplete line list used to generate the theoretical PHOENIX model grid, variation of the elemental abundances in the different stars, or imperfect interpolation of the model grid. To estimate the magnitude of these effects, we apply the same method of stellar parameter measurements on stars of the X-SHOOTER spectral library \citep[XSL,][]{2014A&A...565A.117C,2020A&A...634A.133G}, and compare the results to the stellar parameter measurements of \cite{2019A&A...627A.138A}, which were obtained by fitting the optical region of the XSL spectra. 

To estimate systematic uncertainties, we convolved the XSL spectra to the spectral resolution of the F2 data, which is a 2. degree polynomial function of wavelength. We use the average of the six slitlet regions (see Sect. \ref{sec:res}). We set the same bounds for \logg\space as in \cite{2022MNRAS.513.5920F}. Further, we excluded peculiar stars or Carbon stars. We applied similar cuts on the fit quality (statistical uncertainties $\sigma_{\mh}$\textless 0.25\,dex, $\sigma_{T_\text{eff}}$\textless 250\,K, and $\sigma_{\logg}$\textless 1\,dex) and stellar parameter ranges ([Fe/H]$_{A19}$ and \mh$_\text{starkit}$ \textgreater $-1.3$\,dex, 2800\,K \textless \teff \textless 6000\,K, \logg\textless 4.5\,dex) as in \cite{2022MNRAS.513.5920F}, to ensure we compare stars with similar stellar parameters as the Galactic Centre data, leaving us with 232 spectra. 

The mean difference to the reference values after 3$\sigma$ clipping are $\langle\Delta\text{\teff}\rangle$=17\,K, $\langle\Delta\text{\logg}\rangle$=-0.4\,dex, and $\langle\Delta\text{\mh}\rangle$ = -0.13\,dex. The standard deviation of these differences are $\sigma_{\Delta \teff}$=271\,K, $\sigma_{\Delta \logg}$=0.9\,dex, $\sigma_{\Delta \mh}$=0.32\,dex. As the standard deviations are larger than the mean differences, our measurements have no strong biases, and we use the standard deviations as systematic uncertainties for our stellar parameter measurements. These values are very similar to those found by \cite{2022MNRAS.513.5920F}, probably because the spectral resolution in the fit region is very similar.

For a detailed discussion of the XSL sample and its limitations to estimate systematic uncertainties for our method, see \cite{2022MNRAS.513.5920F}. We also refer to \cite{2021A&A...649A..97L} for a detailed comparison of the XSL SEDs with PHOENIX spectra. They found stellar parameter discrepancies for the three different X-SHOOTER arms at \teff$\lesssim$5000\,K, which may explain why our $K$-band results deviate from the \cite{2019A&A...627A.138A} results obtained in the X-SHOOTER VIS arm wavelength region, in particular for \logg.

\section{Moving average profiles of stellar parameters}

\begin{figure}
 \resizebox{\hsize}{!}{\includegraphics{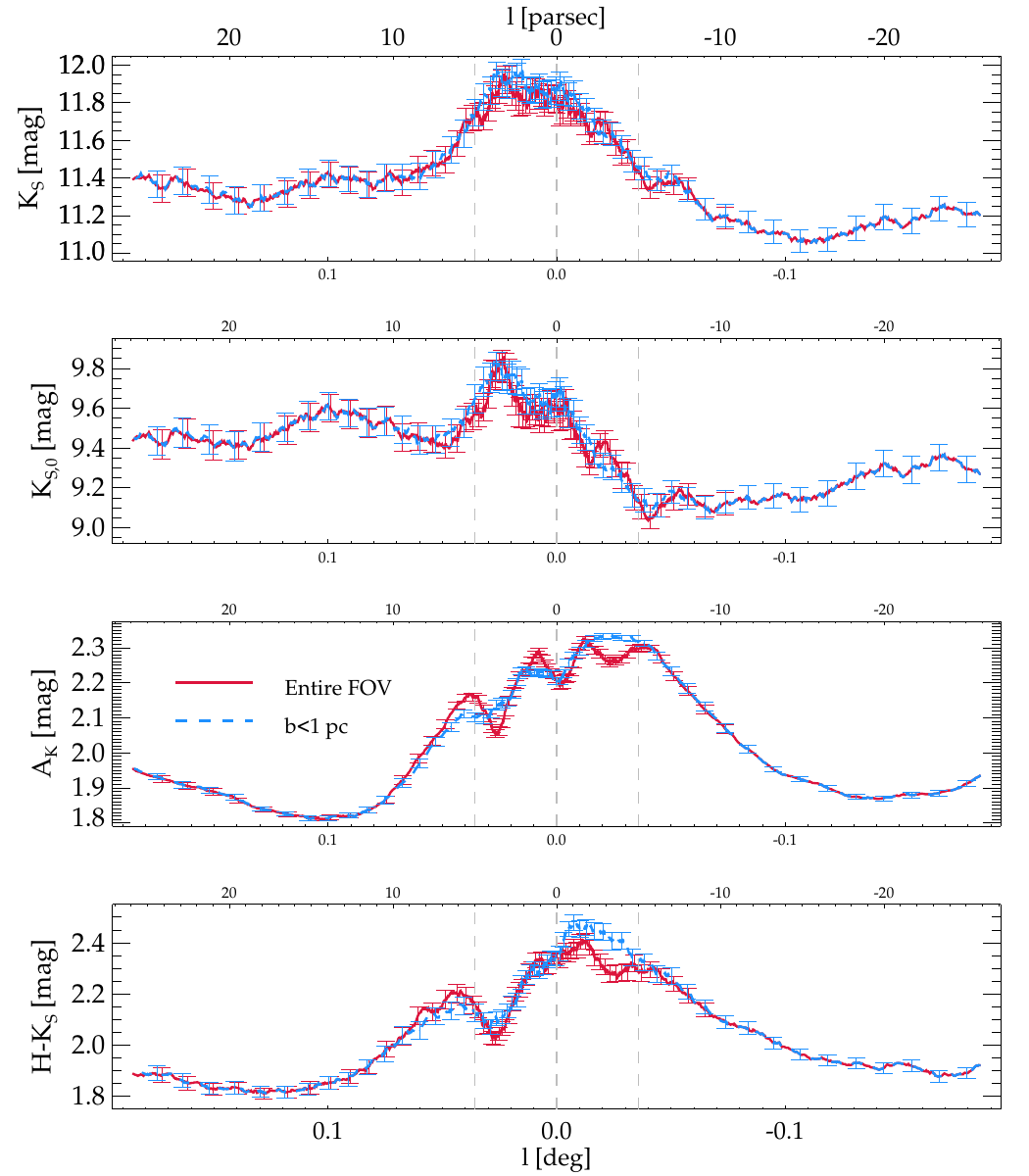}}
 \caption{Moving average profile of Galactic Centre late-type stars along Galactic longitude, centred on \sgra. Red lines denote the average over the entire FOV, and blue dashed lines for stars at $b$\textless 1\,pc, i.e. without stars located in the northern region of the FOV. The average is computed on 200 stars, uncertainties are Poisson errors.
 Top panel: Observed $K_S$ band photometry; Second panel: Extinction-corrected $K_{S,0}$ photometry; Third panel: Extinction in the $K_S$ band A$_\text{K}$; Bottom panel: Observed colour \col. Vertical dashed lines denote the NSC \re=5\,pc, and the centre at 0\,pc.
} \label{fig:lphotav}
 \end{figure}

\begin{figure}
 \resizebox{\hsize}{!}{\includegraphics{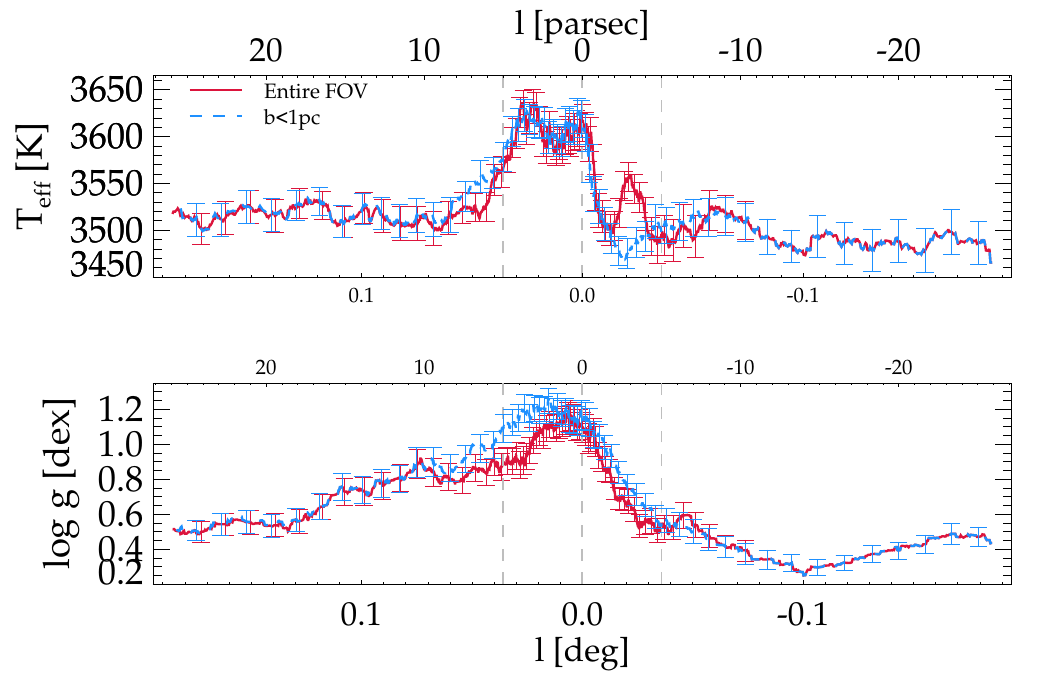}}
 \caption{Same as Fig.~\ref{fig:lphotav}, but for stellar parameters.
 Top panel: \teff;  Bottom panel: \logg.
} \label{fig:lspav2}
 \end{figure}

We show the moving average of various photometric properties as a function of Galactic longitude $l$ in Fig. \ref{fig:lphotav}, and in addition stellar parameters \teff and \logg in Fig.~\ref{fig:lspav2}.  These plots correspond to Figs.~\ref{fig:vorK} and \ref{fig:vorSP}, but they neglect any dependence on Galactic latitude $b$ and are computed for a larger number of stars per bin (200 rather than 40), and neighbouring data points are correlated. We show the Poisson errors only for a subset of the data points, which improves visibility. We show two regions, stars from the entire FOV, and only stars at $b$\textless1\,pc, i.e. we cut off the region in the north that covers only in the central $l\sim\pm$8\,pc

\begin{figure}
 \resizebox{\hsize}{!}{\includegraphics{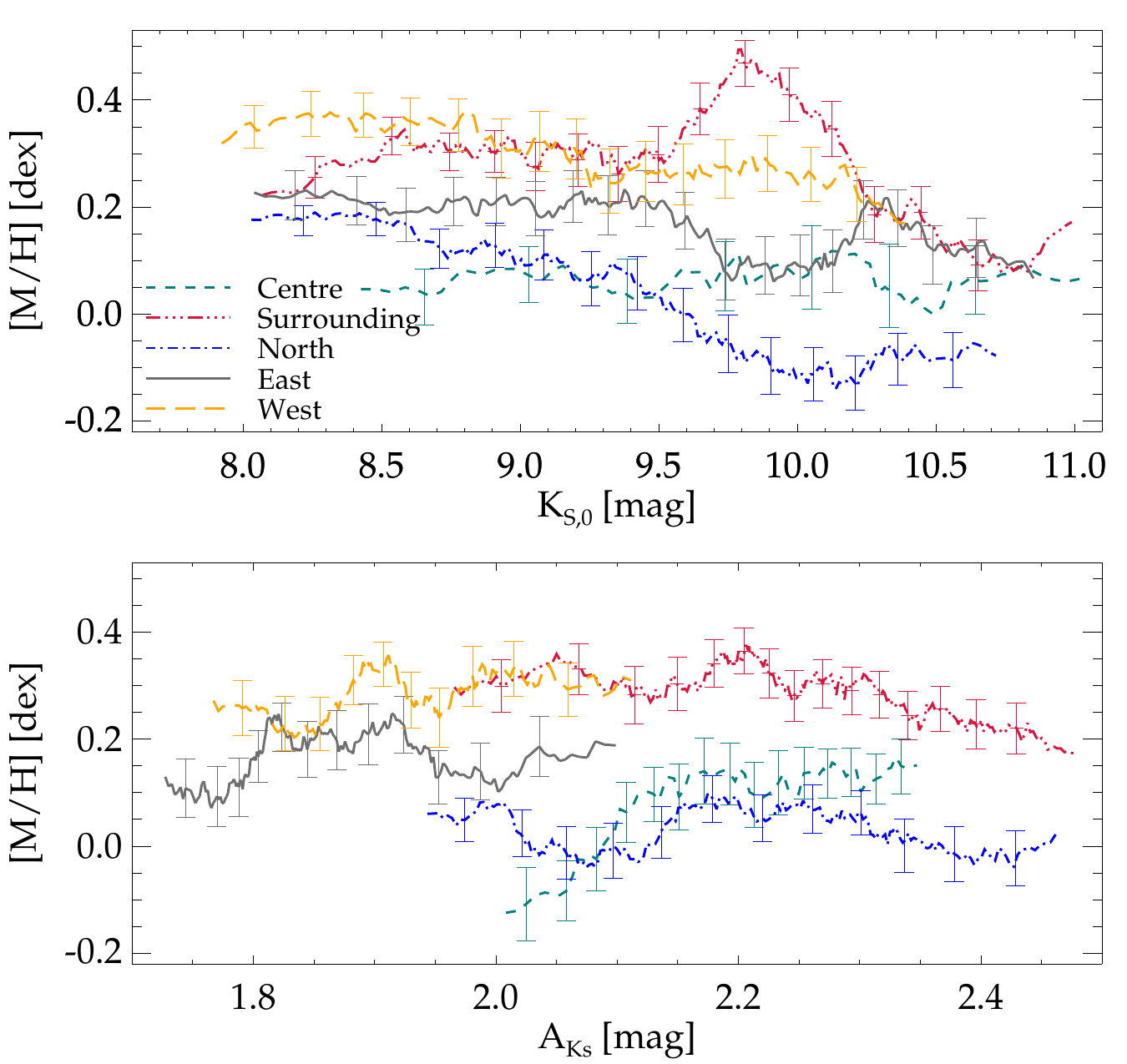}}
 \caption{Moving average \mh profile of Galactic Centre late-type stars in several regions as a function of different photometric properties.
 The panels show, from top to bottom, extinction corrected $K_{S,0}$, extinction \ak. 
 The regions are as defined in Table \ref{tab:gmm} and shown with colours and symbols as described in the upper panel legend. The average is computed on 50 (centre), 80 (East and West), or 100 stars (surrounding and north regions), as the number of stars varies in these regions. Uncertainties of \mh are Poisson errors.
} \label{fig:mhphot}
 \end{figure}

In Fig.~\ref{fig:mhphot}, we show the moving average of \mh as a function of photometric properties for various regions, as defined in Table~\ref{tab:gmm}. Though we see fluctuations of $\mean{\rm \mh}$ as a function of the stellar brightness $K_{S,0}$, this behaviour is not identical in the regions, suggesting there is no systematic trend with stellar brightness. 
The bottom panel shows no consistent variation with \ak in the various regions, but rather random fluctuations of $\mean{\rm \mh}$. The east and west regions have overall lower values of \ak, but they also overlap with the more central regions at $\sim$1.95--2.1\,mag.

We observe a higher or at most similar value of $\mean{\rm \mh}$ in the surrounding region compared to the centre and north regions irrespective of $K_{S,0}$ or \ak. The same is found for the west region compared to the east.
 When we remove stars with \mh\textgreater 0.5\,dex, as done by \cite{2023A&A...680A..75N}, or replace those \mh values with a floor of 0.5\,dex, the surrounding region still has higher $\mean{\rm \mh}$ than the centre, though the difference is less. 

\section{\mh distributions in various regions}

\begin{figure}
 \centering
 \includegraphics[width=\columnwidth]{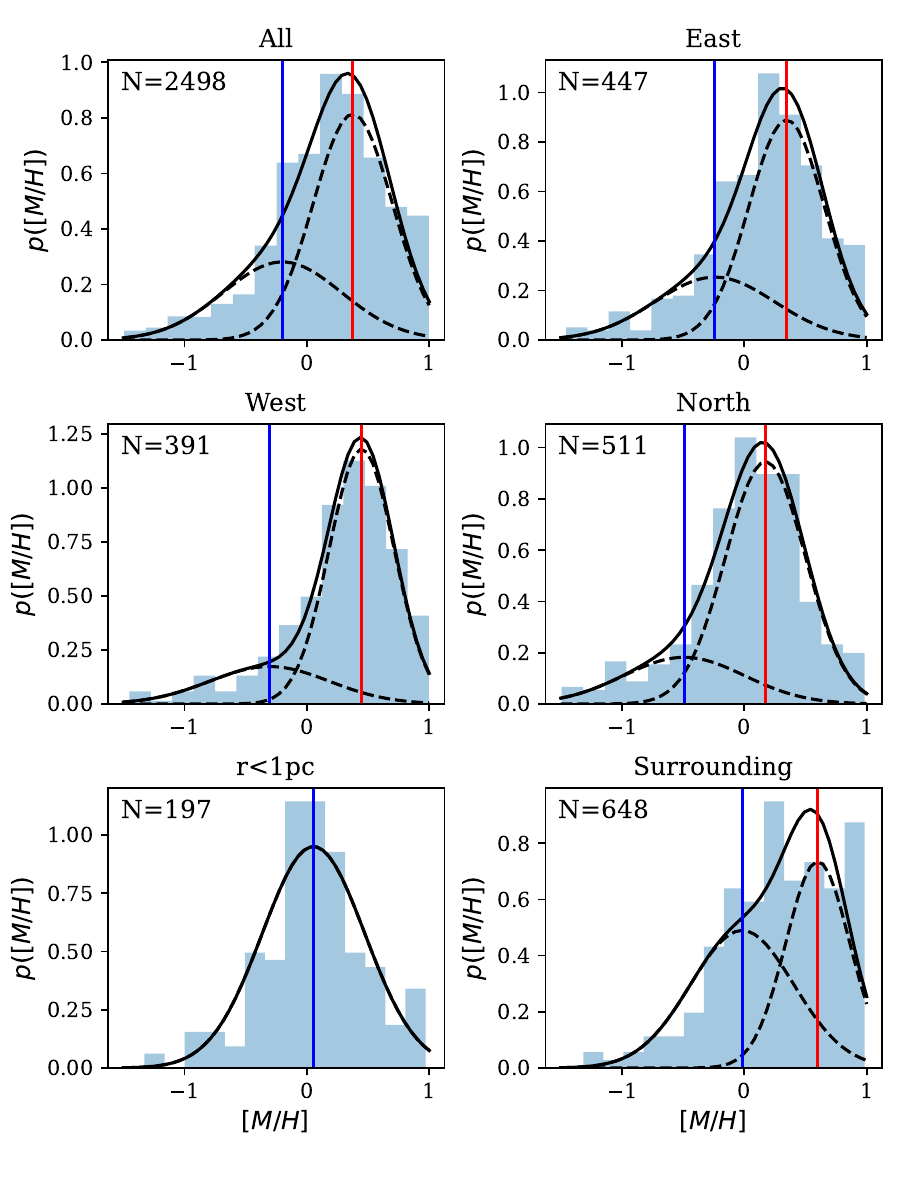}
 \caption{Gaussian mixture models of the \mh distribution in different regions, as detailed in Table \ref{tab:gmm}. We removed likely foreground stars and high-velocity stars. The blue and red vertical lines denote the centres of the Gaussian components, and the number of stars N is noted on the top left of each panel.}
 \label{fig:gmm}
 \end{figure}

We show histograms of the \mh distribution in various regions, as listed in Table~\ref{tab:gmm}, and the respective Gaussian mixture models in Fig.~\ref{fig:gmm}. These histograms do not include foreground stars or high-velocity stars.

%-------------------------------------------------------------------

%-------------------------------------------------------------------
\section{Hot star candidates with proper motion matches in the literature}
In this section, we present the list of hot star candidates with matches in proper motion catalogues, namely \cite{2022A&A...662A..11S} in Table \ref{tab:pmhot}, \cite{2021MNRAS.500.3213L} in Table \ref{tab:pmhotlib}, and \cite{2022ApJ...939...68H} in Table \ref{tab:pmhothosek}. In Table \ref{tab:hotcand} we list all our hot star candidates. 
\begin{table*}
\caption{List of hot star candidates with proper motion measurement. }
 \label{tab:pmhot}
\begin{tabular}{@{}lrrrrrrrr@{}}
\noalign{\smallskip}
\hline
\noalign{\smallskip}
Source name&pm$_l$ &$\sigma$ pm$_{l}$& pm$_b$ & $\sigma$ pm$_{b}$ &\col&x$_{\rm offset}$&y$_{\rm offset}$ \\
&[mas$\cdot$ yr$^{-1}$]&[mas$\cdot$ yr$^{-1}$]&[mas$\cdot$ yr$^{-1}$]&[mas$\cdot$ yr$^{-1}$]&[mag]&[arcsec]&[arcsec]\\
 \noalign{\smallskip}
\hline
\noalign{\smallskip}

 F2\_26630347-29161867\_Ks12.97& -7.57976 & 0.43751 & -0.65124 & 0.37540 &1.93& -659.12 & 15.16 \\
 F2\_26636584-29096720\_Ks12.81& -4.31695 & 0.49181 & 0.43328 & 0.46925 &1.77& -356.72 & -29.87 \\
 F2\_26639297-29048702\_Ks13.22& -3.29715 & 0.44726 & -2.12941 & 0.65878 &1.97& -164.70 & -12.62 \\
 F2\_26640100-29016754\_Ks12.85& -6.95449 & 0.46801 & -3.79296 & 0.46173 &1.53& -53.37 & 25.74 \\
 F2\_26641345-29012257\_Ks13.11& 5.26893 & 0.81193 & 4.58466 & 0.65169 &1.77& -19.13 & 0.72 \\
 F2\_26640979-29010162\_Ks12.83& 2.51382 & 0.37853 & -2.41210 & 0.41530 &2.33& -18.69 & 14.48 \\
 F2\_26641458-29010031\_Ks11.83& -6.07076 & 1.10348 & -1.46760 & 0.93555 &1.74& -10.43 & 1.85 \\
 F2\_26641415-29009539\_Ks12.13& -4.05693 & 0.98551 & 3.53581 & 0.83244 &1.69& -9.63 & 3.93 \\
 F2\_26641397-29009415\_Ks10.90& -0.65542 & 0.99264 & -6.00427 & 0.89523 &1.87& -9.54 & 4.65 \\
 F2\_26641885-29006378\_Ks10.15& 3.35352 & 0.44717 & 6.32263 & 0.56343 &3.24& 7.80 & -2.77 \\
 F2\_26641107-28999369\_Ks11.50& 3.10693 & 0.37106 & -0.54336 & 0.35236 &1.88& 16.57 & 31.28 \\
 F2\_26642368-29002531\_Ks12.29& -1.91629 & 0.55541 & 0.04954 & 0.52567 &1.97& 27.54 & -8.54 \\
 F2\_26640643-28988766\_Ks11.85& 2.08185 & 0.50043 & 3.29429 & 0.45478 &1.37& 41.54 & 63.64 \\
 % F2\_26642279-28990896\_Ks11.70& -0.52487 & 0.32179 & -1.56844 & 0.52730 &2.10& 61.83 & 15.67 \\
 F2\_26643866-28966084\_Ks12.46& 0.42095 & 0.37018 & 1.33551 & 0.34400 &1.41& 164.12 & 19.53 \\
 % F2\_26649176-28895479\_Ks11.59& -4.35922 & 0.37537 & 1.85773 & 0.26630 &1.86& 468.23 & 9.01 \\
 F2\_26650085-28895227\_Ks12.15& 2.76172 & 0.38619 & -0.99954 & 0.29024 &1.42& 483.92 & -14.98 \\
 F2\_26652917-28857695\_Ks12.95& -6.01836 & 0.43309 & 3.70225 & 0.34873 &2.03& 645.75 & -20.93 \\
 F2\_26652484-28854546\_Ks12.20& -2.00436 & 0.38303 & 8.36375 & 0.41923 &1.34& 648.33 & -3.38 \\
 F2\_26653299-28843451\_Ks11.39& 5.23811 & 0.51540 & 1.14242 & 0.50952 &1.51& 695.81 & -4.55 \\
 F2\_26652979-28836376\_Ks12.48& 1.01451 & 0.33587 & 0.40466 & 0.30085 &1.42& 712.32 & 17.32 \\
 F2\_26654034-28827679\_Ks11.74& -1.56756 & 0.33822 & -0.16292 & 0.38063 &1.64& 756.36 & 5.18 \\
 F2\_26654547-28825665\_Ks11.71& 2.05280 & 0.37501 & -1.13780 & 0.65325 &1.66& 770.97 & -4.87 \\
 F2\_26654773-28826262\_Ks12.68& 2.58119 & 0.48971 & -1.26814 & 0.66280 &1.60& 772.84 & -12.07 \\
 F2\_26654877-28813469\_Ks12.39& -3.82819 & 0.68378 & 1.27034 & 0.40760 &1.63& 813.89 & 9.08 \\

\hline 
\end{tabular}
\tablefoot{The columns denote the source name, proper motion from \cite{2022A&A...662A..11S} along Galactic longitude $l$, associated uncertainty, proper motion along Galactic latitude $b$, associated uncertainty, colour \col, and the offset coordinates from \sgra, as shown in Fig.~\ref{fig:hotpm}.}
\end{table*}
%---------------------------
\begin{table*}
\caption{List of hot star candidates with proper motion measurement.}
 \label{tab:pmhotlib}
\begin{tabular}{@{}lrrrrrrrr@{}}
\noalign{\smallskip}
\hline
\noalign{\smallskip}
Source name&pm$_l$ &$\sigma$ pm$_{l}$& pm$_b$ & $\sigma$ pm$_{b}$ &\col&x$_{\rm offset}$&y$_{\rm offset}$ \\
&[mas$\cdot$ yr$^{-1}$]&[mas$\cdot$ yr$^{-1}$]&[mas$\cdot$ yr$^{-1}$]&[mas$\cdot$ yr$^{-1}$]&[mag]&[arcsec]&[arcsec]\\
 \noalign{\smallskip}
\hline
\noalign{\smallskip}

 F2\_26628195-29192692\_Ks12.37 & 5.24955 & 0.13227 & -1.19721 & 0.17863 &0.13& -789.08 & 14.95\\
 F2\_26629854-29181913\_Ks11.79 &-2.53096 & 0.19062 & -0.57642 & 0.23858 &0.10& -728.75 & -9.30\\
 F2\_26630038-29173820\_Ks11.47 &-2.44452 & 0.50339 & -1.32795 & 0.48355 &0.26& -700.89 & 0.99\\
 F2\_26630347-29161867\_Ks12.97 &-9.35104 & 0.33850 & -0.23914 & 0.39156 &1.93& -659.12 & 15.16\\
 F2\_26648660-28902523\_Ks11.00 &-1.56492 & 1.69571 & -0.06578 & 1.29770 &0.36& 438.12 & 9.70\\
 F2\_26648845-28894819\_Ks9.96 &-1.75364 & 0.67338 & -0.13754 & 0.89820 &1.15& 464.85 & 19.13\\
 F2\_26650085-28895227\_Ks12.15 &-4.50275 & 0.39608 & -1.08558 & 0.35254 &1.42& 483.92 & -14.98\\
 F2\_26649948-28885368\_Ks11.78 &-1.64393 & 1.77768 & 0.32466 & 1.20549 &1.38& 511.98 & 7.17\\
 F2\_26651227-28873440\_Ks12.64 &-2.14175 & 2.43056 & -1.47325 & 2.63909 &0.87& 569.63 & -4.91\\
\hline 
\end{tabular}
\tablefoot{ The columns denote the source name, proper motion from \cite{2021MNRAS.500.3213L} along Galactic longitude $l$, associated uncertainty, proper motion along Galactic latitude $b$, associated uncertainty, colour \col, and the offset coordinates from \sgra. Note the different proper motions of F2\_26630347-29161867\_Ks12.97 and F2\_26650085-28895227\_Ks12.15 compared to Table \ref{tab:pmhot}, which may be due to different reference frames.}
\end{table*}
 % -----------------------
 \begin{table*}
\caption{List of hot star candidates with proper motion measurement.}
 \label{tab:pmhothosek}
\begin{tabular}{@{}lrrrrrrrrr@{}}
\noalign{\smallskip}
\hline
\noalign{\smallskip}
Source name&pm$_l$ &$\sigma$ pm$_{l}$& pm$_b$ & $\sigma$ pm$_{b}$ &\col&x$_{\rm offset}$&y$_{\rm offset}$ &$P_{Q}$\\
&[mas$\cdot$ yr$^{-1}$]&[mas$\cdot$ yr$^{-1}$]&[mas$\cdot$ yr$^{-1}$]&[mas$\cdot$ yr$^{-1}$]&[mag]&[arcsec]&[arcsec]&\\
 \noalign{\smallskip}
\hline
\noalign{\smallskip}
 F2\_26653311-28833078\_Ks12.48 & -2.64131 & 0.17644 & -2.06506 & 0.18685 & 0.14 & 727.90 & 14.55&0.0\\
 F2\_26653436-28831776\_Ks99.00 & -2.22163 & 0.15722& -0.15861 & 0.16876& ...& 733.96 & 13.62&0.68\\
 F2\_26653812-28830481\_Ks13.48 &-3.25798 & 0.10660 & 0.96662 & 0.11696 &0.12 & 744.11 & 5.92&0.0\\
 F2\_26654034-28827679\_Ks11.74 &-4.26246 & 0.08474 & 0.09750 & 0.09223 &1.64 & 756.36 & 5.18&0.0\\
 F2\_26654547-28825665\_Ks11.71 &-2.56479 & 0.06036 & -0.40243 & 0.06097 &1.66 & 770.97 & -4.87&0.92\\
 F2\_26654773-28826262\_Ks12.68 &-2.48197 & 0.05071 & -0.48074 & 0.05188 &1.60 & 772.84 &-12.07&0.93\\
 F2\_26655209-28824980\_Ks10.63 &-2.63055 & 0.06036 & -0.31353 & 0.06097 &1.60 & 783.93 &-21.42&0.86\\
 F2\_26655139-28821077\_Ks12.61 &-2.56704 & 0.07745 & -0.32171 & 0.07290 &1.49 & 794.79 &-12.23&0.91\\
 F2\_26654932-28818583\_Ks10.55 &-2.57111 & 0.09681 & -0.48814 & 0.09113 &1.54 & 799.06 & -1.98&0.88\\
 F2\_26655266-28819681\_Ks11.80 &-2.56853 & 0.08713 & -0.35771 & 0.08202 &1.49 & 801.18 &-13.06&0.91\\
 F2\_26655383-28816727\_Ks8.90 &-2.51840 & 0.14313 & -0.15156 & 0.13062 &2.07 & 812.17 &-10.66&0.80\\
 F2\_26655600-28816496\_Ks9.86 &-2.73455 & 0.13345 & -0.41186 & 0.12151 &1.73 & 816.44 &-16.07&0.71\\

\hline 
\end{tabular}
\tablefoot{ The columns denote the source name, proper motion from \cite{2022ApJ...939...68H} along Galactic longitude $l$, associated uncertainty, proper motion along Galactic latitude $b$, associated uncertainty, colour \col, offset coordinates from \sgra, and Quintuplet membership probability $P_{Q}$ assigned by \cite{2022ApJ...939...68H}. Note the systematic offset to Table \ref{tab:pmhot} since the proper motions are in the absolute Gaia reference frame. The motion of Quintuplet is pm$_l$=-2.45528\,mas$\cdot$ yr$^{-1}$, pm$_b$=-0.3705\,mas$\cdot$ yr$^{-1}$.}
\end{table*}
%---------------------------

\onecolumn
\setcounter{table}{3}
\renewcommand{\thetable}{E.\arabic{table}}

\begin{longtable}{lcccccrrrr}
\caption{Sample of hot star candidates. }\\
\label{tab:hotcand} 
    % \hline   
    Id & RA & Dec & $H$& $K_S$& location & $EW_{CO}$&  $EW_{Br\gamma}$ &&FK ID\\
    \noalign{\smallskip}
\hline    
\noalign{\smallskip}
\endfirsthead
\caption{continued.}\\
\hline 
\noalign{\smallskip}
    Id & RA & Dec & $H$& $K_S$& location & $EW_{CO}$&  $EW_{Br\gamma}$ &&FK ID\\
    \noalign{\smallskip}
    \hline
     \noalign{\smallskip}
    \endhead
    \hline
    
    \endfoot
    F2\_26627740-29199083\_Ks8.90&  266.27740 & -29.199083 & 8.86 & 8.90&      fg & -2.648&   3.093 &  vis \\
  F2\_26630038-29173820\_Ks11.47&  266.30038 & -29.173820 &11.73 &11.47&      fg & -1.516&   6.634 &  vis \\
  F2\_26629854-29181913\_Ks11.79&  266.29855 & -29.181913 &11.89 &11.79&      fg &  0.116&   6.179 &  vis \\
  F2\_26628195-29192692\_Ks12.37&  266.28195 & -29.192692 &12.50 &12.37&      fg &  1.516&   1.214 &  vis \\
  F2\_26630347-29161867\_Ks12.97&  266.30347 & -29.161867 &14.90 &12.97&  GC &  3.782&   0.572 &  vis \\
  F2\_26638467-29044554\_Ks12.35&  266.38467 & -29.044554 &12.53 &12.35&      fg &  0.157&   5.541 &  vis \\
  F2\_26638434-29070555\_Ks12.57&  266.38434 & -29.070555 &12.90 &12.57&      fg &  7.292&   1.470 &  vis \\
  F2\_26638516-29043568\_Ks13.37&  266.38516 & -29.043568 &13.67 &13.37&      fg &-15.454&   7.960 &  vis \\
  F2\_26638837-29043842\_Ks11.76&  266.38837 & -29.043842 &11.90 &11.76&      fg &  7.374&   1.485 &  vis \\
  F2\_26639279-29046192\_Ks11.41&  266.39279 & -29.046192 &11.70 &11.41&      fg &  0.860&   3.724 &  vis \\
  F2\_26639297-29048702\_Ks13.22&  266.39297 & -29.048702 &15.18 &13.22&  GC &  0.591&   0.802 &  vis \\
  F2\_26639532-29033218\_Ks12.39&  266.39532 & -29.033218 &12.57 &12.39&      fg &  4.712&   0.655 &  vis \\
  F2\_26639774-29040318\_Ks11.52&  266.39774 & -29.040318 &12.98 &11.52&  GC & -1.364&   0.914 &  vis \\
  F2\_26640100-29016754\_Ks12.85&  266.40100 & -29.016754 &14.38 &12.85&  GC &  4.621&   0.363 &  vis \\
  F2\_26640826-29026239\_Ks99.00&  266.40826 & -29.026239 &... &...& unknown & -3.563&   0.482 &  vis &\tablefootmark{f}\\
  F2\_26641235-28997864\_Ks13.27&  266.41235 & -28.997864 &13.83 &13.27&      fg &  3.480&  -0.468 &  vis \\
    F2\_26640979-29010162\_Ks12.83&  266.40979 & -29.010162 &15.16 &12.83&  GC &  0.462&  -3.292 &  vis &982\tablefootmark{a} \\
  F2\_26641397-29009415\_Ks10.90&  266.41397 & -29.009415 &12.77 &10.90&  GC & -1.504&   1.968 &  vis &166\tablefootmark{a} \\
  F2\_26641415-29009539\_Ks12.13&  266.41415 & -29.009539 &13.82 &12.13&  GC &  1.172&  -0.820 &  vis &617\tablefootmark{a} \\
  F2\_26641422-29008635\_Ks12.32&  266.41422 & -29.008635 &14.45 &12.32&  GC &  1.175&  -0.242 &  vis &726\tablefootmark{a} \\
  F2\_26641434-29007421\_Ks10.76&  266.41434 & -29.007421 &12.78 &10.76&  GC & -2.008&  -6.830 &  vis &96\tablefootmark{a} \\
  F2\_26641516-29007589\_Ks10.17&  266.41516 & -29.007589 &13.27 &10.17&  GC & -2.304&  -5.273 &  vis &283\tablefootmark{a} \\
  F2\_26641681-29004972\_Ks11.25&  266.41681 & -29.004972 &12.94 &11.25&  GC &  2.690&  -0.998 &  vis &273\tablefootmark{a} \\
  F2\_26641683-29007483\_Ks10.29&  266.41684 & -29.007483 &11.97 &10.29&  GC & -0.718&  -2.148 &  vis &9\tablefootmark{a} \\
  F2\_26641721-28999716\_Ks99.00&  266.41721 & -28.999716 &14.61 &...& unknown & -0.593& -12.123 &  vis &477\tablefootmark{a}  \\
  F2\_26641742-29007648\_Ks10.61&  266.41742 & -29.007648 &12.65 &10.61&  GC & -0.878&  -4.149 &  vis &64\tablefootmark{a} \\
  F2\_26641753-29008579\_Ks10.47&  266.41754 & -29.008579 &14.00 &10.47&      bg & -0.207& -13.485 &  vis &106\tablefootmark{a} \\
  F2\_26641846-29007645\_Ks99.00&  266.41846 & -29.007645 &12.32 &...& unknown &  1.317& -21.867 &  vis &25347\tablefootmark{a} \\
  F2\_26641885-29006378\_Ks10.15&  266.41885 & -29.006378 &13.40 &10.15&  GC &  2.791& -19.654 &  vis &88\tablefootmark{a} \\
  F2\_26641458-29010031\_Ks11.83&  266.41458 & -29.010031 &13.57 &11.83&  GC &  3.817&  -1.403 &  vis  &511\tablefootmark{a} \\
    F2\_26642368-29002531\_Ks12.29&  266.42368 & -29.002531 &14.27 &12.29&  GC &  4.400&  -3.269 &  vis &663\tablefootmark{a} \\
  F2\_26642334-28992006\_Ks12.60&  266.42334 & -28.992006 &12.76 &12.60&      fg & -1.951&   3.167 &  vis &\\
  F2\_26643228-28957920\_Ks12.24&  266.43228 & -28.957920 &12.38 &12.24&      fg &  3.782&   2.027 &  vis &\\
  F2\_26644479-28980379\_Ks13.13&  266.44479 & -28.980379 &13.54 &13.13&      fg &  4.736&   1.041 &  vis &\\
  F2\_26647960-28930998\_Ks11.46&  266.47961 & -28.930998 &11.62 &11.46&      fg &  1.793&   2.996 &  vis &\\
  F2\_26648660-28902523\_Ks11.00&  266.48660 & -28.902523 &11.36 &11.00&      fg & -3.328&   3.386 &  vis &\\
  F2\_26648077-28914286\_Ks12.63&  266.48077 & -28.914286 &14.12 &12.63&  GC &  5.976&  -4.319 &  vis \\
   F2\_26648845-28894819\_Ks9.96&  266.48846 & -28.894819 &11.11 & 9.96&  GC &  0.316&   2.582 &  vis \\
  F2\_26649762-28880814\_Ks10.57&  266.49762 & -28.880814 &12.02 &10.57&  GC & -2.471&   1.080 &  vis &900175\tablefootmark{c,f} \\
  F2\_26649948-28885368\_Ks11.78&  266.49948 & -28.885368 &13.16 &11.78&  GC & -1.338&   0.361 &  vis &900191\tablefootmark{c}\\
  F2\_26650085-28895227\_Ks12.15&  266.50085 & -28.895227 &13.57 &12.15&  GC & -5.776&   1.146 &  vis \\
  F2\_26650887-28872196\_Ks10.27&  266.50888 & -28.872196 &10.39 &10.27&      fg & -1.369&   5.283 &  vis \\
  F2\_26651227-28873440\_Ks12.64&  266.51227 & -28.873440 &13.51 &12.64&      fg &  3.880&   1.051 &  vis \\
  F2\_26652258-28864065\_Ks12.51&  266.52258 & -28.864065 &13.98 &12.51&  GC & -0.810&  -1.096 &  vis \\
  F2\_26652341-28858862\_Ks12.74&  266.52341 & -28.858862 &... &12.74& unknown & -2.361&   0.679 &  vis \\
  F2\_26652466-28851357\_Ks12.48&  266.52466 & -28.851357 &12.68 &12.48&      fg &  1.515&  -0.103 &  vis \\
  F2\_26652484-28854546\_Ks12.20&  266.52484 & -28.854546 &13.55 &12.20&  GC & -3.523&  -1.918 &  vis \\
  F2\_26652979-28836376\_Ks12.48&  266.52979 & -28.836376 &13.91 &12.48&  GC &  2.021&   0.534 &  vis \\
  F2\_26653299-28843451\_Ks11.39&  266.53299 & -28.843451 &12.90 &11.39&  GC & -2.416&   2.334 &  vis \\
  F2\_26653311-28833078\_Ks12.48&  266.53311 & -28.833078 &12.62 &12.48&      fg &  0.486&   2.606 &  vis \\
  F2\_26653436-28831776\_Ks99.00&  266.53436 & -28.831776 &11.49 &...& unknown & -1.572&  -0.301 &  vis &\tablefootmark{e}\\
  F2\_26653452-28831758\_Ks10.13&  266.53452 & -28.831758 &11.45 &10.13&  GC &  0.010&  -0.631 &  vis &\tablefootmark{e}\\
  F2\_26653812-28830481\_Ks13.48&  266.53812 & -28.830481 &13.60 &13.48&      fg &  5.852&   1.235 &  vis \\
  F2\_26654034-28827679\_Ks11.74&  266.54034 & -28.827679 &13.37 &11.74&  GC & -1.281&   0.534 &  vis \\
  F2\_26654547-28825665\_Ks11.71&  266.54547 & -28.825665 &13.37 &11.71&  GC &  0.520&   1.419 &  vis \\
  F2\_26654568-28812784\_Ks11.07&  266.54568 & -28.812784 &11.40 &11.07&      fg &  4.458&  -1.472 &  vis \\
  F2\_26654773-28826262\_Ks12.68&  266.54773 & -28.826262 &14.29 &12.68&  GC &  2.741&   0.635 &  vis \\
  F2\_26654877-28813469\_Ks12.39&  266.54877 & -28.813469 &14.03 &12.39&  GC & -0.009&  -0.478 &  vis \\
  F2\_26654932-28818583\_Ks10.55&  266.54932 & -28.818583 &12.09 &10.55&  GC & -1.132&   3.815 &  vis \\
  F2\_26655139-28821077\_Ks12.61&  266.55139 & -28.821077 &14.10 &12.61&  GC & -0.204&  -0.913 &  vis \\
  F2\_26655209-28824980\_Ks10.63&  266.55209 & -28.824980 &12.23 &10.63&  GC & -0.841&   2.828 &  vis \\
  F2\_26655266-28819681\_Ks11.80&  266.55267 & -28.819681 &13.29 &11.80&  GC & -1.978&   0.044 &  vis \\
   F2\_26655383-28816727\_Ks8.90&  266.55383 & -28.816727 &10.97 & 8.90&  GC & -0.446&   4.020 &  vis \\
   F2\_26655600-28816496\_Ks9.86&  266.55600 & -28.816496 &11.60 & 9.86&  GC & -0.436&  -1.757 &  vis& \tablefootmark{d}\\
  F2\_26634747-29121044\_Ks12.26&  266.34747 & -29.121044 &12.90 &12.26&      fg &  4.126&   0.237 & CO \\
  F2\_26636584-29096720\_Ks12.81&  266.36584 & -29.096720 &14.58 &12.81&  GC &  1.818&   0.742 & CO \\
  F2\_26637106-29089252\_Ks12.33&  266.37106 & -29.089252 &13.01 &12.33&      fg & -0.302&   0.940 & CO \\
  F2\_26639999-29005692\_Ks13.06&  266.39999 & -29.005692 &14.17 &13.06&  GC &  4.264&  -1.509 & CO \\
  F2\_26640643-28988766\_Ks11.85&  266.40643 & -28.988766 &13.22 &11.85&  GC &  4.153&  -0.751 & CO \\
  F2\_26641107-28999369\_Ks11.50&  266.41107 & -28.999369 &13.38 &11.50&  GC &  4.211&   1.761 & CO \\
  F2\_26641314-29011974\_Ks13.88&  266.41315 & -29.011974 &15.60 &13.88&  GC &  4.732&  -7.683 & CO \\
  F2\_26641345-29012257\_Ks13.11&  266.41345 & -29.012257 &14.89 &13.11&  GC &  3.428&  -3.423 & CO \\
  F2\_26641806-29007921\_Ks13.00&  266.41806 & -29.007921 &14.65 &13.00&  GC &  4.555& -28.845 & CO &1387\tablefootmark{b}\\
  F2\_26641840-29005163\_Ks13.10&  266.41840 & -29.005163 &15.09 &13.10&  GC &  4.556& -21.124 & CO &1374\tablefootmark{b}\\
  F2\_26642648-29003094\_Ks11.18&  266.42648 & -29.003094 &13.57 &11.18&  GC &  4.442&  -1.398 & CO \\
  F2\_26643866-28966084\_Ks12.46&  266.43866 & -28.966084 &13.87 &12.46&  GC &  4.485&   0.693 & CO \\
  F2\_26647446-28927660\_Ks12.43&  266.47446 & -28.927660 &13.57 &12.43&  GC &  3.023&  -1.049 & CO \\
  F2\_26652917-28857695\_Ks12.95&  266.52917 & -28.857695 &14.98 &12.95&  GC &  4.597&  -1.108 & CO \\
\hline
\end{longtable}
\tablefoot{We list the name, coordinates RA and Dec, and observed $H$ and $K_s$ photometry. Our classification as GC star, foreground (fg) star, background (bg) star or unknown is based on the \col colour. The classification as hot stars was done via visual inspection of the spectra. Some of these stars have values of $\left|EW_{CO}\right|$\textgreater5\,\AA\space due to bad pixels. We also list stars with spectra where weak CO absorption is visible, but only low ($\left|EW_{CO}\right|$\textless5\,\AA.). We matched the list with the KMOS observations of \cite{2015A&A...584A...2F,2017MNRAS.464..194F,2020MNRAS.494..396F,2022MNRAS.513.5920F} and note the corresponding Id in the last column. \\
\tablefoottext{a}{Early-type star in \cite{2015A&A...584A...2F} }
\tablefoottext{b}{Late-type star in \cite{2017MNRAS.464..194F} }
\tablefoottext{c}{Early-type star in \cite{2022MNRAS.513.5920F} }
\tablefoottext{d}{qF 381 in \cite{2018A&A...618A...2C}}
\tablefoottext{e}{F2\_26653436-28831776\_Ks99.00 and F2\_26653452-28831758\_Ks10.13 may be the same star}
\tablefoottext{f}{Isolated massive star candidate in \cite{2021A&A...649A..43C}}
}

\end{appendix}

\end{document}